\documentclass[prd,aps,twocolumn,nofootinbib,floatfix,10pt]{revtex4}
\usepackage{graphicx,epsfig,mathtools}
\usepackage[usenames]{color}
\usepackage{slashed}
\usepackage{epstopdf}

\begin{document}
\title{Transverse-momentum-dependent  wave functions and Soft functions at one-loop in Large Momentum Effective Theory}

\author{Zhi-Fu Deng}
\author{Wei Wang} 
\author{Jun Zeng~\footnote{Corresponding author: zengj@sjtu.edu.cn}} 
\affiliation{ 
$^1$ INPAC, Key Laboratory for Particle Astrophysics and Cosmology (MOE),  Shanghai Key Laboratory for Particle Physics and Cosmology, School of Physics and Astronomy, Shanghai Jiao Tong University, Shanghai 200240, China
}
\date{\today}

\begin{abstract}
In large-momentum effective theory (LaMET), the transverse-momentum-dependent (TMD) light-front wave functions and  soft functions can be extracted from the simulation  of a four-quark form factor and equal-time correlation functions. In this work,  using expansion by regions we provide a one-loop proof of TMD factorization of the form factor. For the one-loop validation, we also present a detailed calculation of  ${\cal O}(\alpha_s)$ perturbative corrections to these quantities, in which we adopt a modern  technique for the calculation of TMD form factor  based  the integration by part and differential equation.  The  one-loop hard functions are then extracted. Using lattice data from Lattice Parton Collaboration on quasi-TMDWFs, we estimate the effects from the one-loop matching kernel and find that the perturbative corrections depend on the operator to define the form factor, but are less sensitive to the transverse separation. These results will be helpful to precisely extract the soft functions and TMD wave functions from the first-principle in future. 
\end{abstract}

\maketitle

\section{Introduction}
The exploration  of   underlying  structures of hadrons has always been one of the most important  frontiers in particle and nuclear physics. The one-dimensional and three-dimensional  light-front wave functions (LFWFs) are  important physical quantities describing the distributions of constituents' momentum in the hadron, and  reflect  the non-perturbative internal structure of hadrons~\cite{Lepage:1979zb, Lepage:1980fj, Brodsky:1997de}.  As for a light  Nambu-Goldstone boson, the LFWFs   also help us to understand the chiral symmetry breaking~\cite{Politzer:1982mf, Miransky:1985wzx, Alkofer:2008tt, Serna:2018dwk}.

It was firstly noticed that in an exclusive process~\cite{Lepage:1979zb, Lepage:1980fj}  the non-perturbative LFWFs for a given Fock state are required. As an inevitable  input, LFWFs also  play  an important role in theoretical analyses of $B$ meson weak decays~\cite{Li:1994iu,Keum:2000wi,Keum:2000ph,Lu:2000em,Beneke:2001ev}, which are of great values for the test of the standard model (SM) and the search for new physics beyond the SM.  The unprecedented high precision of measurements at the current and forthcoming  experimental facilities strongly  request the improvement of  the accuracy of these non-perturbative physical quantities from quantum chromodynamics (QCD). 

In 1970s, the formation of  one-dimensional hadronic wave function, namely, light-cone distribution amplitudes (LCDAs) was established  in the large momentum limit under light-front quantization~\cite{Lepage:1979zb}. Results for a few  lowest moments of the LCDAs were firstly obtained from  QCD sum rules~\cite{Shifman:1978bx, Chernyak:1983ej}, and since then many progresses have been made in extracting the moments of LCDAs in the past decades.   Despite of these progresses, a complete knowledge   of meson wave functions from the first principle is not well-established yet.

An obvious  difficulty  in  calculating LCDAs  lies in the fact that it is inherently non-perturbative and needs to be treated by  methods  such as lattice QCD (LQCD). However,   LCDAs belong to light-cone correlations of  quark/gluon field operators, and thus contain explicit time dependence,  which cannot be  directly calculated  by lattice field  theory defined in the Euclidean spacetime. In this respect, only the moments of LCDAs, namely matrix elements of the local operators,  can be performed  in the traditional   LQCD approach~\cite{Gockeler:2005jz,Braun:2006dg,Boyle:2006pw,Arthur:2010xf,Braun:2015axa,Bali:2017ude,RQCD:2019osh}. 

A  remarkable  approach  to circumvent  the above problem is proposed in Ref.~\cite{Ji:2013dva,Ji:2014gla}, which is now systematically  formulated as  large momentum effective theory  (LaMET). In LaMET, one can construct the   directly computable  hadron matrix elements with non-local operators, named as  quasi-distributions,  on the lattice. Through a perturbative matching,  the corresponding LCDAs can be accessed~\cite{Ji:2013dva, Ji:2014gla, Cichy:2018mum, Ji:2020ect}.  Many inspiring results on LCDAs were reported in recent years, and    reviews of recent developments can be found in Refs.~\cite{Cichy:2018mum,Ji:2020ect}.

Compared with LCDAs and parton distribution functions (PDFs),  transverse-momentum-dependent wave functions (TMDWFs) and TMD parton distribution functions (TMDPDFs) provide more versatile  information on the internal three-dimensional structure of hadrons, which are also relevant for observables with transverse momentum  dependent (TMD) distributions of final-state particles in high-energy experiments. For instance, TMDWFs have been applied to calculate various transition form factors such as the pion electromagnetic form factor~\cite{Li:1992nu,Efremov:1979qk}, the proton form factors ~\cite{Aznaurian:1979zz,Li:1992ce,Duncan:1979hi,Lepage:1979za} and  exclusive  $B$ decays~\cite{Li:1994iu,Li:2012nk}. Therefore, it is highly prerequisite to further investigate the three-dimensional LFWFs from the first-principle QCD.
 
A very important progress in LaMET is that TMD distributions can be accessible  through the Euclidean equal-time correlations~\cite{Ebert:2019okf,Ebert:2019tvc,Ji:2019sxk,Ji:2019ewn,Ji:2021znw}. In Ref.~\cite{Ji:2019sxk}, it has been demonstrated that the form factor of a bi-local four-quark operator, calculable on the lattice, can be factorized into TMDWFs, a universal soft factor (function) and the matching kernel through QCD factorization at large momentum transfer. A combined  analysis of quasi-TMDWFs on lattice  allows  a direct extraction of  the universal soft function and TMDWFs. Based on these proposals, Lattice determinations of the rapidity evolution anomalous dimension, namely Collins-Soper (CS) kernel~\cite{Collins:1981va}, can be found in Refs.~\cite{Shanahan:2020zxr,LatticeParton:2020uhz,Schlemmer:2021aij,Li:2021wvl,Shanahan:2021tst,LPC:2022ibr}.   

It is anticipated that in the large momentum limit, the form factor can be expressed as a convolution of TMDWFs, soft functions and a hard kernel. In this work, we aim to present a complete one-loop analysis of these quantities and provide the necessary details to the proposal in Refs.~\cite{Ji:2019sxk,Ji:2021znw}. We use the expansion by regions and provide a proof of the TMD factorization for the form factor.  In the explicit calculation factor,  we adopt a modern technique based on the integration by part (IBP). With these results, we will demonstrate the cancellation of the infrared divergences, and explicitly validate the TMD factorization. Finally, we extract the hard kernels for the form factor and quasi-TMDWFs through the factorization, which will be useful for a precision determination of TMDWFs and soft functions.  We also make use of the lattice data from Lattice Parton Collaboration on quasi-TMDWFs~\cite{LPC:2022ibr} and show that the perturbative corrections to soft functions depend on the Lorentz structures, and the magnitude can reach  the order  $(10-30)\%$.  The results are found to be less sensitive to the transverse separation, which may imply a factorized form for the quasi-TMDWFs. As a comparison, we also give a phenomenological parametrization which contains explicit dependence on   the transverse separation.

The rest  of this work is organized as follows.  In Sec.~\ref{sec:TMDWFs} we will briefly introduce the concept of TMDWFs. In Sec.~\ref{sec:one-loop_LaMET}, the one-loop perturbative results for TMDWFs, soft functions, form factors and Wilson loops will be presented in order. To regularize the rapidity divergence in TMDWFs and soft function,  the delta regulator will be used in the calculation.  Based on these results, we will explicitly validate the TMD factorization of four quark form factors and extract the short-distance hard kernel at one-loop level. In Sec.~ \ref{sec:QuasiTMDWF}, the one-loop perturbative results for quasi-TMDWFs will be presented. In Sec.~\ref{application}, we use the hard kernel and calculate the effects to extract the soft functions. We conclude this work in Sec.~\ref{conslusion}. In the appendix, we collect some details in the calculation.

\section{TMD wave functions}
\label{sec:TMDWFs}

In this section, we will follow the spirit of Ref.~\cite{Ji:2021znw} and give a self-contained description of TMDWFs. 
 
An intuitive identification of   LFWFs is the light front   correlation functions between hadron state and QCD vacuum, in which light-like gauge links extending to infinities are required to maintain the gauge invariance. This allows the identification of LF divergences as rapidity divergences, known in the literature of TMD physics. 

In high energy limit, light quarks and gluons inside a hadron move on the lightcone, and are generally  named as partons. In parton physics, light-front quantization (LFQ) is a useful formalism to handle the hadron states. It provides a Hamiltonian description of QCD similar with the diagonalized Hamiltonian in non-relativistic quantum mechanics as
\begin{equation}
	\hat P^-|\Psi_n\rangle = \frac{M^2_n}{2P^+}|\Psi_n\rangle \ ,
\end{equation}
where $ |\Psi_n\rangle $ denotes a  QCD bound state~\cite{Brodsky:1997de}, and $ P^+=(P^0+P^z)/\sqrt{2} $. The wave functions obtained in this picture can  in principle  be used to calculate all   partonic densities and correlations functions.

In the infinity momentum frame (IMF),  making an IR cut-off on the longitudinal momentum scale $k^+=\epsilon$ and taking all physics below it into renormalization constants, one can  get an effective Hilbert space and obtain an effective LF theory with trivial vacuum,
\begin{equation}
	a_i|0\rangle =0 \ .
\end{equation}
Here $|0\rangle $ is the vacuum of LFQ, and $a_i $ denotes  the annihilation operator of all kinds of possible partons.
Therefore, in the LF gauge $A^+=0$ the hadron can be expanded in terms of the superposition of all kinds of possible Fock states~\cite{Brodsky:1997de},
\begin{align}
	|P\rangle=\sum_{n=1}^{\infty} \int d\Gamma_n \psi_n(x_i,\vec{k}_{i\perp})\prod a^{\dagger}_i(x_i, \vec{k}_{i\perp})|0\rangle, 
\end{align}
where $a^\dagger_i$ is the creation operator of partons on the light-front, and the phase-space integral  takes a light-core decomposition $d\Gamma_n=\prod \frac{dk^+d^2k_\perp}{2k^+(2\pi)^3}$. The $\psi_n(x_i,\vec{k}_{i\perp})$
is the LFWF, where $x_i$ denotes the set of momentum fractions of each parton, and $k_{i\perp}$ is the corresponding transverse momentum. The summation over index $n$ sums all possible partons   of hadron state, and the multiplication over index $i$ multiplies all partons which that in the state.

With the truncation $k^+\ge \epsilon$, we can write the above expansion in form of invariant matrix elements,
\begin{align}
	\psi_n(x_i,\vec{k}_{i\perp})=\langle 0|\prod a_i(x_i,\vec{k}_{i\perp})|P\rangle \ .
\end{align}
With the inclusion of  gauge-invariance and regularizations, this invariant matrix element will become the correlator matrix elements.

To get the correlations, we   define the hardon momentum $P^{\mu}=(P^z,0,0,P^z)$. The light-cone unit vector $n^\mu=(1,0,0,-1)/\sqrt{2}$ is anti-collinear with the hardon momentum, and $\bar{n}^\mu = (1,0,0,1)/\sqrt2$ is the collinear light-cone unit vector. The covariant derivative   is $D_\mu=\partial_{\mu}-igA_{\mu}$.

For a generic notation $\phi_i$ denotes all kinds of partons including the quark fields $\psi$ and gluon fields $A^\mu$, with the index `$i$' to label the field. We introduce a gauge-invariant field $\Phi_i$ which contains gauge-link along the light-cone direction $n$, pointing to positive or negative infinity:
\begin{align}
\Phi^{\pm}_{i}(\xi)=W_{n}^{\pm}(\xi)\phi(\xi) \ ,
\end{align}
with the light-like Wilson line $W_{n}^{\pm}(\xi)$
\begin{align}
   W_{n}^{\pm}(\xi)= {\cal P}e^{ig\int_{0}^{\pm \infty} ds n\cdot A(\xi+s n)} \ ,
\end{align}
where ${\cal P}$ is a path order.  
Then the generic naive three-dimensional  LFWFs, namely TMDWFs,  are  written as
\begin{align}\label{eq:naiveLFa}
&\psi^{\pm}_{N}(x_i,\vec{b}_{i\perp},\mu)=\int \bigg(\prod^N_{i=1} d\lambda_i e^{i\lambda_i x_{i}}\bigg) \times e^{i\lambda_0 x_{0}}\nonumber \\ &\times\langle 0 | {\cal P}_N\bigg(\prod^N_{i=1}\Phi^{\pm}_{i}(\lambda_{i}n\!+\!\vec{b}_{i\perp})\bigg) \times \Phi^{\pm}_{0}(\lambda_{0}n\!+\!\vec{b}_{0\perp}) |P\rangle . 
\end{align}
In the above $\sum_{k=0}^{N}x_k=1$, $\sum_{k=0}^{N}\lambda_k=0$, and each $x_{i\ge1}$ are longitudinal momentum fractions carried by partons satisfying  $0<x_k<1$. Likewise, when $P$ has no transverse component, the transverse coordinate $\vec{b}_\perp$ can be shifted by an overall constant without any effect.

The ultraviolet and infrared divergences from these amplitudes can be regulated in dimensional regularization (DR) with modified minimal subtraction ($\overline {\rm MS}$)-scheme. However, in TMDWFs,  there is a new type of 
divergence from light-like gauge-links extending to infinities, which is called rapidity divergence. It arises from the collinear gluons radiation with the momentum fraction approaching zero but cannot be regulated by dimensional regularization. There are multiple  methods to regulate the rapidity divergence,  and an option  is the  so-called $\delta$ regulator~\cite{Echevarria:2015usa,Echevarria:2015byo}. In this scheme, the gauge-link is modified as~\cite{Ji:2021znw}
\begin{align}
&W^{\pm}_n(\xi)\rightarrow W^{\pm}_{n}(\xi)|_{\delta^-}\nonumber \\
&={\cal P}{\rm exp}\left[ig\int_0^{\pm \infty} ds n\cdot A(\xi+s n)e^{-\frac{\delta^-}{2} |s|}\right] \,,
\end{align}
where $\delta^-$ is a positive quantity  to characterize the rapidity divergence. It breaks the gauge-invariance, but the breaking  effects approach zero when $\delta^-\rightarrow 0$. The regularization for the other light-cone direction is similar with the regulator $\delta^+$. Then the TMDWFs are  written as
\begin{align}\label{eq:naivefullam}
&\psi^{\pm}_{N}(x_i,\vec{b}_{i\perp},\mu,\delta^-)=\int \prod^N_{i=1} d\lambda_i e^{i\lambda_i x_{i}} \nonumber\\
&\;\;\;\; \;\;\;\;  \times \langle 0 | {\cal P}_N\prod^N_{i=1}\Phi^{\pm}_{i}(\lambda_{i}n\!+\!\vec{b}_{i\perp};\delta^-) |P\rangle \ . 
\end{align}
where the  fields $\Phi_i$ are now defined as
\begin{align}
\Phi^{\pm}_{i}(\xi;\delta^-)= W^{\pm}_{n}(\xi)|_{\delta^-}\phi(\xi).
\end{align}

As $\delta^- \rightarrow 0$, TMDWFs and TMDPDFs diverge logarithmically, and  the remanent  finite part also depends on the rapidity regulator. Therefore, the naive TMDWFs in Eq.~(\ref{eq:naivefullam}) can not  solely absorb all  nonperturbative dynamics  in the factorization for physical observables. One must remove all divergences and rapidity regularization scheme dependencies in $\psi$, in a way similar with  removing UV divergences in physical quantities. These are accomplished with the help of soft functions to be introduced in the next section.

For a pseudoscalar pion,  the TMDWF is defined as
\begin{align}\label{eq:lc-wave-function-define}
&\psi^{\pm} \left(x, b_{\perp}, \mu, \delta^{-}\right)=\frac{1}{-if_{\pi}P^{+}} \int \frac{d (\lambda P^{+})}{2 \pi} e^{-i (x-\frac{1}{2})P^{+} \lambda}  \nonumber\\
&\times\left\langle 0\left|\overline{\Psi}_{n}^{\pm}\left(\lambda n / 2+b\right) \gamma^{+} \gamma^{5} \Psi_{n}^{\pm}(-\lambda n / 2)\right| P\right\rangle|_{\delta^{-}},
\end{align}
where $b^\mu=(0,\vec{b}_\perp,0)$ is the transverse space coordinate of the light-quark field, $f_{\pi}$ is the decay constant of  pion. The factor $\frac{1}{-if_{\pi}P^+}$ comes from a  normalization in terms of a hadronic local operator matrix element,
\begin{align}
\left\langle 0\left|\overline{\psi}\left(0\right) \gamma^{+} \gamma^{5} \psi(0)\right| \pi\right\rangle =-if_{\pi}P^+.
\end{align}
$\Psi_{n}^{\pm}(\xi)|_{\delta^{-}}$ is the field with a delta regulator $\delta^{-}$
\begin{eqnarray}
\Psi_{n}^{\pm}(\xi)|_{\delta^{-}} 
={\cal P}e^{i g \int_{0}^{\pm \infty} ds n \cdot A(\xi+sn) e^{-\frac{\delta^{-}}{2}|s|}}\psi(\xi).
\end{eqnarray}

\begin{figure}
\centering
	\includegraphics[width=0.45\textwidth]{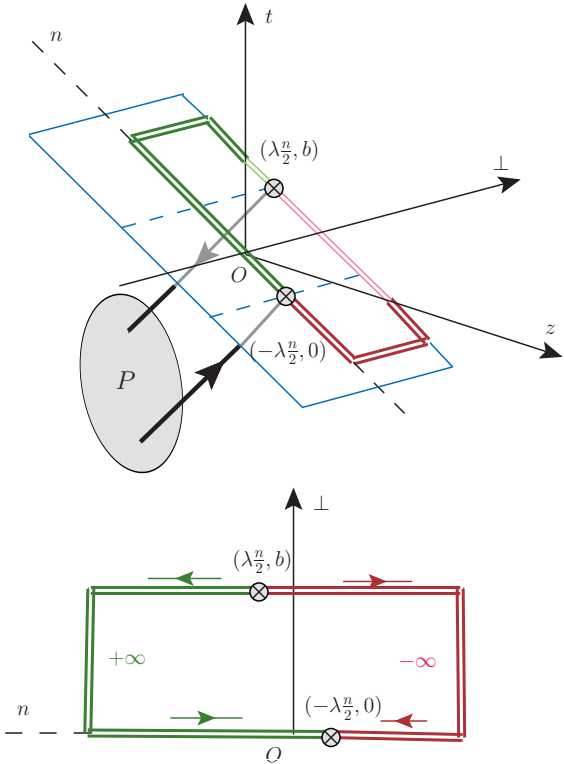}
	\caption{\label{fig:TMD_LCWF} The Wilson line structure in the TMDWF, where the red line show the Wilson line in the $-$ direction, and the green line show the Wilson line in the $+$ direction. All the Wilson lines are in the $n-\perp$ plane which show in blue, and the two end points of Wilson line are given by the position $(+,\perp)$.}
\end{figure}
 
\section{TMDWFs in LaMET}
\label{sec:one-loop_LaMET}

At present, a most  systematic approach to solve non-perturbative QCD is lattice field theory~\cite{Wilson:1974sk}. 
Though quantities on the lightcone can not be straightforwardly implemented on the lattice,   
LaMET offers a practical way to carry out the program of light-front quantization (LFQ)~\cite{Ji:2021znw}. In a certain sense, the quantization using tilted light-cone coordinates~\cite{Lenz:1991sa} is similar to the spirit of  LaMET~\cite{Ji:2021znw}. Therefore, a practical implementation of LaMET can be done through lattice calculations. While LFQ may provide an attractive physical picture for the proton, the Euclidean equal-time formulation is more practical for carrying out the calculations, and LaMET serves to bridge them.

The relation  between partonic observables on the LF and  the properties of a hadron with a large momentum is not one to one. There are infinite possible Euclidean operators in the large-momentum proton that generate the same LF
observable. This is because the large-momentum physical
states have built-in collinear (as well as soft) parton modes, and upon acting on a Euclidean operator they help to project out the leading LF physics. All operators projecting out the same LF physics form a universality class. In the
operator formulation for parton physics such as soft collinear effctive theory (SCET)~\cite{Bauer:2000yr}, one uses LF operators to project out parton physics off the external states of any momentum, including $P=0$. Concepts such as the universality class have been explored in critical phenomena in condensed matter physics, where systems with different microscopic Hamiltonians can have the same scaling properties near their critical points. Critical
phenomena correspond to the infrared fixed points of the scale transformation and are dominated by physics at long-distance scales. In this case, parton physics arises from the infinite momentum limit, which is a ultraviolet fixed point of the momentum renormalization equations (RGEs). It is the longitudinal short-distance physics that is relevant at the fixed point. However, the short distance here does not mean that everything is perturbative. The part that is nonperturbative characterizes the partonic structure of the meson. The critical region $P\to \infty$ acts as a filter to select only the physics that is relevant, so universality classes emerge. 

It has been pointed out that the soft function can be obtained from a form factor of a pseudoscalar light-meson state~\cite{Ji:2019sxk}: 
\begin{align}\label{eq:formfactor}
	F(b_\perp,P_1,P_2,\mu)= \frac{\left\langle P_2\left|\left( \bar{\psi}_a\Gamma\psi_b \right)(b) \left( \bar{\psi}_c\Gamma^{\prime}\psi_d \right)(0) \right|P_1\right\rangle}{f^2_{\pi}P_1 \cdot P_2},
\end{align}
where $\psi_{a,b,c,d}$ are light quark fields of different flavors. The factor $\frac{1}{f^2_{\pi}P_1\cdot P_2}$ comes from the normalization of two local hadronic operator matrix elements:
\begin{align}
\left\langle 0\left|\overline{\psi}\left(0\right) \gamma^{\mu} \gamma^{5} \psi(0)\right| P_1\right\rangle=-if_{\pi}P_1^{\mu},\\
\left\langle P_2\left|\overline{\psi}\left(0\right) \gamma_{\mu} \gamma^{5} \psi(0)\right| 0\right\rangle=if_{\pi}{P_2}_{\mu},
\end{align}
where $P_1^\mu=(P^z,0,0,P^z)$ and $P_2^\mu=(P^z,0,0,-P^z)$ are two  momenta which approach two opposite light-like directions in the limit $P^z\to\infty$. It should be warned that this choice of normalization  is not equivalent with the local matrix element  $\left\langle P_2\left|\left( \bar{\psi}_a\Gamma\psi_b \right) \left( \bar{\psi}_c\Gamma^{\prime}\psi_d \right)(0) \right|P_1\right\rangle$. $\Gamma$ and $\Gamma'$ are Dirac gamma matrices, which can be chosen as $\Gamma=\Gamma'=I$, $\gamma_5$ or
$\gamma_\perp$ and $\gamma_\perp\gamma_5$, so that the quark fields have leading power components
on the respective light-cones.  Here $\gamma_\perp = \gamma_x$ or $\gamma_y$. In principle, the combination $\Gamma=\sigma_{\mu\nu\perp}$ and $\Gamma'=\sigma^{\mu\nu}_\perp$ also gives the leading power contribution, but their matrix elements between the pion state vanish.

\begin{figure}
\centering
\includegraphics[width=0.45\textwidth]{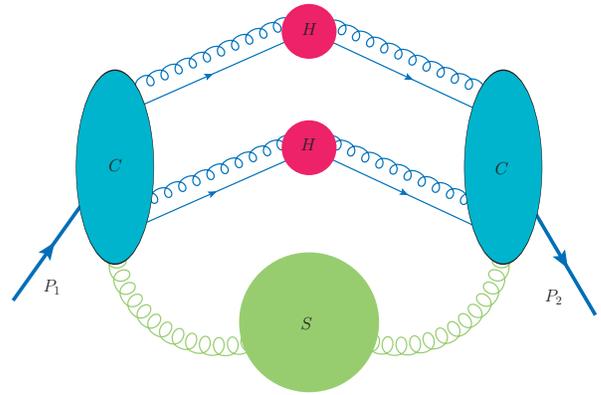}
\caption{ The leading-power reduced diagram for the large-momentum form factor $F$ of
a meson. Two $H$ denote the two hard cores separated in the transverse space by $\vec{b}_\perp$, $C$ are collinear sub-diagrams
and $S$ denotes the soft sub-diagram. It should be noted that for TMD factorization the separation of collinear and soft contribution can be achieved, however, there is no equivalence between the collinear modes and TMDWFs.  }\label{fig:reduced_form}
\end{figure}

At large momentum transfer, the form factor factorizes through TMD factorization into TMDWFs. To motivate the factorization,
one needs to consider the leading region of IR divergences in a similar way with SIDIS and Drell-Yan~\cite{Ji:2004wu,Collins:2011ca},
and the leading reduced diagram is shown in Fig.~\ref{fig:reduced_form}.
There are two collinear sub-diagrams responsible for collinear modes in $+$ and $-$ directions, and a soft sub-diagram responsible for soft
contributions. Besides, there are two IR-free hard cores localized around $(0,0,0,0)$ and $(0,\vec{b}_\perp,0)$. In the covariant gauge, there are arbitrary numbers of longitudinally-polarized collinear and soft gluons that can connect to   hard and collinear sub-diagrams, respectively. Based on the region decomposition, we now follow the standard procedure for the factorization~\cite{Collins:2011ca} .

The soft divergences can be incorporated into the soft function $S(b_\perp,\mu,\delta^+,\delta^-)$. It resums the soft gluon radiations from fast-moving color-charges. Intuitively, soft gluons have no impact on the velocity of the fast-moving color charged partons, and the propagators of partons eikonalize to straight gauge links along their moving trajectory.

For the incoming hadron, the collinear divergences are captured by the TMDWFs for the incoming parton $\psi_{\bar q q}(x,b_\perp,\mu,\delta^{'-})$. However, the naive TMDWFs contain soft divergences as well, and to avoid double counting, one must subtract the soft contribution from the bare collinear amplitude with the soft function $S(b_\perp,\mu,\delta^+,\delta^{'-})$. This leads to the collinear function for the incoming direction: $\psi_{\bar q q}(x,b_\perp,\mu,\delta^{'-})/S(b_\perp,\mu,\delta^+,\delta^{'-})$. Similarly, for the out-going direction one obtains the collinear function $\psi^{\dagger}(x',b_\perp,\mu,\delta^{'+})/S(b_\perp,\mu,\delta^{'+},\delta^{-})$.

Thus  the explicit factorization form   is conjectured as
\begin{eqnarray}\label{eq:form_fac_bare}
&&F(b_\perp,P_1,P_2,\mu)=\int dx_1 dx_2 H_F(Q^2,\bar Q^2,\mu^2)  \nonumber \\ 
&& \times \left[\frac{\psi^{\pm}_{\bar{q}q}(x_2,b_{\perp},\mu,\delta^{'+})}{\sqrt{S^{\pm}(b_{\perp},\mu,\delta^{'+},\delta^-)}}\right]^{\dagger} \left[\frac{\psi^{\pm}_{\bar{q}q}(x_1,b_{\perp},\mu,\delta^{'-})}{\sqrt{S^{\pm}(b_{\perp},\mu,\delta^{+},\delta^{'-})}}\right] \nonumber\\
&& \times \frac{S^\pm(b_\perp,\mu,\delta^+,\delta^-)}{\sqrt{S^{\pm}(b_{\perp},\mu,\delta^{'+},\delta^-)S^{\pm}(b_{\perp},\mu,\delta^{+},\delta^{'-})} }
\end{eqnarray}
Here $H_F(Q^2,\bar Q^2,\mu^2)$ is the hard kernel, $\psi^{\pm}_{\bar q q}$ is the TMDWF, $S$ is the TMD soft function, $Q^2=x_1 x_2P_1\cdot P_2$, $\bar Q^2=\bar x_1\bar x_2 P_1\cdot P_2$. An integral over the momentum fractions $x_1$,$x_2$ is assumed. 

Here we briefly comment on  the gauge-link directions in soft functions and TMDWFs. The gauge-links along the $n$ direction can be past-pointing. However, similar with the arguments in~\cite{Collins:2004nx} for the SIDIS process, based on the space-time picture of collinear divergences, one can choose future-pointing gauge-links along $n$ direction as well. With all the gauge-links being future pointing, the soft function equals to $S^-$ which is manifestly real, and the TMDWFs  for the incoming and outgoing hadrons are in  complex conjugation with each other. All rapidity regulators in TMDWFs and the soft functions are cancelled.

In the following, we will perform the one-loop perturbative calculation of TMDWFs, soft function, and form factor at the partonic level, and the results are presented in order.

\subsection{TMDWFs}
\begin{figure*}[htb!]
\centering
\includegraphics[width=0.9\textwidth]{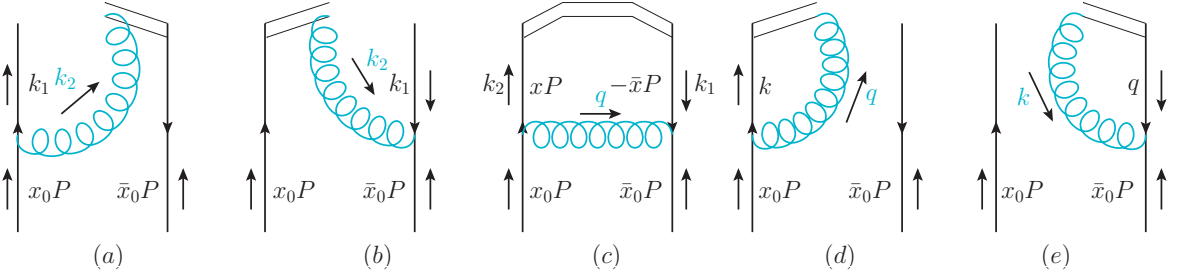}
\caption{One-loop diagrams for   TMDWFs. The meson state is replaced by  a pair of   quark and anti-quark. The first and second panels  represent the real diagrams, and the third one represents the vertex diagram. The last two are virtual diagrams.}
\label{fig:TMDWF-one-loop}
\end{figure*}

Since the short-distance coefficient is insensitive to the hadrons, in the calculation of TMDWFs one can replace the hadron by the partonic state. Therefore, we  replace the hadron state $|P\rangle$ by a pair of quark and anti-quark, and give the normalized definition on quark level:
\begin{align}\label{eq:lc-wave-function-quark-define}
&\psi_{\bar{q} q}^{\pm} \left(x, b_{\perp}, \mu, \delta^{-}\right)=\frac{1}{2P^{+}}\int \frac{d (\lambda P^+)}{2 \pi} e^{-i (x-\frac{1}{2})P^{+} \lambda}  \nonumber\\
&\times \left\langle 0\left|\overline{\Psi}_{n}^{\pm}\left(\lambda n / 2+b\right) \gamma^{+} \gamma^{5} \Psi_{n}^{\pm}(-\lambda n / 2)\right| q\bar{q}\right\rangle|_{\delta^{-}},
\end{align}
where the factor $\frac{1}{2P^{+}}$ is derived from the tree-level result of the local operator matrix element,
\begin{align}\label{eq:WFlocaloperator}
\left\langle 0\left|\overline{\psi}_{\bar q}\left(0\right) \gamma^{+} \gamma^{5} \psi_{q}(0)\right| q\bar{q}\right\rangle|_{\rm tree}=2P^{+}.
\end{align}
Here, the quark pair   is chosen to have the same $J^{PC}$ with the pion, and  the spin average with a Clebsch-Gordon coefficient and color average are assumed
in this calculation. In  Appendix~\ref{sec:Fierz_transformation}, we  provide a detailed explanation of Eq.~(\ref{eq:WFlocaloperator}), and the corresponding trace formalism to derive this convention. 
It is necessary to mention that due to the partial conservation of axial-vector current,  matrix elements in  Eq.~\eqref{eq:WFlocaloperator} are not  affected by   loop corrections. 

According to the definition of the normalized TMDWF, one can calculate directly at the tree level, 
\begin{eqnarray}
\psi_{\overline{q} q}^{\pm (0)}=\delta(x-x_0),
\end{eqnarray}
where $x_0$ is the momentum fraction of quark in initial state. Here, the spin average and color average are considered in this calculation. 

At the one-loop order, all Feynman diagrams are shown in Fig.~\ref{fig:TMDWF-one-loop}.  We choose the dimensional regularization $d=4-2\epsilon$ to regularize the UV and IR divergences. The real diagram  shown in Fig.~\ref{fig:TMDWF-one-loop}(a) can be obtained as follows:
\begin{eqnarray}
\psi_{\overline{q} q}^{\pm (1,a)}&=&\mu_{0}^{2\epsilon}\frac{i g^2C_F}{2}  \int \frac{d^dq}{(2\pi)^d} e^{-iq\cdot b} \delta\bigg[(x-x_0)P^++q^+\bigg] \nonumber\\
&&\times  \frac{\bar{v}\gamma^+\gamma^5 (x_0\slashed{P}-\slashed{q})\slashed{n} u}{(-q^+\pm i\frac{\delta}{2})[(x_0P-q)^2+i\epsilon](q^2+i\epsilon)} \nonumber\\
&=&\frac{\alpha_s C_F }{2 \pi} \frac{ \theta(x_0-x)  x}{  x_0(x-x_0\pm i\frac{\delta^{-}}{2P^{+} }) }\left(\frac{1}{\epsilon_{\rm IR}} + L_b\right),
\end{eqnarray}
The virtual diagram Fig. \ref{fig:TMDWF-one-loop} (d) gives 
\begin{eqnarray}
\psi_{\overline{q} q}^{\pm (1,d)}
&=&-\mu_{0}^{2\epsilon}\frac{i g^2C_F}{2P^+} \delta (x-x_0) \int\frac{d^dq}{(2\pi)^d}  \nonumber \\&& \times \frac{\bar{v}\gamma^+\gamma^5 (x_0\slashed{P}-\slashed{q})\slashed{n} u}{(-q^+\pm i\frac{\delta}{2})[(x_0P-q)^2+i\epsilon](q^2+i\epsilon)}  \nonumber\\
&=&\delta (x-x_0) \frac{\alpha_s C_F }{2\pi}  \times\nonumber \\&&\int_0^{x_0} d y \frac{\theta(x_0-y)y}{x_0(y-x_0\pm i\frac{\delta^{-}}{2P^+})}\left(\frac{1}{\epsilon_{\rm UV}}-\frac{1}{\epsilon_{\rm IR}}\right),\nonumber \\
\end{eqnarray}
where $L_b=\ln \frac{\mu^2 b_\perp^2}{4e^{-2\gamma_E}}$ with $b_\perp \equiv |\vec{b}_\perp|$, and $\mu=\mu_0 e^{\left(\ln(4\pi)-\gamma_E\right)/2}$ is the renormalization scale which is defined in the $\overline{\rm MS}$ scheme.  
After the UV renormalization, the remanent result can be written   in the form of plus function:
\begin{eqnarray}
\psi_{\overline{q} q}^{\pm (1,a+d)}&=&\frac{\alpha_s C_F }{2 \pi}\bigg\{\bigg[ \frac{ x \theta(x_0-x)  }{  x_0(x-x_0) }\left(\frac{1}{\epsilon_{\rm IR}} + L_b\right)\bigg]_+\nonumber\\
&& +\delta(x-x_0) \left(1+\frac{1}{2}\ln \frac{- {\delta^-}^2\mp i0}{4x^2{P^+}^2}\right)\nonumber\\
&& \times
\left(\frac{1}{\epsilon_{\rm UV}}+L_b\right)\bigg\},  \label{eq:TMDWFs_ad}
\end{eqnarray}
where   the plus function is 
\begin{eqnarray}
\bigg[g(x,x_0)\bigg]_{+}=g(x,x_0)-\delta(x-x_0)\int_0^1 dy g(y,x_0).
\end{eqnarray}
The summation of two Feynman diagrams exactly cancel out the infrared divergence in the delta function term.

The vertex diagram Fig. \ref{fig:TMDWF-one-loop} (c) gives
\begin{eqnarray}
\psi_{\overline{q} q}^{\pm (1,c)}&=&	-\frac{\alpha_s C_F}{2\pi } \Bigg(\frac{\bar{x}}{\bar{x}_0}\theta(x-x_0)+\frac{x}{x_0}\theta(x_0-x)\Bigg)\nonumber\\
&&\times\left(\frac{1}{\epsilon_{\rm{IR}}} +L_b-1\right),
\end{eqnarray}
which can be rewritten as:
\begin{eqnarray}
\psi_{\overline{q} q}^{\pm (1,c)}&=&\frac{\alpha_s C_F}{2\pi }\bigg\{\bigg[- \Bigg(\frac{\bar{x}}{\bar{x}_0}\theta(x-x_0)+\frac{x}{x_0}\theta(x_0-x)\Bigg)\nonumber\\
&&\times\left(\frac{1}{\epsilon_{\rm{IR}}} +L_b-1\right)\bigg]_+\nonumber\\
&& - \delta(x-x_0)\frac{1}{2}  \left(\frac{1}{\epsilon_{\rm{IR}}} +L_b-1\right)\bigg\}. 
\end{eqnarray}
The quark self-energy will turn the IR divergence in the second term into the UV divergence.

The other two diagrams in Fig.~\ref{fig:TMDWF-one-loop} can be obtained from Eq.~\eqref{eq:TMDWFs_ad} with the exchange $x\leftrightarrow 1-x$,  and the details are collected in the appendix~\ref{appendixTMDWF}. Summing the results in Eq.~(\ref{eq:wave-func-c}, \ref{eq:wave-func-ad}), and Eq.~(\ref{eq:wave-func-be}), one can obtain the one loop TMDWFs
\begin{align}\label{eq:unrenorTMDWF}
&\psi^\pm_{\bar{q}q} (x,b_{\perp},\mu,\delta^-)=\delta(x-x_0)+\frac{\alpha_s C_F}{2\pi}\bigg[f(x,x_0,b_{\perp},\mu)\bigg]_+ \nonumber\\
&+\frac{\alpha_s C_F}{2\pi}\delta(x-x_0)\bigg[L_b\bigg(\frac{3}{2}+\ln\frac{-{\delta^-}^2\mp i0}{4\bar{x} x  P^{+2}} \bigg) +\frac{1}{2}\bigg],
\end{align}
where 
\begin{align}
&f(x,x_0,b_{\perp},\mu)= \Bigg[\Bigg(\frac{x}{x_0(x-x_0)} -\frac{x}{x_0} \bigg)\Big(\frac{1}{\epsilon_{\rm IR}}+L_b \Big)\nonumber\\
&+\frac{x}{x_0} \Bigg]\theta(x_0-x)+\{x \to 1-x,x_0 \to 1-x_0\}. 
\label{eq:f_tmdwf}
\end{align}
All the UV divergence term have been eliminated   by composite operator renormalization. The imaginary part in Eq.~(\ref{eq:TMDWFs_ad}) comes from the contribution of the Wilson line with $\delta^-$ regulator, namely Fig.~\ref{fig:TMDWF-one-loop}(a)(b)(d)(e), which makes opposite imaginary parts for $``+"$ direction and $``-"$ direction in TMDWF.

\subsection{Soft functions and Rapidity Divergences}

With a rapidity scale $\zeta$, the rapidity divergences of the TMDWF showing in Eq.~(\ref{eq:naivefullam}) can be renormalized by the on-light-cone soft functions. The soft function is defined with two on-light-cone Wilson-line cusps explicitly as
\begin{align}\label{define_soft_function}
S^{\pm}(b_\perp,\mu,\delta^+,\delta^-)&=\frac{1}{N_c} {\rm tr } \langle 0|\mathcal{T}W_{\bar{n}}^{-\dagger}(b_{\perp})|_{\delta^{+}}W_{n}^{\pm}(b_{\perp})|_{\delta^{-}}\nonumber\\
&\times W_{n}^{\pm\dagger}(0)|_{\delta^{-}}W_{\bar{n}}^{-}(0)|_{\delta^{+}}| 0 \rangle.
\end{align}
where $W_{\bar{n}}$ is defined as
\begin{align}
&W^{\pm}_{\bar{n}}(\vec{b}_\perp)={\cal P}{\rm exp}\left[ig\int_{0}^{\pm\infty} ds \bar{n} \cdot A(s \bar{n}+\vec{b}_\perp)\right]  \ .
\end{align}
Here the subscript $\bar{n}$ give the direction of the Wilson-line, and the superscript $\pm$ in $W_n^\pm$ should be chosen the same as that of the WF amplitudes, and ${\cal T}$ gives the time-ordered product for quantum fields. 

The soft function will  be used to remove the rapidity divergence. It is interesting to notice that the soft functions can be obtained from TMDWFs  with an eikonal approximation on the incoming parton lines, which re-sum the soft-gluon radiations and suffer from rapidity divergences. To ensure the scheme independence of physical TMDWFs, one needs to introduce a square root on the soft function. Since it contains two light-like directions, one can  define the ``physical'' TMDWFs amplitudes as
\begin{align}
&\Psi^{\pm}_{\bar{q}q}(x,b_{\perp},\mu,\zeta)=\lim_{\delta^- \rightarrow 0}\frac{\psi^{\pm}_{\bar{q}q}(x,b_{\perp},\mu,\delta^-)}{\sqrt{S^{\pm}(b_{\perp},\mu,\delta^-e^{2y_n},\delta^-)}} \ ,\label{eq:physicalWF}
\end{align}
where $y_n$ is a dimensionless rapidity parameter  for the renormalized TMDWFs. The rapidity divergences cancel between the bare TMDWFs  and the soft function, which leaves a dependence of rapidity scales $\zeta$ in TMDWFs as $\zeta=2(xP^+)^2 e^{2y_n}$ with $e^{2y_n}=\delta^+/\delta^-$.

\begin{figure}[h]
\centering
	\includegraphics[width=0.5\textwidth]{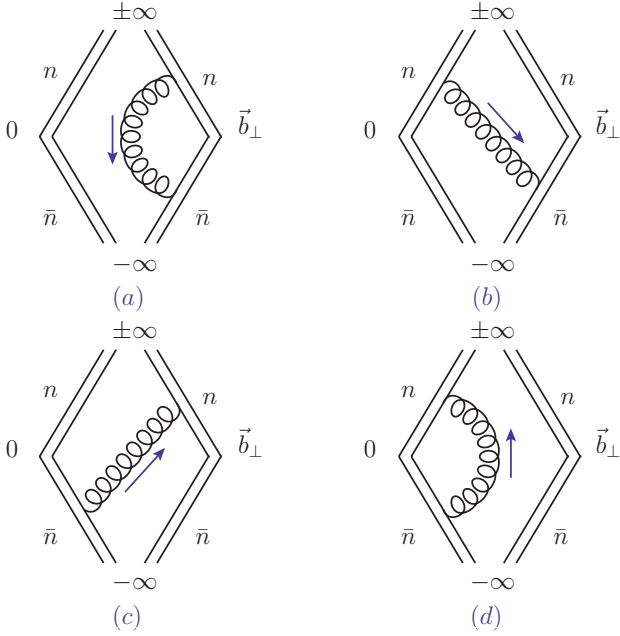}
	\caption{One-loop diagrams for the soft funtion. Diagram (a)(d) give the virtual diagram, and diagram (b)(c) give the real diagram.}
	\label{soft-function}
\end{figure}

At tree-level, the matrix element    in Eq.~(\ref{define_soft_function}) only involves a $N_c \times N_c$ unit   matrix. Therefore,   it is easy to obtain  $S^{(0)\pm}(b_\perp,\mu,\delta^+,\delta^-)=1$.

One-loop Feynman   diagrams for soft functions are shown in Fig. \ref{soft-function}.  Based on the exchange symmetry,  Fig. \ref{soft-function} (a) and Fig. \ref{soft-function} (d) give the same contributions, and similar for   Fig. \ref{soft-function} (b) and Fig. \ref{soft-function} (c). The results of  these diagrams are given as
\begin{eqnarray}
S^{(a)\pm}&=&S^{(d)\pm}\nonumber\\
&=&-\mu_{0}^{2\epsilon}ig^2C_F\int\frac{d^dq}{(2\pi)^d}\frac{\bar{n}^\mu}{q^-+i\frac{\delta^+}{2}} \frac{n_\mu}{q^+\pm i\frac{\delta^-}{2}}\frac{1}{q^2+i\epsilon}\nonumber\\
&=&\frac{\alpha_s C_F}{4\pi}\Bigg[-\frac{2}{\epsilon_{\rm{UV}}^2}+\frac{2}{\epsilon_{\rm{UV}}}\ln \frac{\mp\delta ^- \delta ^+-i0}{2\mu^2} \nonumber\\
&&- \ln^2 \bigg( \frac{\mp\delta ^- \delta ^+-i0}{2\mu^2}\bigg)-\frac{\pi ^2}{2}\Bigg],\label{eq:softfunctionad}\\
S^{(b)\pm}&=&S^{(c)\pm}\nonumber\\
&=&\mu_{0}^{2\epsilon}ig^2C_F\int\frac{d^dq}{(2\pi)^d}\frac{\bar{n}^\mu}{q^-+i\frac{\delta^+}{2}} \frac{n_\mu}{q^+\pm i\frac{\delta^-}{2}}\frac{e^{-iq \cdot b}}{q^2+i\epsilon} \nonumber\\
&=&\frac{\alpha_s C_F}{4\pi} \bigg[
L_b^2+2L_b \ln \frac{\mp\delta ^- \delta ^+-i0}{2\mu^2} \nonumber\\
&&+ \ln^2 \bigg(\frac{\mp\delta ^- \delta ^+-i0}{2\mu^2}\bigg)  +\frac{ 2\pi ^2}{3} 
\bigg].\label{eq:softfunctionbc}
\end{eqnarray} 

These results in Eq.~(\ref{eq:softfunctionad}) and Eq.~(\ref{eq:softfunctionbc}) do not contain the infrared divergence, because the   $\delta^+$ and $\delta^{-}$ act as the infrared regulators. When $q^+\to \infty$ and $q_\perp\to \infty$, the soft function in Eq.~(\ref{eq:softfunctionad}) contains a UV divergence, which is manifested as $1/\epsilon$. This is similar for the kinematic region $q^-\to \infty$ and $q_\perp\to \infty$. The overlap of the above two kinematic regions gives the $1/\epsilon^2$ divergences. For the real diagrams, the results in Eq.~(\ref{eq:softfunctionbc})  do not have UV divergence. This is due to the factor that the transverse  momentum of gluon is limited by the $1/b_{\perp}$.

After UV renormalization, the renormalized  soft functions are
\begin{eqnarray}
&& S^{\pm}(b_\perp,\mu,\delta^+,\delta^-)=1+\frac{\alpha_s C_F}{2\pi}\bigg(
L_b^2 \nonumber\\
&& \;\;\;\; \;\;\;\;  \;\;\;\;  + 2L_b\ln \frac{\mp\delta ^- \delta ^+-i0}{2\mu^2}+\frac{\pi ^2}{6}
\bigg),
\end{eqnarray}
where $S^+$ contains an imaginary part. In the $``+"$ direction, all  Feynman diagrams in Fig. \ref{soft-function} contribute the imaginary part. The $S^-$ can be obtained by changing the sign in front of   $\delta^-\delta^+$  from $S^+$. Our results are in agreement with those in the literature~\cite{Echevarria:2012js,Echevarria:2011epo}.

Combining the above results, we obtain the one-loop TMDWFs as 
\begin{eqnarray}\label{RenorWF}
&&\Psi^{\pm}_{\bar{q}q}(x,b_\perp,\mu,\zeta)=\delta(x-x_0)+\frac{\alpha_sC_F}{2\pi}[f(x,x_0,b_\perp,\mu)]_+\nonumber\\
&&+\frac{\alpha_s C_F}{2\pi}\delta(x-x_0)     \bigg\{-\frac{L_b^2}{2}+L_b\bigg( \frac{3}{2}+\ln \frac{\mu^2}{\pm\sqrt{\zeta\bar{\zeta}}-i0}\bigg)\nonumber \\
&& \;\;\; +\frac{1}{2}-\frac{\pi^2}{12}   \bigg\},
\end{eqnarray}
where $\bar \zeta=2(\bar xP^+)^2 e^{2y_n}$. The renormalized TMDWFs in Eq.~(\ref{RenorWF}) satisfies the rapidity evolution equation
\begin{eqnarray}\label{CSkernel}
2 \zeta \frac{d}{d \zeta} \ln \Psi_{\bar{q} q}^{\pm}\left(x, b_{\perp}, \mu, \zeta\right)=K_{1}\left(b_{\perp}, \mu\right).
\end{eqnarray}
Substituting Eq.~(\ref{RenorWF}) into Eq.~(\ref{CSkernel}), the one-loop Collins-Soper  kernel can be determined as
\begin{eqnarray}
K_1(b_\perp,\mu)=-\frac{\alpha_s C_F}{\pi}L_b.
\end{eqnarray}
In the above evolution equation, the CS kernel $K_{1}\left(b_{\perp}, \mu\right)$ is the same with that determined from   TMDDPFs. 

\subsection{Four-Quark  Form Factor}
\label{sec:others}

In this subsection, we aim to give a complete calculation of the four-quark form factors that can be used to validate the TMD factorization scheme. 
At the quark level, we define the four-quark form factor as
\begin{widetext}
\begin{eqnarray}\label{eq:quarkstateformfactor}
	F(b_\perp,P_1,P_2,\mu)= \frac{\left\langle\bar{q}_{d}\left(\bar{x}_{2} P_{2}\right) q_{a}\left(x_{2} P_{2}\right)\left|\left(\bar{\psi}_{a} \Gamma \psi_{b}\right)(b)  (\bar{\psi}_{c} \Gamma \psi_{d} )(0)\right| q_{b}\left(x_{1} P_{1}\right) \bar{q}_{c}(\bar{x}_{1} P_{1})\right\rangle}{4P_1\cdot P_2},
\end{eqnarray}
\end{widetext}
where the denominator is a normalization from two tree-level  matrix elements 
\begin{eqnarray} \label{eq:formal_normalization}
\left\langle 0\left|
\bar{\psi}_{c} \gamma^{\mu}\gamma^5 \psi_{b} \right| q_{b}\left(x_{1} P_{1}\right) \bar{q}_{c}(\bar{x}_{1} P_{1})\right\rangle|_{\rm tree}=2P_1^{\mu},\\
\left\langle\bar{q}_{d}\left(\bar{x}_{2} P_{2}\right) q_{a}\left(x_{2} P_{2}\right)\left|
\bar{\psi}_{a} \gamma_{\mu}\gamma^5 \psi_{d} \right|0\right\rangle|_{\rm tree}=2{P_2}_{\mu}.
\end{eqnarray}
Here, the spin average and color average are   employed. Actually the spinor calculation can be implemented with a trace formalism that is described in Appendix~\ref{sec:Fierz_transformation}. 

At tree level, the form factor can be directly evaluated as: 
\begin{widetext}
\begin{eqnarray}
F^{0} &=&-\frac{1}{4 N_{c}P_1 \cdot P_2}   \bar{u}_{a}\left(x_{2} P_{2}\right) e^{i x_{2} P_{2} \cdot b} \Gamma u_{b}\left(x_{1} P_{1}\right) e^{-i x_{1} P_{1} \cdot b} \bar{v}_{c}\left(\bar{x}_{1} P_{1}\right) \Gamma v_{d}\left(\bar{x}_{2} P_{2}\right) \nonumber\\
&=&\frac{1}{16N_c P_1 \cdot P_2}{\rm tr}\bigg[\gamma_5 \slashed{P}_2 \Gamma  \slashed{P}_1 \gamma_5 \Gamma \bigg]= \left\{ \begin{aligned}
\frac{1}{4N_c},~~~{\rm{for}}~~~~\Gamma=I~~~~~~~~~~~~~~~~~~~~~~\\
-\frac{1}{4N_c},~~~{\rm{for}}~~~~\Gamma=\gamma_5,~\gamma_\perp~\rm{or}~ \gamma_\perp \gamma_5.
\end{aligned}\right.
\end{eqnarray}
\end{widetext}

\begin{figure}[ht!]
\centering
\includegraphics[width=0.5\textwidth]{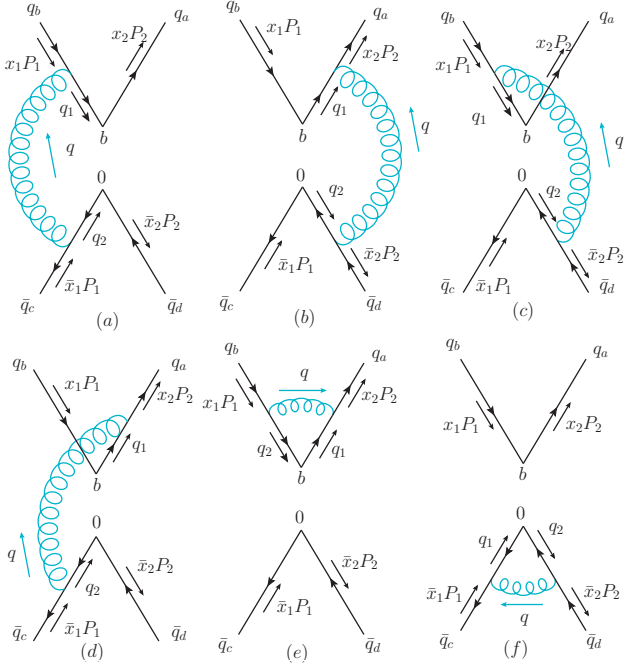}
\caption{One-loop Feynman diagrams to the form factor. The quark self-energy corrections are not shown. }
	\label{fig:form_factor}
\end{figure}

One-loop Feynman  diagrams for the form factor  are shown in Fig.~\ref{fig:form_factor}. The contribution of Fig.~\ref{fig:form_factor} ($a$) and Fig. \ref{fig:form_factor} ($b$) give the same contributions: 
\begin{eqnarray}
\label{eq:form-factor-i-ii}
&& F^{(1,a)}=\mu_{0}^{2\epsilon} \frac{ig^2C_F}{4 P_1 \cdot P_2} \int\frac{d^dq}{(2\pi)^d}e^{-iq\cdot b}  \nonumber\\
&& \;\;\; \times  \frac{1}{[(q+x_1P_1)^2+i\epsilon][(q-\bar{x}_1P_1)^2+i\epsilon](q^2+i\epsilon)}\nonumber\\
&& \;\;\; \times \bar{u}_a(x_2P_2)\Gamma   (\slashed{q}+x_1\slashed{P}_1)  \gamma_\mu u_b(x_1P_1) 
\nonumber\\
&&  \;\;\;\times  \bar{v}_c(\bar{x}_1P_1) \gamma^\mu (\slashed{q}-\bar{x}_1\slashed{P}_1)  \Gamma v_d(\bar{x}_2P_2)\nonumber\\
&&= \;\;\;-F^0\times \frac{\alpha_s C_F}{4\pi}\bigg(\frac{1}{\epsilon_{\rm {IR}}}+L_b-1\bigg).
\end{eqnarray} 
It is  interesting to notice that there is no UV divergence in the above equation.  In addition,  theses contributions  are independent of the Lorentz structure $\Gamma$  and the momentum fraction $x_1$ and $x_2$.

For  Fig. \ref{fig:form_factor} ($c$), the amplitude is given as:
\begin{eqnarray}
F^{(1,c)}&=&\frac{-\mu_{0}^{2\epsilon}ig^2C_F}{4 P_1 \cdot P_2}  \int\frac{d^dq}{(2\pi)^d}  e^{-iq\cdot b}   \nonumber\\
&&\times \frac{ \bar{u}_a(x_2P_2)\Gamma   (\slashed{q}+x_1\slashed{P}_1)  \gamma_\mu u_b(x_1P_1)}{[(q+x_1P_1)^2+i\epsilon][(q+\bar{x}_2P_2)^2+i\epsilon](q^2+i\epsilon)} \nonumber\\
&&  \times \bar{v}_c(\bar{x}_1P_1) \Gamma ({q\!\!\!\slash}+\bar{x}_2{P\!\!\!\!\slash}_2) \gamma^\mu   v_d(\bar{x}_2P_2). 
\label{eq:form_factor_iii}
\end{eqnarray} 
There is a three-point loop integral in this amplitude:
\begin{align}\label{three-pointInt}
\int\frac{d^dq}{i\pi^{d/2}}\frac{\mu_0^{2\epsilon}e^{-iq\cdot b} }{\left[(q+x_1P_1)^2+i\varepsilon\right]\left[(q+\bar{x}_2P_2)^2+i\varepsilon\right](q^2+i\varepsilon)}, 
\end{align}
which is rather difficult to evaluate in a brutal force way. 

In the past   decades, the study of mathematical properties of Feynman integrals has received increasing attention, and significant progress has been made in understanding the analytical behavior of multiloop Feynman integrals. One of the most powerful and advanced tools to evaluate the master integrals analytically is the method of differential equations (DEs) \cite{Kotikov:1990kg,Kotikov:1991pm,Remiddi:1997ny,Gehrmann:1999as,Argeri:2007up}. With the development of recent decades \cite{Henn:2013pwa,Henn:2013nsa,Argeri:2014qva,Henn:2014qga}, this method has been widely used in various processes. Ref.~\cite{Henn:2013pwa} points out that a suitable principal integral basis (canonical basis) can be chosen in a general multiloop calculation. With the canonical basis, the corresponding DEs are greatly simplified and their iterative solutions are derived in the form of the dimensional regularization parameter. In addition, the  boundary conditions of DEs can be straightforwardly determined. 

To compute the  integral in Eq.~(\ref{three-pointInt}), we consider the following one-loop triangle integral family
\begin{widetext}
\begin{eqnarray}\label{eq:MaterIntegration_1}
&&G_{\alpha_{1}, \alpha_{2}, \alpha_{3}, \alpha_{4}} \equiv \int \frac{d^{d} q}{i \pi^{d / 2}} \frac{\left(-i q \cdot b\right)^{-\alpha_{4}} e^{-i q \cdot b}}{\left(q^{2}+i \varepsilon\right)^{\alpha_{1}}\left[\left(q+\bar x_2P_{2}\right)^{2}+i \varepsilon\right]^{\alpha_{2}}\left[\left(q+x_1P_{1}\right)^{2}+i \varepsilon\right]^{\alpha_{3}}},
\end{eqnarray}
\end{widetext}
and the integral we need is $G_{1,1,1,0}$. With the integration-by-parts (IBP) technique,  one-loop QCD corrections to the real diagram of form factor are reduced into a set of integrals, named as master integrals, which are then solved using the method of DEs.  The   $G_{1,1,1,0}$  is evaluated as~\cite{Kai_Yan}:
\begin{eqnarray}
G_{1,1,1,0}&=&\frac{e^{-\epsilon \gamma_E}}{(Q'^2)^{1+\epsilon}} \bigg(-\frac{1}{\epsilon^2}+\frac{1}{2} \ln^2 \frac{Q'^2b_{\perp}^2}{4}  \nonumber \\&& +2\gamma_E \ln \frac{Q'^2b_{\perp}^2}{4} +2\gamma_E^2+\frac{\pi^2}{12} \bigg), 
\end{eqnarray}
with $Q'^2=2x_1 \bar x_2P_1\cdot P_2 $.
When the integration variable $q$ in Eq.~(\ref{eq:MaterIntegration_1}) goes to infinity, the exponential oscillation in $e^{-iq \cdot b_\perp}$ indicates the power suppression and thus there is no UV divergence in $G_{1,1,1,0}$. The divergence in the above result is infrared.

Then the contribution from  Fig.~(\ref{fig:form_factor}$c$) is evaluated as: 
\begin{widetext}
\begin{eqnarray}\label{eq:result-form-factor-iii}
F^{(1,c)}&=&-F^0\times \frac{\alpha_s C_F}{2\pi}   \bigg[-\frac{1}{\epsilon^2_{\rm{IR}}}+\frac{1}{\epsilon_{\rm{IR}}
   }\bigg(\ln \frac{4x_1 \bar{x}_2
   P^{z2}}{\mu^2}-2     \bigg)+\frac{L_b^2}{2}+L_b\bigg( \ln \frac{4x_1\bar{x}_2
   P^{z2}}{\mu^2}-2\bigg)+\frac{\pi ^2}{12}\bigg].  
\end{eqnarray}
\end{widetext}
Result for Fig.~\ref{fig:form_factor} ($d$) can be obtained with the replacement $x_1 \rightarrow -\bar{x}_1$,  $\bar{x}_2 \rightarrow -x_2$ from Eq.~(\ref{eq:result-form-factor-iii}): 
\begin{widetext}
\begin{eqnarray}\label{eq:form-factor-iv}
F^{(1,d)} 
&=&-F^0\times \frac{\alpha_s C_F}{2\pi}   \bigg[-\frac{1}{\epsilon_{\rm{IR}}^2}+\frac{1}{\epsilon_{\rm{IR}}
   }\bigg(\ln \frac{4\bar x_1
   x_2P^{z2}}{\mu^2}-2     \bigg) +\frac{L_b^2}{2}+L_b\bigg( \ln \frac{4\bar x_1
   x_2P^{z2}}{\mu^2}-2\bigg)+\frac{\pi ^2}{12}\bigg].
\end{eqnarray}
\end{widetext}

The vertex diagram Fig. \ref{fig:form_factor} ($e$) gives
\begin{align}
&& F^{(1,e)}= \mu_{0}^{2\epsilon}\frac{ig^2C_F}{4 P_1 \cdot P_2}  \int\frac{d^dq}{(2\pi)^d}   \bar{v}_c(\bar{x}_1P_1) \Gamma     v_d(\bar{x}_2P_2) \nonumber\\
&& \times  \frac{\bar{u}_a(x_2P_2)\gamma_\mu   (x_2\slashed{P}_2-\slashed{q})  \Gamma (x_1\slashed{P}_1-\slashed{q}) \gamma^\mu u_b(x_1P_1)}{[(x_2P_2-q)^2+i\epsilon][(x_1P_1-q)^2+i\epsilon](q^2+i\epsilon)}. 
\end{align} 
The result for this diagram depends on the Lorentz structure structure.  If $\Gamma=\gamma_5$ or $\Gamma=I$, we obtain
\begin{eqnarray}\label{vertexdigram}
&&F^{(1,e)}_{(1)}=-F^0\times \frac{\alpha_s C_F}{2\pi}  \bigg\{  \frac{1}{\epsilon_{\rm{IR}}^2}+\frac{1}{\epsilon_{\rm{IR}} }\ln\frac{\mu ^2}{2Q^2} \nonumber \\&&+\frac{1}{2 }\ln^2 \frac{\mu ^2}{2Q^2}-\frac{\pi ^2}{12}+1 -2\bigg( \frac{1}{\epsilon_{\rm{UV}}}-\frac{1}{\epsilon_{\rm{IR}}}\bigg)\bigg\},
\end{eqnarray}
where $Q^2=x_1x_2 P_1\cdot P_2 $. Here, we use subscript `$(1)$' to represent the results of (pseudo) scalar structures.
After absorbing the contribution from the quark self-energy
\begin{eqnarray}
Z_\psi= 1-\frac{\alpha_s C_F}{4\pi}\left(\frac{1}{\epsilon_{\rm{UV}}}-\frac{1}{\epsilon_{\rm{IR}}}\right) , 
\end{eqnarray}
one has 
\begin{eqnarray}\label{eq:form-factor-v1}
F^{e'}_{(1)} 
&=&-F^0\times \frac{\alpha_s C_F}{2\pi}  \bigg\{  \frac{1}{\epsilon_{\rm{IR}} ^2}+\frac{1}{\epsilon_{\rm{IR}} }\bigg(\ln\frac{\mu ^2}{2Q^2}+\frac{3}{2}\bigg)\nonumber \\&&+\frac{1}{2 }\ln^2 \frac{\mu ^2}{2Q^2}-\frac{\pi ^2}{12}+1-\frac{3}{2\epsilon_{\rm{UV}}} \bigg\}.
\end{eqnarray}
If $\Gamma=\gamma_\perp$ or $\Gamma=\gamma_\perp\gamma_5$, the contribution is given as
\begin{eqnarray} 
F^{e}_{(2)}&=&-F^0\times \frac{\alpha_s C_F}{2\pi}  \bigg\{ \frac{1}{\epsilon _{\rm{IR}}^2}+\frac{1}{\epsilon
   _{\rm{IR}}}\ln \frac{\mu ^2}{2Q^2}\nonumber \\&& +\frac{1}{2} \ln \frac{\mu ^2 }{2Q^2} \bigg(\ln \frac{\mu ^2 }{2Q^2}+3\bigg) -\frac{\pi ^2}{12}\nonumber\\
&&+4+\frac{3}{2\epsilon_{\rm{UV}}}-2\bigg( \frac{1}{\epsilon_{\rm{UV}}}-\frac{1}{\epsilon_{\rm{IR}}}\bigg)  \bigg\}.
\end{eqnarray}
Here, we use subscript `$(2)$' to represent the results of (pseudo) vector structures.
After absorbing the contribution from the quark self-energy, we have 
\begin{eqnarray}\label{eq:form-factor-v2}
F^{e'}_{(2)}&=&-F^0\times \frac{\alpha_s C_F}{2\pi}  \bigg\{ \frac{1}{\epsilon _{\rm{IR}}^2}+\frac{1}{\epsilon
   _{\rm{IR}}}\bigg(\ln \frac{\mu ^2}{2Q^2}+\frac{3}{2}\bigg)\nonumber \\
   &&+ \frac{1}{2} \ln \frac{\mu ^2 }{2Q^2} \bigg(\ln \frac{\mu ^2 }{2Q^2}+3\bigg)  -\frac{\pi ^2}{12}+4\bigg\}. 
\end{eqnarray}

Results for   Fig.~\ref{fig:form_factor} ($f$) can be obtained from the previous results with the replacement $x_1 \rightarrow -\bar{x}_1$ and $x_2 \rightarrow -\bar{x}_2$.  If $\Gamma=\gamma_5$ or $\Gamma=I$ we obtain
\begin{eqnarray}\label{eq:form-factor-vi1}
F^{f'}_{(1)}&=&-F^0\times \frac{\alpha_s C_F}{2\pi}  \bigg[  \frac{1}{\epsilon_{\rm{IR}} ^2}+\frac{1}{\epsilon_{\rm{IR}} }\left(\ln\frac{\mu ^2}{2\bar{Q}^2}+\frac{3}{2}\right)\nonumber \\&&+\frac{1}{2 }\ln^2 \frac{\mu ^2}{2\bar{Q}^2}-\frac{\pi ^2}{12}+1-\frac{3}{2\epsilon_{\rm{UV}}} \bigg], 
\end{eqnarray}
where $\bar{Q}^2=\bar{x}_1\bar{x}_2P_1\cdot P_2$. 
If $\Gamma=\gamma_\perp$ or $\Gamma=\gamma_\perp\gamma_5$ we obtain
\begin{eqnarray}\label{eq:form-factor-vi2}
F^{f'}_{(2)}&=&-F^0\times \frac{\alpha_s C_F}{2\pi}  \bigg[ \frac{1}{\epsilon _{\rm{IR}}^2}+\frac{1}{\epsilon
   _{\rm{IR}}}\bigg(\ln \frac{\mu ^2}{2\bar{Q}^2} +\frac{3}{2}\bigg) \nonumber \\
   &&+\frac{1}{2} \ln \frac{\mu ^2 }{2\bar{Q}^2} \bigg(\ln \frac{\mu ^2 }{2\bar{Q}^2}+3\bigg)  -\frac{\pi ^2}{12}+4\bigg]. \end{eqnarray}

Combining the above results,  we obtain the complete result for the form factor  with $\Gamma=I, \gamma_5$
\begin{eqnarray}
F(b_\perp,P_1,P_2,\mu)&=&F^0 \bigg\{ 1-\frac{\alpha_s C_F}{2\pi}  \bigg[L_b^2+L_b \bigg(\ln \frac{4 Q^2 \bar{Q}^2}{\mu^4} \nonumber\\
&&-3\bigg)+\frac{1}{2} \ln^2 \frac{2Q^2}{\mu^2} +\frac{1}{2}
   \ln^2 \frac{2\bar{Q}^2}{\mu^2} \nonumber\\
   &&+1-\frac{3}{\epsilon_{\rm{UV}}} \bigg] \bigg\}.
\end{eqnarray}
If $\Gamma=\gamma_{\perp},~\gamma_{\perp}\gamma_5$, the result is given as:
\begin{eqnarray}\label{vectorGamma}
F(b_\perp,P_1,P_2,\mu)&=&F^0 \bigg\{1- \frac{\alpha_s C_F}{2\pi}  \bigg[L_b^2+ L_b\bigg( \ln \frac{4Q^2\bar{Q}^2}{\mu^4}\nonumber\\
&&-3\bigg)-\frac{3}{2}\ln \frac{4Q^2\bar{Q}^2}{\mu^4}+\frac{1}{2} \ln^2 \frac{2Q^2 }{\mu^2}\nonumber\\
&& +\frac{1}{2}
   \ln^2 \frac{2\bar{Q}^2}{\mu^2} +7 \bigg]\bigg\}.
\end{eqnarray}

A few remarks are given in order. 
\begin{itemize}

\item The UV divergence in the $I$ and $\gamma_5$ form factor.
can be removed by the renormalization constant of scalar density  operator 
\begin{eqnarray}
Z_{S}=1+\frac{\alpha_s C_F}{4\pi} \frac{3}{\epsilon_{\rm UV}} . 
\end{eqnarray} 
Therefore, the renormalized   form factor is 
\begin{eqnarray}\label{eq:renormalized_scalar_form_factor}
F(b_\perp,P_1,P_2,\mu)&=&F^0 \bigg\{ 1-\frac{\alpha_s C_F}{2\pi}  \bigg[L_b^2+L_b \bigg(\ln \frac{4 Q^2 \bar{Q}^2}{\mu^4} \nonumber\\
&&-3\bigg)+\frac{1}{2} \ln^2 \frac{2Q^2}{\mu^2} +\frac{1}{2}
   \ln^2 \frac{2\bar{Q}^2}{\mu^2} +1 \bigg] \bigg\}. \nonumber\\
\end{eqnarray}

\item There is no UV divergence in the $\gamma_\perp$ and $\gamma_\perp\gamma_5$ form factor. 
After some simplifications, Eq.~(\ref{vectorGamma})  gives
\begin{eqnarray}
F(b_\perp,P_1,P_2,\mu)&=&F^0 \bigg[1- \frac{\alpha_s C_F}{2\pi}  \bigg( 7-\frac{3}{2}\ln  \frac{ Q^2\bar{Q}^2b_\perp^4}{4e^{-4\gamma_E}}
\nonumber\\
&&  +\frac{1}{2}\ln^2 \frac{Q^2b_\perp^2}{2e^{-2\gamma_E}} +\frac{1}{2}\ln^2 \frac{ \bar{Q}^2b_\perp^2 }{2e^{-2\gamma_E}}          \bigg) \bigg].  \nonumber
\end{eqnarray}
This is due to the fact that there is no UV divergence between the nonlocal operators, and the local ones are also free of renormalization due to the vector/axial-vector current conservation. 

\item An observation is that although there are infrared divergences in every diagram, summing all the results will cancel out all infrared divergences. As a result,  the form factor is an infrared-safe quantity at one-loop order. 

\end{itemize}

\subsection{TMD factorization for the form factor}

It has been conjectured that the form factor can be factorized  into hard, collinear, and soft functions~\cite{Ji:2019sxk,Ji:2021znw}:
\begin{eqnarray}
&&F(b_\perp,P_1,P_2,\mu)=\int dx_1dx_2 H_F(Q^2,\bar{Q}^2,\mu^2)\nonumber\\
&&\times \left[\frac{\psi^{\pm}_{\bar{q}q}(x_2,b_{\perp},\mu,\delta^{'+})}{\sqrt{S^{\pm}(b_{\perp},\mu,\delta^{'+},\delta^-)}}\right]^{\dagger}  \left[\frac{\psi^{\pm}_{\bar{q}q}(x_1,b_{\perp},\mu,\delta^{'-})}{\sqrt{S^{\pm}(b_{\perp},\mu,\delta^{+},\delta^{'-})}}\right] \nonumber\\
&& \times \frac{S^\pm(b_\perp,\mu,\delta^+,\delta^-)}{\sqrt{S^{\pm}(b_{\perp},\mu,\delta^{'+},\delta^-)S^{\pm}(b_{\perp},\mu,\delta^{+},\delta^{'-})} }. \label{eq:factorization_form_factor}
\end{eqnarray}
A rigorous  proof of the factorization  requests a thorough analysis of the behaviors of different dynamical modes, and  in particular the cancellation of collinear and soft divergences. Though the all-order analysis is not yet given in the literature, one can use the previous results and explore the factorization at ${\cal O}(\alpha_s)$.

At ${\cal O}(\alpha_s)$, the form factor does not contain any infrared divergence, as shown in Eq.~\eqref{vectorGamma} and \eqref{eq:renormalized_scalar_form_factor}.  There are infrared  divergences in the TMDWFs at ${\cal O}(\alpha_s)$, but these divergences only appear  the plus function.  To match the four factor,  one must expand the TMDWFs and soft-functions on the right-hand side of Eq.~\eqref{eq:factorization_form_factor} and integrate   over the momentum fraction. At ${\cal O}(\alpha_s)$, if the plus function term contributes in one of the two TMDWFs, the other quantities should take tree-level result. Thereby this term vanishes when one integrates over the momentum fraction since the hard kernel and soft function at tree-level are constants, and the other TMDWFs is a delta function. As a result, the infrared divergence  on the right-hand side also vanishes.  This indicates that the TMD factorization for the form-factor  is valid at ${\cal O}(\alpha_s)$.

At tree level, one has the factorization formula: 
\begin{eqnarray}
&&F^{(0)}(b_\perp,P_1,P_2) =\int dx_1dx_2 H_F^{(0)}(Q^2,\bar{Q}^2)\nonumber\\
&& \;\;\;\times\Psi^{(0)\dagger}_{\bar{q}q}(x_2,b_\perp,\mu)  \Psi^{(0)}_{\bar{q}q}(x_1,b_\perp,\mu)\nonumber\\
&& \;\;\;\times \bigg[ \frac{S(b_\perp,\mu,\delta ^+,\delta ^-)}{\sqrt{S(b_\perp,\mu,\delta '^+,\delta^-)S(b_\perp,\mu,\delta ^+,\delta '^-)}}     \bigg]^{(0)},
\end{eqnarray}
from which one can obtain:
\begin{eqnarray}
&& H_F^{(0)}=\left\{ \begin{aligned}
\frac{1}{4N_c},~~~ \Gamma=I~~~~~~~~~~~~~~~~~~~~~~\\
-\frac{1}{4N_c},~~~ \Gamma=\gamma_5,~\gamma_\perp~\rm{or}~ \gamma_\perp\gamma_5. 
\end{aligned}\right.
\end{eqnarray}
In the above  the  arguments   in $H_F$ are omitted  whenever  there is no confusion. It is interesting to note that tree level hard kernel can also be obtained by   Fierz transformation of four-quark operators, which is detailed in Appendix \ref{sec:Fierz_transformation}.

In a similar way, the one-loop form factor has the factorization expansion:
\begin{eqnarray}
&&H_F^{(1)}(Q^2,\bar{Q}^2)=F^{(1)}(b_\perp,P_1,P_2)-H_F^{(0)}(Q^2,\bar{Q}^2)\nonumber\\
&& \times \frac{\alpha_s C_F}{2\pi}\bigg[-L_b^2+L_b\bigg(3-\ln \frac{4Q^2\bar{Q}^2 }{\mu^4}\bigg) +1-\frac{\pi^2}{6}\bigg].\nonumber
\end{eqnarray}
For $\Gamma=I$ or $\Gamma= \gamma_5$, we have the hard kernel
\begin{eqnarray}
H_F(Q^2,\bar{Q}^2)&=&H_F^{(0)}\bigg[1+ \frac{\alpha_s C_F}{2\pi}\bigg(-\frac{1}{2} \ln^2 \frac{2Q^2}{\mu^2} 
\nonumber\\
&&-\frac{1}{2}
   \ln^2 \frac{2\bar{Q}^2}{\mu^2} +\frac{\pi ^2}{6}-2      \bigg)\bigg].
\end{eqnarray}
For $\Gamma=\gamma_{\perp}$ or $\Gamma=\gamma_{\perp}\gamma_5$, the hard kernel is calculated as
\begin{eqnarray}
&&H_F(Q^2,\bar{Q}^2)=H_F^{(0)}\bigg[1+ \frac{\alpha_s C_F}{2\pi}\bigg( \frac{3}{2}\ln \frac{4Q^2\bar{Q}^2}{\mu^4}\nonumber\\
&&\;\;\;\;\;\;\;\;\;\; -\frac{1}{2} \ln^2 \frac{2Q^2 }{\mu^2} -\frac{1}{2}
   \ln^2 \frac{2\bar{Q}^2}{\mu^2} +\frac{\pi ^2}{6}-8\bigg)\bigg].
\end{eqnarray}

The validation of   TMD factorization requests a full calculation of form factors, but an interesting observation on the hard kernel can be found based on the expansion by regions. This hard function arises from the exchanges of highly offshell gluons, with a typical momentum $q^\mu \sim (1, 1,1)P^z$. 
Since the  amplitudes in  Fig.~\ref{fig:form_factor} (a, b, c, d) contain an exponential factor $e^{iq_\perp\cdot b_\perp}$,   these amplitudes are suppressed if the momentum is hard since the  factor $e^{iq_\perp\cdot b_\perp}$ is highly oscillating in the region $\Lambda_{\rm QCD} \ll1/b_\perp\ll P^z$. The hard function only arises from Fig.~\ref{fig:form_factor} ($e$) and ($f$), which are identical with the vertex corrections to the scalar/pseudoscalar, and vector/axial-vector current. Then it can be found that the hard kernel $H_F(Q^2,\bar{Q}^2)$ is determined by the spacelike Sudakov form factor as follows:
\begin{eqnarray}
H_F(Q^2,\bar{Q}^2)&=&H^{Sud}(-Q^2)H^{Sud}(-\bar{Q}^2),
\end{eqnarray}
where $H^{Sud}(-Q^2)$ is as given by~\cite{Collins:2017oxh}.

\subsection{Factorization  Analysis  based on expansion by regions}

In this subsection we will adopt the expansion by regions technique and give a factorization analysis of the form factor. 
The analysis requests  the multipole  expansions of the form factor, TMDWFs, and soft function. There are a few  remarks given as follows.
\begin{itemize}
\item Contributions from  three  modes are  at leading power, which are  hard, collinear, and soft modes with the typical momentum  as:
\begin{eqnarray}
p_h^\mu &\sim& (Q, Q, Q),\nonumber\\  
p_c^\mu&\sim& (Q, \Lambda,  \Lambda^2/Q), \nonumber\\
 p_s^\mu&\sim&  (\Lambda, \Lambda,  \Lambda), 
\end{eqnarray}
with $p^\mu = (p^+,p_\perp, p^-)$,  $Q\sim P^z$, and $\Lambda \sim \Lambda_{\rm QCD}$. 
\item When the amplitudes are expanded in different regions, the lightcone divergence will show up.   However, in this analysis, the $\delta$ regulator will not show up, and thus the rapidity divergence is not properly accounted for.  A realistic proof in future must properly regularize the rapidity divergence. 
\item 
While the soft-function is homogeneously expanded, there are entangled contributions in TMDWFs. Taking Fig.~\ref{fig:TMDWF-one-loop}(a) as an example, this diagram contains contributions from both collinear and soft modes. When $q$ is soft, the amplitude is simplified as:
\begin{eqnarray}
\psi_{\overline{q} q}^{(1,a)}|_{\rm soft}&=&\mu_{0}^{2\epsilon}\frac{i g^2C_F}{2}  \int \frac{d^dq}{(2\pi)^d} e^{-iq\cdot b } \delta\bigg[(x-x_0)P^++q^+\bigg] \nonumber\\
&&\times  \frac{\bar{v}\gamma^+\gamma^5 (x_0\slashed{P}-\slashed{q})\slashed{n} u}{-q^+[(x_0P-q)^2+i\epsilon](q^2+i\epsilon)}|_{soft} \nonumber\\
&=&  \delta\bigg[(x-x_0)P^+\bigg] \mu_{0}^{2\epsilon}\frac{i g^2C_F}{2}  \int \frac{d^dq}{(2\pi)^d} e^{-iq\cdot b } \nonumber\\
&&\times  \frac{\bar{v}\gamma^+\gamma^5 x_0P^+\bar n\!\!\!\slash\slashed{n} u}{-q^+ (-2x_0P^+ q^-)(q^2+i\epsilon)} \nonumber\\
&=&  \delta(x-x_0) \frac{\bar{v}\gamma^+\gamma^5 u }{2P^+}   \nonumber\\
&&\times  \mu_{0}^{2\epsilon} i g^2C_F  \int \frac{d^dq}{(2\pi)^d} e^{-iq\cdot b }  \frac{1}{q^+   q^-(q^2+i\epsilon)}  \nonumber\\
&=& \psi_{\overline{q} q}^{(0)}\times  S^{(1, b)}. 
\end{eqnarray}
This contribution also contains a collinear contribution, and thus the 
\begin{eqnarray}
\psi_{\overline{q} q}^{(1,a)} = \psi_{\overline{q} q}^{(1,a)} |_{\rm collinear} + \psi_{\overline{q} q}^{(0)}\times  S^{(1, b)}. 
\end{eqnarray}
\end{itemize}

Now one can perform the expansion by region for the form factor in order. 
For Fig.~\ref{fig:form_factor} (a), one can directly find this diagram is related to the vertex correction to TMDWFs, and in particular only the collinear mode contributes in this diagram: 
\begin{eqnarray}\label{eq:factorization}
F^{(1,a)}  &=& H_F^{(0)} \otimes \psi_{\overline{q} q}^{  (1,c)}    \otimes  (\psi_{\overline{q} q}^{  (0)})^\dagger \times \left(\frac{1}{S}\right)^{(0)}.  
\end{eqnarray}
This can be derived   as follows. One can make the Fierz transformation of the amplitude in Eq.~\eqref{eq:form-factor-i-ii}:
\begin{eqnarray} 
&& F^{(1,a)}=\mu_{0}^{2\epsilon} \frac{ig^2C_F}{4N_c P_1 \cdot P_2} \int\frac{d^dq}{(2\pi)^d}e^{-iq\cdot b}  \nonumber\\
&& \;\;\; \times  \frac{1}{[(q+x_1P_1)^2+i\epsilon][(q-\bar{x}_1P_1)^2+i\epsilon](q^2+i\epsilon)}\nonumber\\
&& \;\;\; \times c_\Gamma \bar{u}_a(x_2P_2)\gamma_\nu \gamma_5 v_d(\bar{x}_2P_2)  \nonumber\\
&&    \;\;\; \times \bar{v}_c(\bar{x}_1P_1) \gamma^\mu (\slashed{q}-\bar{x}_1\slashed{P}_1) \gamma^\nu\gamma_5  (\slashed{q}+x_1\slashed{P}_1)  \gamma_\mu u_b(x_1P_1)\nonumber\\
&& =  \mu_{0}^{2\epsilon} \frac{ig^2C_F}{4 P_1 \cdot P_2} \int\frac{d^dq}{(2\pi)^d}e^{iq\cdot b}  \nonumber\\
&& \;\;\; \times  \frac{1}{[(q-x_1P_1)^2+i\epsilon][(q+\bar{x}_1P_1)^2+i\epsilon](q^2+i\epsilon)}\nonumber\\
&& \;\;\; \times  (-H_F^{(0)}) \bar{u}_a(x_2P_2)\gamma_\nu \gamma_5 v_d(\bar{x}_2P_2)  \nonumber\\
&&    \;\;\; \times \bar{v}_c(\bar{x}_1P_1) \gamma^\mu (\slashed{q}+\bar{x}_1\slashed{P}_1) \gamma^\nu\gamma_5  (\slashed{q}-x_1\slashed{P}_1)  \gamma_\mu u_b(x_1P_1). \nonumber\\
\end{eqnarray} 
The structure $\bar{u}_a(x_2P_2)\gamma_\nu \gamma_5 v_d(\bar{x}_2P_2)$ is evaluated as $\bar{u}_a(x_2P_2)\gamma^- \gamma_5 v_d(\bar{x}_2P_2)= 2P_2^-$, and thus the $\gamma^\mu$ in the last line of the above equation becomes $\gamma^+$. Accordingly the amplitude is written as:
\begin{eqnarray} 
&& F^{(1,a)}  =  H_F^{(0)} \mu_{0}^{2\epsilon} \frac{ig^2C_F}{2  P_1^+} \int\frac{d^dq}{(2\pi)^d}e^{iq\cdot b}  \nonumber\\
&& \;\;\; \times  \frac{1}{[(q-x_1P_1)^2+i\epsilon][(q+\bar{x}_1P_1)^2+i\epsilon](q^2+i\epsilon)}\nonumber\\
&& \;\;\; \times    \nonumber\\
&&    \;\;\; \times \bar{v}_c(\bar{x}_1P_1) \gamma^\mu (\slashed{q}+\bar{x}_1\slashed{P}_1) \gamma^\nu\gamma_5  (x_1\slashed{P}_1-\slashed{q})  \gamma_\mu u_b(x_1P_1) \nonumber\\
&&=  H_F^{(0)} \times  \int dx \psi_{\overline{q} q}^{(1,c)} (x). 
\end{eqnarray} 
Since at tree-level the soft function and the TMDWFs are trivial, the above amplitude takes a factorized form as Eq.~(\ref{eq:factorization})
with $\otimes$ denoting the convolution over the longitudinal momentum fractions.

\begin{figure}[ht!]
\centering
\includegraphics[width=0.5\textwidth]{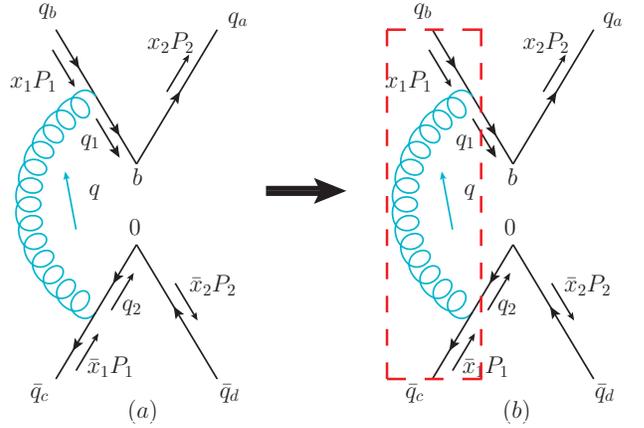}
\caption{Factorization  of form factor shown in  Fig.~\ref{fig:form_factor} (a). Only collinear mode contributes in this diagram, while both hard and soft contributions are power suppressed. }
	\label{fig:Fig_a_fact}
\end{figure}

It is similar for the conjugate diagram: 
\begin{eqnarray}
F^{(1,b)}  &=& H_F^{(0)} \otimes \psi_{\overline{q} q}^{  (0)}    \otimes  (\psi_{\overline{q} q}^{  (1,c)})^\dagger \times \left(\frac{1}{S}\right)^{(0)}. 
\end{eqnarray}

\begin{figure}[ht!]
\centering
\includegraphics[width=0.5\textwidth]{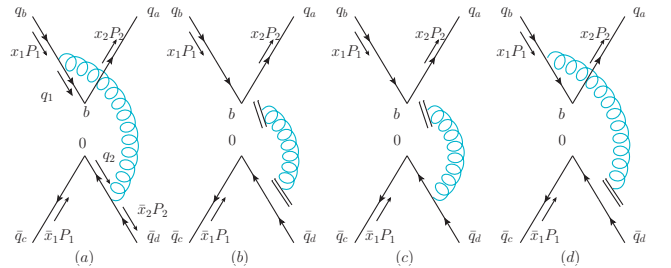}
\caption{Factorization  of form factor shown in  Fig.~\ref{fig:form_factor} (c).   The collinear, and soft modes contribute in this diagram, while the hard mode's contribution is power suppressed.    }
\label{fig:Fig_c_fact}
\end{figure}

\begin{figure}[ht!]
\centering
\includegraphics[width=0.5\textwidth]{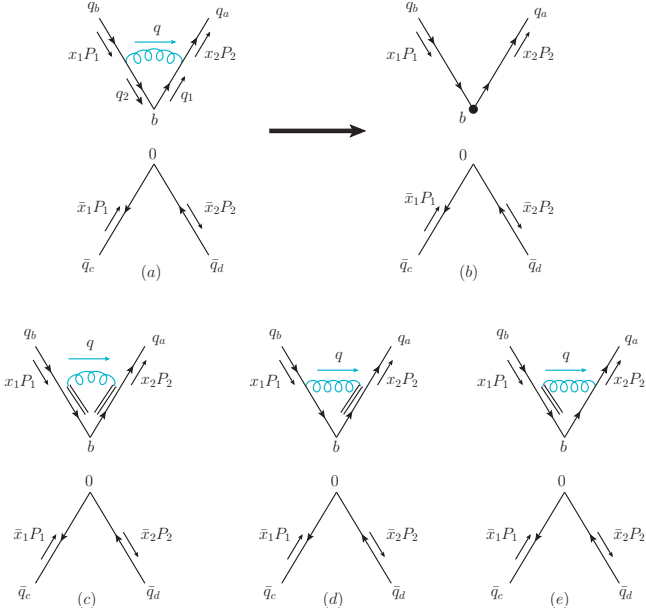}
\caption{Factorization  of the form factor shown in  Fig.~\ref{fig:form_factor} (e). The panel (b) denotes the hard kernel function due to the exchange of a hard gluon, which corresponds to a vertex correction to the vector current or scalar density operator.   }
\label{fig:Fig_e_fact}
\end{figure}

The amplitude for  Fig. \ref{fig:form_factor} ($c$) can be incorporated into three different terms and the factorized diagram is shown in Fig.~\ref{fig:Fig_c_fact}. One can firstly expand the amplitude with a soft gluon: 
\begin{eqnarray}
F^{(1,c)}|_{soft}&=&\frac{-\mu_{0}^{2\epsilon}ig^2C_F}{4N_c P_1 \cdot P_2}  \int\frac{d^dq}{(2\pi)^d}  e^{-iq\cdot b}   \nonumber\\
&&\times \frac{ \bar{u}_a(x_2P_2)\Gamma   x_1\slashed{P}_1  \gamma_\mu u_b(x_1P_1)}{[2x_1P_1^+ q^-] [2\bar{x}_2P_2^- q^+](q^2+i\epsilon)} \nonumber\\
&&  \times \bar{v}_c(\bar{x}_1P_1) \Gamma \bar{x}_2{P\!\!\!\!\slash}_2 \gamma^\mu   v_d(\bar{x}_2P_2) \nonumber\\
&=&  -\frac{1}{4N_c P_1 \cdot P_2} \bar{u}_a(x_2P_2)\Gamma    u_b(x_1P_1) \nonumber\\
&& \times \bar{v}_c(\bar{x}_1P_1) \Gamma    v_d(\bar{x}_2P_2)  \nonumber\\
&& \times  \mu_{0}^{2\epsilon}ig^2C_F \int\frac{d^dq}{(2\pi)^d}  e^{-iq\cdot b}   \nonumber\\
&&\times \frac{ 1}{  q^-  q^+ (q^2+i\epsilon)}  \nonumber\\
&=&   H_F^{(0)} \otimes \psi_{\overline{q} q}^{  (0)}   \otimes  (\psi_{\overline{q} q}^{  (0)})^\dagger  \times S^{(1,b)}.
\end{eqnarray} 

When $q$ is collinear with $P_1$, one has the amplitude:
\begin{eqnarray}
&& F^{(1,c)}|_{collinear}=\frac{-\mu_{0}^{2\epsilon}ig^2C_F}{4N_cP_1 \cdot P_2}  \int\frac{d^dq}{(2\pi)^d}  e^{-iq\cdot b}   \nonumber\\
&&\times \frac{ \bar{u}_a(x_2P_2)\Gamma   (\slashed{q}+x_1\slashed{P}_1)  2\bar x_2 P^- \gamma^+ u_b(x_1P_1)}{[(q+x_1P_1)^2+i\epsilon][2q^+\bar{x}_2P_2^-+i\epsilon](q^2+i\epsilon)} \nonumber\\
&&  \times \bar{v}_c(\bar{x}_1P_1) \Gamma   v_d(\bar{x}_2P_2) \nonumber\\
&=&\frac{-\mu_{0}^{2\epsilon}ig^2C_F}{4N_c P_1 \cdot P_2}  \int\frac{d^dq}{(2\pi)^d}  e^{-iq\cdot b}   \nonumber\\
&&\times \frac{ \bar{u}_a(x_2P_2)\Gamma   (x_1\slashed{P}_1-\slashed{q}) \gamma^+ u_b(x_1P_1)}{[(x_1P_1-q)^2+i\epsilon][-q^+](q^2+i\epsilon)} \nonumber\\
&&  \times \bar{v}_c(\bar{x}_1P_1) \Gamma   v_d(\bar{x}_2P_2).   \nonumber\\
&=& \frac{-1}{4N_cP_1\cdot P_2}   \bar{u}_a(x_2P_2)\Gamma   u_b(x_1P_1) \nonumber\\
&& \times \bar{v}_c(\bar{x}_1P_1) \Gamma   v_d(\bar{x}_2P_2) \nonumber\\
&&  \times \mu_{0}^{2\epsilon}ig^2C_F  \int\frac{d^dq}{(2\pi)^d}  e^{-iq\cdot b}   \nonumber\\
&&\times \frac{2 (x_1P_1-{q})^+ }{[(x_1P_1-q)^2+i\epsilon][-q^+](q^2+i\epsilon)}  \nonumber\\
&& =  H_F^{(0)} \otimes \psi_{\overline{q} q}^{  (1,a)} |_{\rm collinear}   \otimes  (\psi_{\overline{q} q}^{  (0)})^\dagger \times \left(\frac{1}{S}\right)^{(0)}. 
\end{eqnarray}  
This is similar with the factorization when $q$ collinear with $P_2$. 
Then this diagram  is factorized as:
\begin{eqnarray}
F^{(1,c)}  &=& H_F^{(0)} \otimes \psi_{\overline{q} q}^{  (1,a)} |_{\rm collinear}   \otimes  (\psi_{\overline{q} q}^{  (0)})^\dagger \times\left(\frac{1}{S}\right)^{(0)}  \nonumber\\
&& + H_F^{(0)} \otimes \psi_{\overline{q} q}^{  (0)}   \otimes  (\psi_{\overline{q} q}^{  (1,a)})^\dagger |_{\rm collinear} \times \left(\frac{1}{S}\right)^{(0)} \nonumber\\
&& +  H_F^{(0)} \otimes \psi_{\overline{q} q}^{  (0)}   \otimes  (\psi_{\overline{q} q}^{  0)})^\dagger |_{\rm collinear} \times S^{(1,b)}    \nonumber\\
&=& H_F^{(0)} \otimes \psi_{\overline{q} q}^{  (1,a)}    \otimes  (\psi_{\overline{q} q}^{  (0)})^\dagger\times  \left(\frac{1}{S}\right)^{(0)}  \nonumber\\
&& + H_F^{(0)} \otimes \psi_{\overline{q} q}^{  (0)}   \otimes  (\psi_{\overline{q} q}^{  (1,a)})^\dagger\times   \left(\frac{1}{S}\right)^{(0)}\nonumber\\
&& +  H_F^{(0)} \otimes \psi_{\overline{q} q}^{  (0)}   \otimes  (\psi_{\overline{q} q}^{  0)})^\dagger  \times \left(\frac{1}{ S}\right)^{(1,b)},
\end{eqnarray}
where we adopted the convention for the perturbative expansion:  $( {1}/{ S})^{(1,b)}=-  { S^{(1,b)}}$. 

This factorization scheme is similar for the  amplitude from  Fig.~\ref{fig:form_factor} ($d$): 
\begin{eqnarray}
F^{(1,d)}   &=& H_F^{(0)} \otimes \psi_{\overline{q} q}^{  (1,b)}    \otimes  (\psi_{\overline{q} q}^{  (0)})^\dagger \times   \left(\frac{1}{S}\right)^{(0)} \nonumber\\
&& + H_F^{(0)} \otimes \psi_{\overline{q} q}^{  (0)}   \otimes  (\psi_{\overline{q} q}^{  (1,b)})^\dagger \times   \left(\frac{1}{S}\right)^{(0)}\nonumber\\
&& +  H_F^{(0)} \otimes \psi_{\overline{q} q}^{  (0)}   \otimes  (\psi_{\overline{q} q}^{  0)})^\dagger    \times \left(\frac{1}{ S}\right)^{(1,c)}. 
\end{eqnarray}

For Fig.~\ref{fig:form_factor} ($e$), there are leading power contribution from the hard modes in which the full amplitude is perturbative. Contributions from the other three modes are similar with Fig.~\ref{fig:form_factor} ($c$). A sketch of the factorization is shown in Fig.~\ref{fig:Fig_e_fact}, and  the factorization formula is given as: 
\begin{eqnarray}
F^{(1,e)}   &=& H_F^{(1,e)} \otimes \psi_{\overline{q} q}^{  (0)}    \otimes  (\psi_{\overline{q} q}^{  (0)})^\dagger\times  \left(\frac{1}{S}\right)^{(0)}\nonumber\\
&& + H_F^{(0)} \otimes \psi_{\overline{q} q}^{  (1,d)}    \otimes  (\psi_{\overline{q} q}^{  (0)})^\dagger\times   \left(\frac{1}{S}\right)^{(0)}\nonumber\\
&& + H_F^{(0)} \otimes \psi_{\overline{q} q}^{  (0)}   \otimes  (\psi_{\overline{q} q}^{  (1,d)})^\dagger  \times   \left(\frac{1}{S}\right)^{(0)} \nonumber\\
&&+  H_F^{(0)} \otimes \psi_{\overline{q} q}^{  (0)}   \otimes  (\psi_{\overline{q} q}^{  0)})^\dagger    \times \left(\frac{1}{ S}\right)^{(1,d)}. 
\end{eqnarray}
This is similar with the factorization of Fig.~\ref{fig:form_factor} ($f$): 
\begin{eqnarray}
F^{(1,f)}   &=& H_F^{(1,f)} \otimes \psi_{\overline{q} q}^{  (0)}    \otimes  (\psi_{\overline{q} q}^{  (0)})^\dagger\times\left(\frac{1}{S}\right)^{(0)} \nonumber\\
&& + H_F^{(0)} \otimes \psi_{\overline{q} q}^{  (1,e)}    \otimes  (\psi_{\overline{q} q}^{  (0)})^\dagger\times   \left(\frac{1}{S}\right)^{(0)}\nonumber\\
&& + H_F^{(0)} \otimes \psi_{\overline{q} q}^{  (0)}   \otimes  (\psi_{\overline{q} q}^{  (1,e)})^\dagger  \times   \left(\frac{1}{S}\right)^{(0)} \nonumber\\
&&+  H_F^{(0)} \otimes \psi_{\overline{q} q}^{  (0)}   \otimes  (\psi_{\overline{q} q}^{  0)})^\dagger    \times \left(\frac{1}{ S}\right)^{(1,a)}. 
\end{eqnarray}

In total, the  TMD factorization of  the form factor at one-loop level is derived as:
\begin{eqnarray}
F = H_F \otimes   \psi_{\overline{q} q} \otimes  (\psi_{\overline{q} q})^\dagger \times  \frac{1}{S}, 
\end{eqnarray}
with the  perturbative kernel up to ${\cal O}(\alpha_s)$: 
\begin{eqnarray}
H_F = H_F^{(0)}+ H_F^{(1,e)} + H_F^{(1,f)}. 
\end{eqnarray}

\section{quasi-TMDWFs}\label{sec:QuasiTMDWF}

The definition of quasi-TMDWFs is similar with the TMDWF, but one should replace the Lorentz structure $\gamma^+\gamma_5$ with $\gamma^z\gamma_5$, and the Wilson line in quasi-TMDWF along  with the $z$ direction: 
\begin{align}\label{eq:hardron-quasiTMDWF}
&\tilde{\Psi}^{\pm}\left(x, b_{\perp}, \mu, \zeta^{z}\right)=\lim _{L \rightarrow \infty}\frac{1}{-if_{\pi}} \int \frac{d \lambda}{2 \pi} e^{-i (x-\frac{1}{2}) (-P^{z}) \lambda} \nonumber\\
&\times\frac{\left\langle 0\left|\overline{\Psi}_{\mp n_{z}}\left(\frac{\lambda n_{z}}{2}+b\right) \gamma^{z} \gamma^{5} \Psi_{\mp n_{z}}\left(-\frac{\lambda n_{z}}{2}\right)\right|P\right\rangle}{\sqrt{Z_{E}\left(2 L, b_{\perp}, \mu\right)}},
\end{align}
where $\zeta^z=(2xP\cdot n_z)^2$ with $n_z^{\mu}=(0,0,0,1)$. $\Psi_{\mp n_{z}}(\xi)$is the field with finite Wilson line 
\begin{eqnarray}\label{eq:quasi-TMDWF_Wilson_line}
\Psi_{\mp n_{z}}(\xi)=\mathcal{P} e^{i g \int_{0}^{\mp L + \xi \cdot n_z} ds  n_{z}\cdot A\left(\xi+sn_{z}\right)} \psi(\xi),
\end{eqnarray}
and $Z_E$ is the Wilson loop which is defined as
\begin{eqnarray}\label{wilsonloop}
Z_{E}\left(2 L, b_{\perp}, \mu\right)&=&\frac{1}{N_c}{\rm tr }\langle 0|\mathcal{T}W(\mathcal{C})| 0 \rangle,
\end{eqnarray}

Before presenting the result, we should mention that it is also feasible to  choose the $\gamma^{t} \gamma^{5}$ instead of $\gamma^{z} \gamma^{5}$ in Eq.~\eqref{eq:hardron-quasiTMDWF}. But actually,  there is no  difference in these two Lorentz structures  up to one-loop, and the  verification of this behavior  is given  in  Appendix~\ref{Appendix:threshold expansion}.

\begin{figure}[h]
	\centering
	\includegraphics[width=0.4\textwidth]{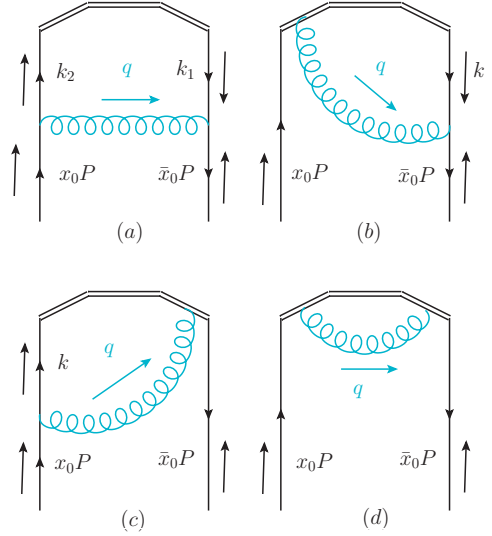}
	\caption{One-loop diagrams for the quasi TMDWF.}
	\label{fig:quasi-WF}
\end{figure}

\begin{figure}[h]
\centering
\includegraphics[width=0.4\textwidth]{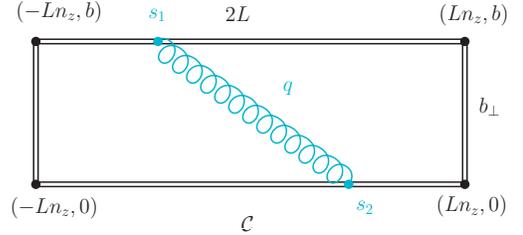}
\caption{One-loop diagrams for the Wilson loop. The coordinates $s_1$ and $s_2$ of gluon propagators are at any point on the Wilson loop. This includes self-energy, real corrections, and virtual corrections.}
\label{fig:Wilson-loop}
\end{figure}

\subsection{One-Loop Calculation}

The calculation of quasi-TMDWFs requests to replace the hadron by a couple of quark and anti-quark state,
\begin{eqnarray}\label{quasiTMDWF}
&\tilde{\Psi}_{q \overline{q}}^{\pm}\left(x, b_{\perp}, \mu, \zeta^{z}\right)=\lim _{L \rightarrow \infty} \int \frac{d \lambda}{4 \pi} e^{-i (x-\frac{1}{2}) (-P^{z}) \lambda} \nonumber\\
&\times\frac{\left\langle 0\left|\overline{\Psi}_{\mp n_{z}}\left(\frac{\lambda n_{z}}{2}+b\right) \gamma^{z} \gamma^{5} \Psi_{\mp n_{z}}\left(-\frac{\lambda n_{z}}{2}\right)\right|q \overline{q}\right\rangle}{\sqrt{Z_{E}\left(2 L, b_{\perp}, \mu\right)}}.
\end{eqnarray}

At tree level, the quasi TMDWF is also a delta function, while the   one-loop Feynman diagrams are shown in Fig.~\ref{fig:quasi-WF}.  The route $\mathcal{C}$ of the Wilson line $W(\mathcal{C})=\mathcal{P}e^{ig\int_{\mathcal{C}} ds_\mu \cdot A^\mu(s)}$ is shown in Fig.~\ref{fig:Wilson-loop}.  
The detailed calculation of these contributions is given in Appendix~\ref{Appendix:quasi-TMDWF}. In the following, we take  Fig.~\ref{fig:quasi-WF}(b,c)  as an example to illustrate the calculation. In dimensional regularization $d=4-2\epsilon$, contributions from Fig.~\ref{fig:quasi-WF} (b) and Fig.~\ref{fig:quasi-WF} (c)  are given  as:
\begin{eqnarray}
&&\tilde{\psi}_{\overline{q} q}^{\pm (1,b+c)}\left(x, b_{\perp}, \mu, \zeta^{z}\right)
=-\mu_{0}^{2\epsilon}\frac{g^2C_F}{2} \int \frac{d \lambda}{2 \pi} e^{i (x-1)P^{z} \lambda} \nonumber \\
&&\times \int_0^1 ds \mathcal{L}_\mu^{\prime}(s) \bigg[ e^{-i \bar{x}_0 \lambda P\cdot  n_{z}}\int \frac{d^d k}{(2\pi)^d}\bar v \gamma^{z} \gamma^{5}\frac{e^{-i (x_0P-k) \cdot \mathcal{L}(s)}}{(x_0P-k)^2+i\epsilon}   \nonumber \\
&&\times \frac{\slashed{k}}{k^2+i\epsilon} \gamma^\mu u +\int \frac{d^d k}{(2\pi)^d}\bar v \gamma^\mu \frac{e^{-i (\bar{x}_0P+k) \cdot \mathcal{L}(s)}}{(\bar{x}_0P+k)^2+i\epsilon}  \frac{\slashed{k}}{k^2+i\epsilon} \nonumber \\
&&\times  e^{-i k \cdot(-\lambda n_{z}-b)} \gamma^{z} \gamma^{5}u\bigg],
\end{eqnarray}
where $\mathcal{L}=\sum_{i=1}^{3}\mathcal{L}_i$ is the Wilson line of quasi-TMDWF: 
\begin{eqnarray}
	\mathcal{L}_1=\left( 
	\begin{array}{ccc} 
		0\\
		\vec{0}_\perp\\
		\mp Ls \\ 
	\end{array}
	\right),~~ 
	\mathcal{L}_2=\left( 
	\begin{array}{ccc} 
		0\\
		s\vec{b}_\perp\\
		\mp L \\ 
	\end{array}
	\right),\nonumber\\
	\mathcal{L}_3=\left( 
	\begin{array}{ccc} 
		0\\
		\vec{b}_\perp\\
		\mp L+(\lambda\pm L)s\\ 
	\end{array}
	\right). \nonumber
\end{eqnarray}
In the large $L$ limit, we have
\begin{eqnarray}
&&\tilde{\psi}_{\overline{q} q}^{\pm (1,b+c)}\left(x, b_{\perp}, \mu, \zeta^{z}\right)\nonumber\\
&=&\frac{\alpha_s C_F}{2\pi }\bigg\{\frac{1}{(x-x_0)P^{z}}\bigg[\frac{x}{x_0} \bigg(\frac{1}{\epsilon_{IR}}+L_b\bigg)\theta(x_0-x)\theta(x) \bigg]_+\nonumber\\
&&+\frac{1}{2}\delta(x_0-x)\left(\frac{1}{\epsilon_{\rm UV}}+L_b\right)\nonumber\\
&&+\frac{\delta(x-x_0)}{4\pi^{1/2} P^{z} }   \bigg[G_{3,5}^{3,3}\left(-\frac{1}{4} (b_\perp P^{z} x_0)^2\pm i0\bigg|
\begin{array}{c}
 \frac{1}{2},1,1 \\
 \frac{1}{2},1,1,-\frac{1}{2},0 \\
\end{array}
\right)\nonumber\\
&&-2 G_{3,5}^{3,3}\left(-\frac{1}{4} (b_\perp P^{z} x_0)^2\pm i0\bigg|
\begin{array}{c}
 1,1,1 \\
 \frac{1}{2},1,1,0,0 \\
\end{array}
\right)\bigg]\nonumber\\
&&+\{x_0\to \bar x_0,x \to \bar x\}\bigg]\bigg\},  
\end{eqnarray}
where the $G$ is the Meijer G-function~\cite{Bateman:1953htf}, and the different sign in $\pm i0$ comes from the  different Wilson line directions   in Eq.~(\ref{eq:quasi-TMDWF_Wilson_line}). We should note that the third term in the above is independent of the renormalization scale.

In the large $P^z$ limit, this diagram gives 
\begin{eqnarray}
&&\tilde{\psi}_{\overline{q} q}^{\pm (1,b+c)}
=\frac{\alpha_s C_F}{2\pi }\bigg\{\frac{1}{x-x_0}\bigg[\frac{x}{x_0} \bigg(\frac{1}{\epsilon_{\rm IR}}+L_b\bigg)\theta(x_0-x)\theta(x) \bigg]_+\nonumber\\
&&+\frac{1}{2}\delta(x_0-x)\left(\frac{1}{\epsilon_{\rm UV}}+L_b\right) -\frac{\delta(x-x_0)}{2  }\Bigg[\frac{L_b^2}{2}\nonumber\\
&&+\left(\ln\frac{-\zeta^z \pm i0}{\mu^2}-1\right)L_b+\frac{1}{2}\Bigg(\ln^2\frac{-\zeta^z \pm i0}{\mu^2}\nonumber\\
&&-2\ln\frac{-\zeta^z \pm i0}{\mu^2}+4\Bigg)+\frac{\pi^2}{2}\Bigg]+\{x_0\to \bar x_0,x \to \bar x\}\bigg\}.
\end{eqnarray}

Summing all contributions  in Eq.~(\ref{eq:quasi-wave-func-sail}, \ref{eq:quasi-wave-func-vertex},\ref{eq:quasi-wave-func-selfenergy}) and Wilson loop  in  Eq.~(\ref{eq:wilson-loop}), we obtain the one-loop renormalized  quasi-TMDWFs:
\begin{eqnarray}
&&\tilde{\Psi}_{q \overline{q}}^{\pm }(x,b_\perp,\mu,\zeta^z)=\delta(x-x_0)+ \frac{\alpha_s C_F}{2\pi}[f(x,x_0,b_{\perp},\mu)]_{+} \nonumber \\
&&\;\;\;\;\;\;\;\;\;\;\;\;\;\;\;\;+\frac{\alpha_s C_F}{2\pi}\delta(x-x_0)A^\pm\left(x,\mu,\zeta^z,\bar \zeta^z\right),
\end{eqnarray}
where
\begin{eqnarray}
A^\pm\left(x, \mu,\zeta^z,\bar \zeta^z\right)&=&-\frac{L_b^2}{2}+ \frac{5}{2} L_b-\frac{3}{2}-\frac{\pi^2}{2}\nonumber \\
&&+\bigg[-\frac{1}{4} \ln ^2\frac{-\zeta^z\pm i0}{\mu
   ^2}\nonumber \\
&&+\frac{1}{2}(1-L_b) \ln \frac{-\zeta^z \pm i0}{\mu ^2}\nonumber \\
&& +\{\zeta^z \rightarrow \bar\zeta^z\}\bigg], 
\end{eqnarray}
with $\bar{\zeta}^z=(2\bar{x}P\cdot n_z)^2$. The imaginary part comes from the  gluon exchange between the quark field and the Wilson line, namely Fig.~(\ref{fig:quasi-WF}b, c). 

With the above results, one can  match the quasi-TMDWFs to the TMDWF as
\begin{eqnarray}
&& \widetilde{\Psi}_{\bar{q} q}^{\pm}\left(x, b_{\perp}, \mu, \zeta^{z}\right) S_{r}^{\frac{1}{2}}\left(b_{\perp}, \mu\right)=H_{1}^{\pm}\left(\zeta^{z}, \bar{\zeta}^{z}, \mu\right)  \nonumber \\
&& \;\;\;\; \times  e^{\frac{1}{2}\ln \frac{\mp \zeta^{z}+i 0}{\zeta} K_1\left(b_{\perp}, \mu\right)} \Psi_{\bar{q} q}^{\pm}\left(x, b_{\perp}, \mu, \zeta\right),
\label{eq:matching}
\end{eqnarray}
where $H_{1}^{\pm}\left(\zeta^{z}, \bar{\zeta}^{z}, \mu\right)$ is the perturbative matching kernel. The $S_r$ is the reduced soft function defined as
\begin{equation}\label{eq:reducedsoftfunction}
S_{r}\left(b_{\perp}, \mu\right)=\lim _{\delta^{+}, \delta^{-} \rightarrow 0} \frac{S^{-}\left(b_{\perp}, \mu, \delta^{+}, \delta^{-}\right)}{S^{-}\left(b_{\perp}, \mu, \delta^{+}\right) S^{-}\left(b_{\perp}, \mu, \delta^{-}\right)}.
\end{equation}
The $S^{-}\left(b_{\perp}, \mu, \delta^{\pm}\right)$ in the denominator is defined similar with  the soft function defined in Eq.~(\ref{define_soft_function}),
but with one on-light-cone gauge-link direction along
$\bar n$ or $n$, and another off-light-cone one along $n_z$.

The reduced soft functions can also be extracted by using off-light-cone soft functions. Both the on-light-cone and off-light-cone soft functions are rapidity dependent, but the reduced soft functions are rapidity independent. The off-light-cone soft functions $S^{\pm}(\vec{b}_{\perp},\mu,Y,Y')$ are composed of two off-light-cone Wilson-line cusps. One can first define the space-like vectors as $\bar n\rightarrow \bar n_Y= \bar n-e^{-2Y}n $, $n \rightarrow n_{Y'}=n-e^{-2Y'}\bar n$ and the off-light-cone Wilson-line cusps ${\cal W}^{\pm}(b_\perp,Y,Y')$:
\begin{align}
&{\cal W}^{\pm}(\vec{b}_\perp,Y,Y')=W^{\pm}_{n_{Y'}}(\vec{b}_\perp)W^{\dagger}_{\bar n_Y}(\vec{b}_\perp) \ ,
\end{align}
where the off-light-cone gauge-links $W_{\bar n_Y}$ and $W_{n_Y'}$ are defined as
\begin{align}
&W_{\bar n_Y}(\vec{b}_\perp)={\cal P}{\rm exp}\left[ig\int_{0}^{-\infty} d\lambda' \bar n_Y \cdot A(\lambda' \bar n_Y+\vec{b}_\perp)\right],\\ 
&W_{n_{Y'}}^{\pm}(\vec{b}_\perp)={\cal P}{\rm exp}\left[ig\int_{0}^{\pm \infty} d\lambda  n_{Y'}\cdot A(\lambda n_{Y'}+\vec{b}_\perp)\right]  \ ,
\end{align}
respectively.

The off-light-cone soft functions are defined in a way similar to the on-light-cone soft function Eq.~(\ref{define_soft_function}).
\begin{align}\label{eq:SNoff}
     S^{\pm}(b_{\perp},\mu,Y,Y')=\frac{1}{N_c}{\rm tr }\frac{\langle 0|{\cal T}{\cal W}^{\pm}(b_{\perp},Y,Y')|0\rangle}{\sqrt{Z_E(Y)Z_E(Y')}}  \ ,
\end{align}
where $Z_E$ is Wilson loop to subtract the pinch singularities and power divergences of the off-light-cone Wilson lines. In the light-cone limit $Y+Y'\rightarrow \infty$ , we have:
\begin{align}\label{eq:Sroff}
      S^{\pm}(b_{\perp},\mu,Y,Y') =e^{ K_1(b_{\perp},\mu) \ln (\mp e^{Y+Y'}-i0) +{\cal D}(b_{\perp},\mu)}, 
\end{align}
where ${\cal D}$ is different from the on-light-cone version.  Similar to the case of $\delta$ regulator, imaginary part appears in the $S^{+}$ case due to analyticity property. In fact, one can show that the the off-light-cone soft function depends only on the (complex) hyperbolic angle for the directions vectors from which the imaginary part can be generated. The rapidity-independent part is defined as the generalized reduced soft function:
\begin{align}\label{eq:reduced-off-LC-soft-function}
S_{r}(b_{\perp},\mu)=e^{-{\cal D}(b_{\perp},\mu)} \ .
\end{align}

According to the properties of off-light-cone soft functions~\cite{Collins:2011zzd,Ebert:2019okf}, the reduced soft function $S_r$ defined in Eq.~(\ref{eq:reducedsoftfunction}) is consistent with that defined by Eq.~(\ref{eq:reduced-off-LC-soft-function}).
At ${\cal O}(\alpha_s)$, the reduced soft function reads 
\begin{eqnarray}
S_{r}\left(b_{\perp}, \mu\right)= 1-\frac{\alpha_s C_F}{\pi}L_b.
\end{eqnarray}

Substituting all the results  known so far into Eq.~(\ref{eq:matching}), the matching kernel  $H_{1}^{\pm}\left(\zeta^{z}, \bar{\zeta}^{z}, \mu\right)$  is  extracted as
\begin{eqnarray}\label{eq:100}
H_{1}^{\pm}\left(\zeta^{z}, \bar{\zeta}^{z}, \mu\right)&=&1+\frac{\alpha_s C_F}{2\pi}\bigg\{  -\frac{5\pi^2}{12}-2 \nonumber\\
&&+\frac{1}{2}\bigg[  \ln\frac{-\zeta^z\pm i0}{\mu^2} -\frac{1}{2}\ln^2\frac{-\zeta^z\pm i0}{\mu^2}  \nonumber\\
&&+\{\zeta^z \rightarrow \bar\zeta^z\} \bigg] \bigg\}.
\end{eqnarray} 
Results in Eq.~\eqref{eq:100} are in agreement  with Ref.~\cite{Ji:2021znw}.

\subsection{Form factor and quasi-TMDWFs}

Substituting  Eq.~(\ref{eq:matching}) into Eq.~(\ref{eq:form_fac_bare}), one can arrive at the factorization of form factor as follows: 
\begin{eqnarray}\label{eq:factorization_form_factor2}
&&F(b_\perp,P_1,P_2,\mu)
=\int dx_1dx_2 H(x_1,x_2) S_r(b_\perp,\mu) \nonumber\\
&&\;\;\;\;\;\;\;\;\;\;\;\;\;\;\;\times\tilde{\Psi}_{q \overline{q}}^{\dagger }(x_2,b_\perp,\mu,\zeta_2^z)\tilde{\Psi}_{q \overline{q}}(x_1,b_\perp,\mu,\zeta_1^z),
\end{eqnarray}
where the hard kernel $ H(x_1,x_2) $ can be written as
\begin{eqnarray}
H(x_1,x_2)= \frac{H_F(Q^2,\bar{Q}^2,\mu^2)}{ \bigg[  H_{1}^{\pm}\left(\zeta_2^{z}, \bar{\zeta}_2^{z}, \mu\right)     \bigg]^{\dagger}\bigg[H_{1}^{\pm}\left(\zeta_1^{z}, \bar{\zeta}_1^{z}, \mu\right)\bigg]}.
\end{eqnarray}
where $\zeta^z_i=(2x_iP\cdot n_z)^2$, $\bar{\zeta}^z_i=(2\bar x_iP\cdot n_z)^2$, and the condition $\zeta_1^{z}\zeta_2^{z}= \zeta_1\zeta_2$ is used. 

For $\Gamma=I$ or $\Gamma=\gamma_5$,  the matching kernel is then derived as:
\begin{align}
&&H(x_1,x_2)=H^{(0)}\bigg\{1+\frac{\alpha_s C_F}{8\pi}\bigg[ 4\pi^2+8+\ln^2\left(\frac{-\zeta_1^z\pm i0}{\mu^2}\right)\nonumber\\
&&+\ln^2\left(\frac{-\bar{\zeta}_1^z\pm i0}{\mu^2}\right)+\ln^2\left(\frac{-\zeta_2^z\mp i0}{\mu^2}\right)\nonumber\\
&&+\ln^2\left(\frac{-\bar{\zeta}_2^z\mp i0}{\mu^2}\right)-\frac{1}{2}\ln^2\left( \frac{\zeta_1^z\zeta_2^z}{\mu^4} \right)\nonumber\\
&&-\frac{1}{2}\ln^2\left(\frac{ \bar{\zeta}_1^z\bar{\zeta}_2^z}{\mu^4} \right)-2\ln\frac{\zeta_1^z\zeta_2^z\bar{\zeta}_1^z\bar{\zeta}_2^z}{\mu^8}      \bigg]\bigg\}\nonumber\\
&&=H^{(0)}\bigg\{1+\frac{\alpha_s C_F}{2\pi}\bigg[ 2+\pi^2+\frac{1}{2}\ln^2\left( -\frac{x_2}{x_1}\mp i0 \right)\nonumber\\
&&+\frac{1}{2}\ln^2\left( -\frac{\bar{x}_2}{\bar{x}_1}  \mp i0 \right)-\ln\frac{16x_1x_2\bar{x}_1\bar{x}_2{P^z}^4}{\mu^4}      \bigg]\bigg\}. 
\end{align}
For $\Gamma=\gamma_{\perp}$ or $\Gamma=\gamma_{\perp}\gamma_5$, we have:
\begin{eqnarray} \label{eq:104}
&& H(x_1,x_2)=H^{(0)}\bigg\{1+\frac{\alpha_s C_F}{8\pi}\bigg[ 4\pi^2-16 \nonumber\\
&& +\ln^2\left(\frac{-\zeta_1^z\pm i0}{\mu^2}\right)\nonumber\\
&&+\ln^2\left(\frac{-\bar{\zeta}_1^z\pm i0}{\mu^2}\right)+\ln^2\left(\frac{-\zeta_2^z\mp i0}{\mu^2}\right)\nonumber\\
&&+\ln^2\left(\frac{-\bar{\zeta}_2^z\mp i0}{\mu^2}\right)-\frac{1}{2}\ln^2\left( \frac{\zeta_1^z\zeta_2^z}{\mu^4} \right)\nonumber\\
&&-\frac{1}{2}\ln^2\left(\frac{ \bar{\zeta}_1^z\bar{\zeta}_2^z}{\mu^4} \right)+\ln\frac{\zeta_1^z\zeta_2^z\bar{\zeta}_1^z\bar{\zeta}_2^z}{\mu^8}      \bigg]\bigg\}\nonumber\\
&=&H^{(0)}\bigg\{1+\frac{\alpha_s C_F}{2\pi}\bigg[ \pi^2-4+\frac{1}{2}\ln^2\left( -\frac{x_2}{x_1}\mp i0 \right) \nonumber\\
&& +\frac{1}{2}\ln^2\left( -\frac{\bar{x}_2}{\bar{x}_1}  \mp i0 \right)+\frac{1}{2}\ln\frac{16x_1\bar{x}_1x_2\bar{x}_2{P^z}^4}{\mu^4}              \bigg]\bigg\}.\nonumber\\
\end{eqnarray}
Results for $\Gamma=\gamma_{\perp}$ in Eq.~\eqref{eq:104} are in agreement  with Ref.~\cite{Ji:2020ect}.

\section{Impact on the reduced soft function}\label{application}

\begin{figure*}[htb!]
\centering
\includegraphics[width=0.45\textwidth]{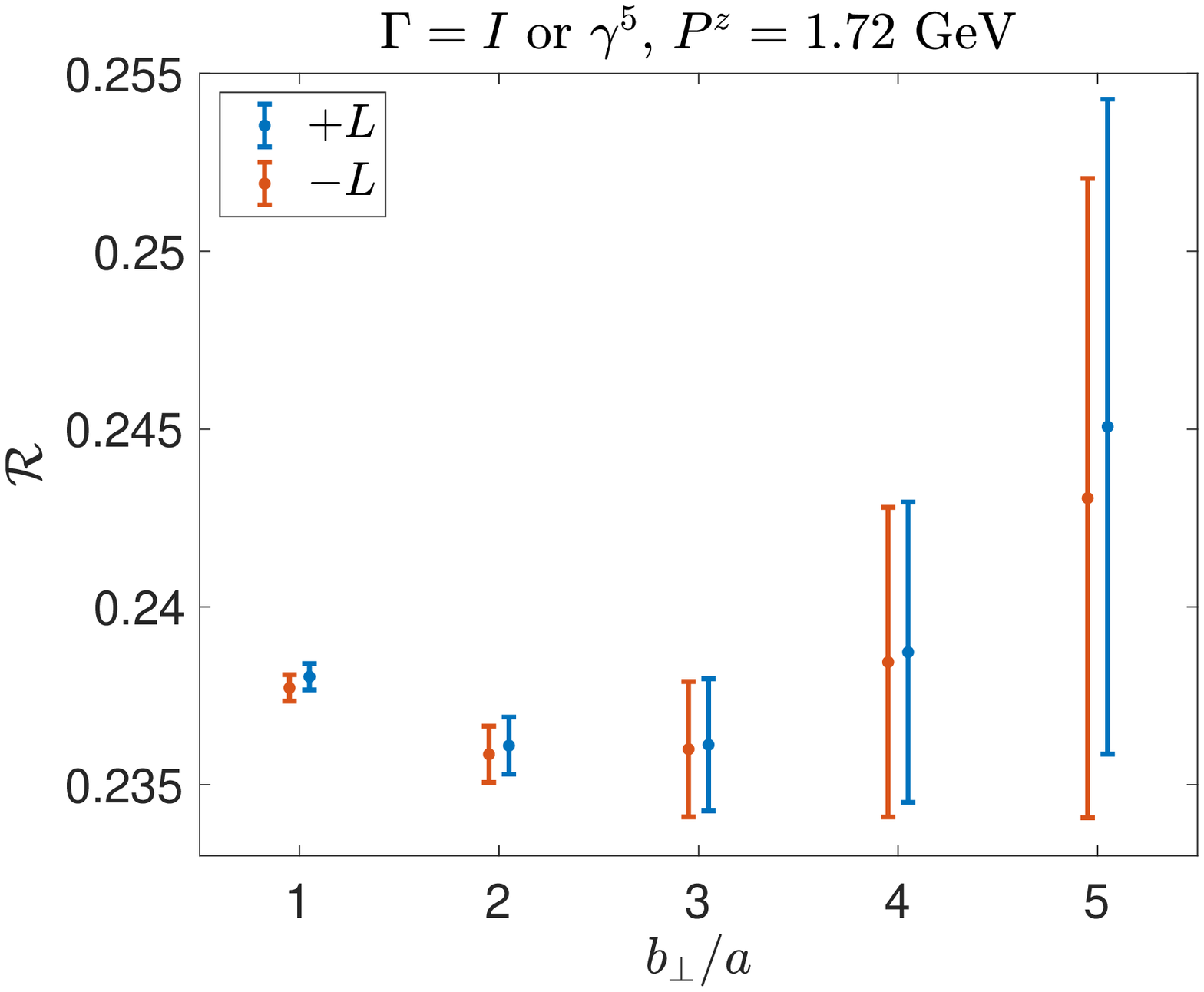}
\includegraphics[width=0.45\textwidth]{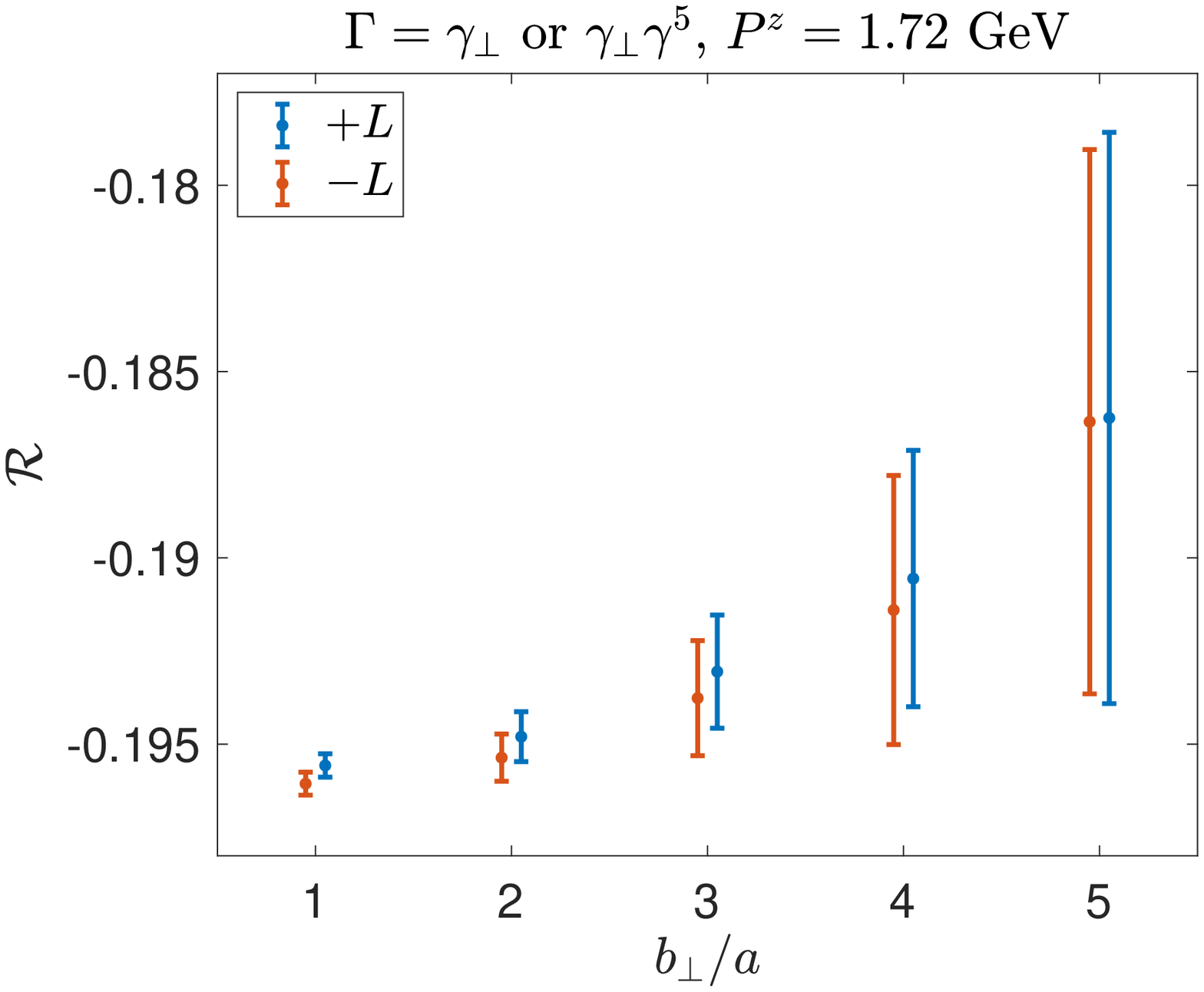}
\includegraphics[width=0.45\textwidth]{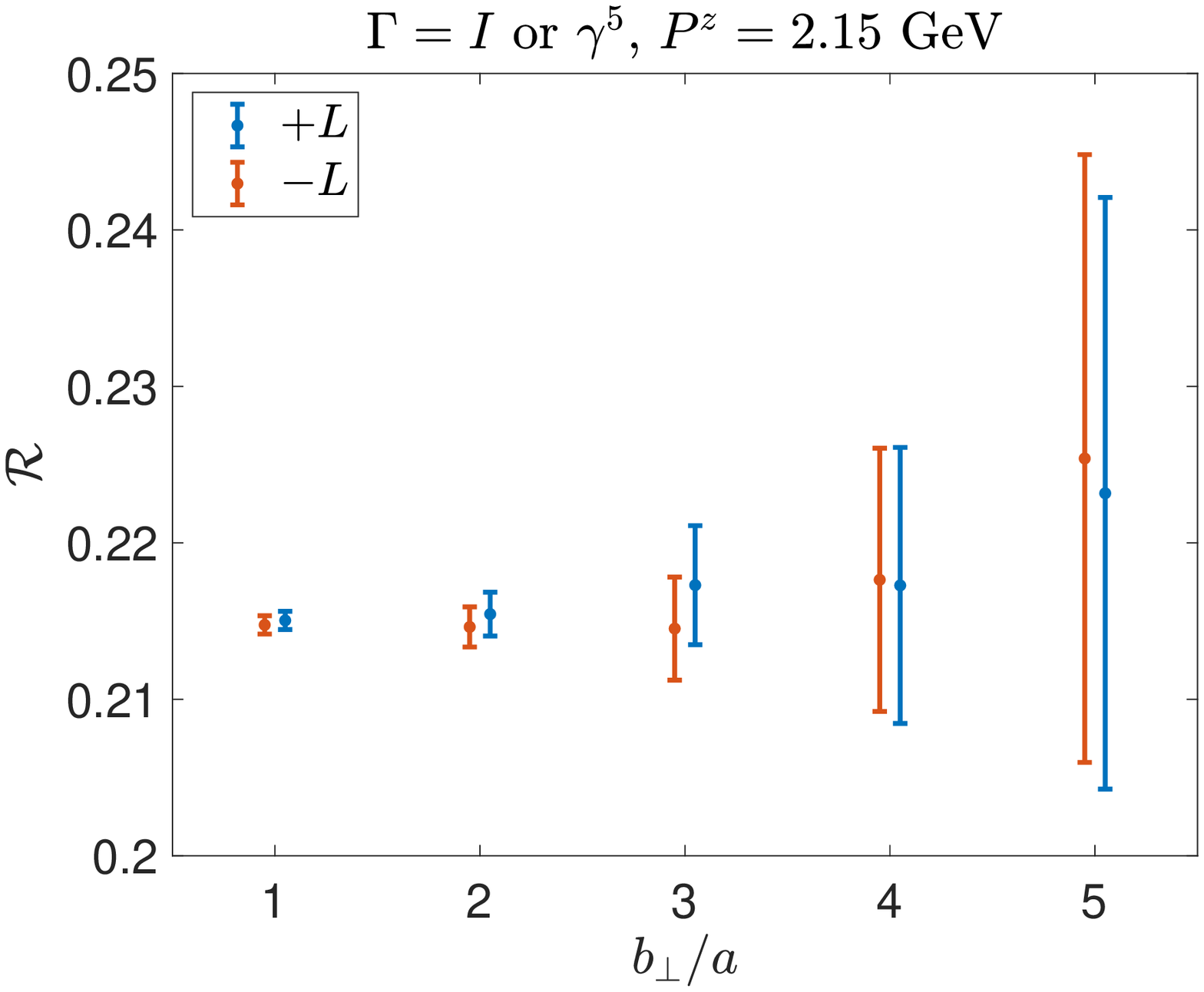}
\includegraphics[width=0.45\textwidth]{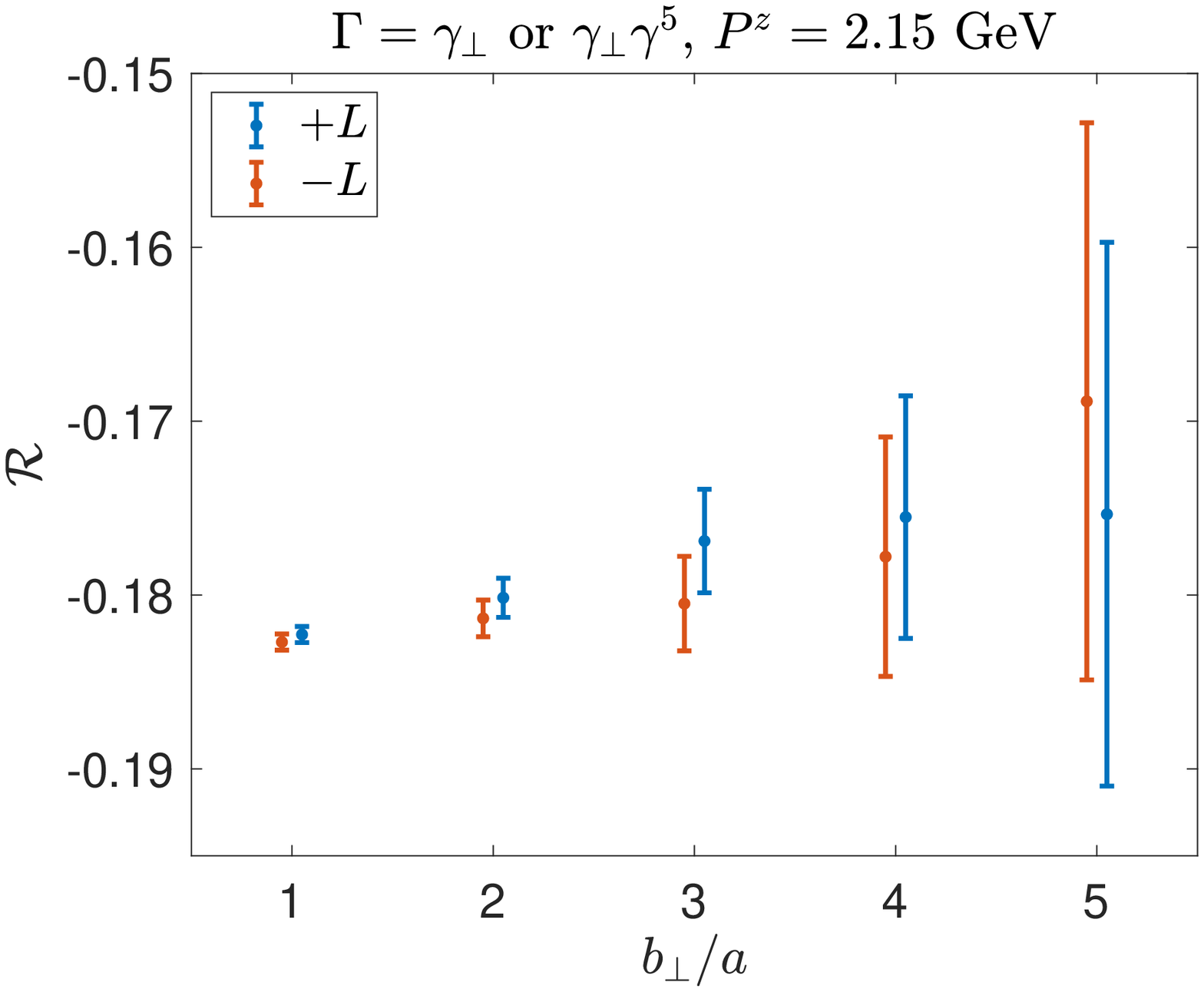}
\includegraphics[width=0.45\textwidth]{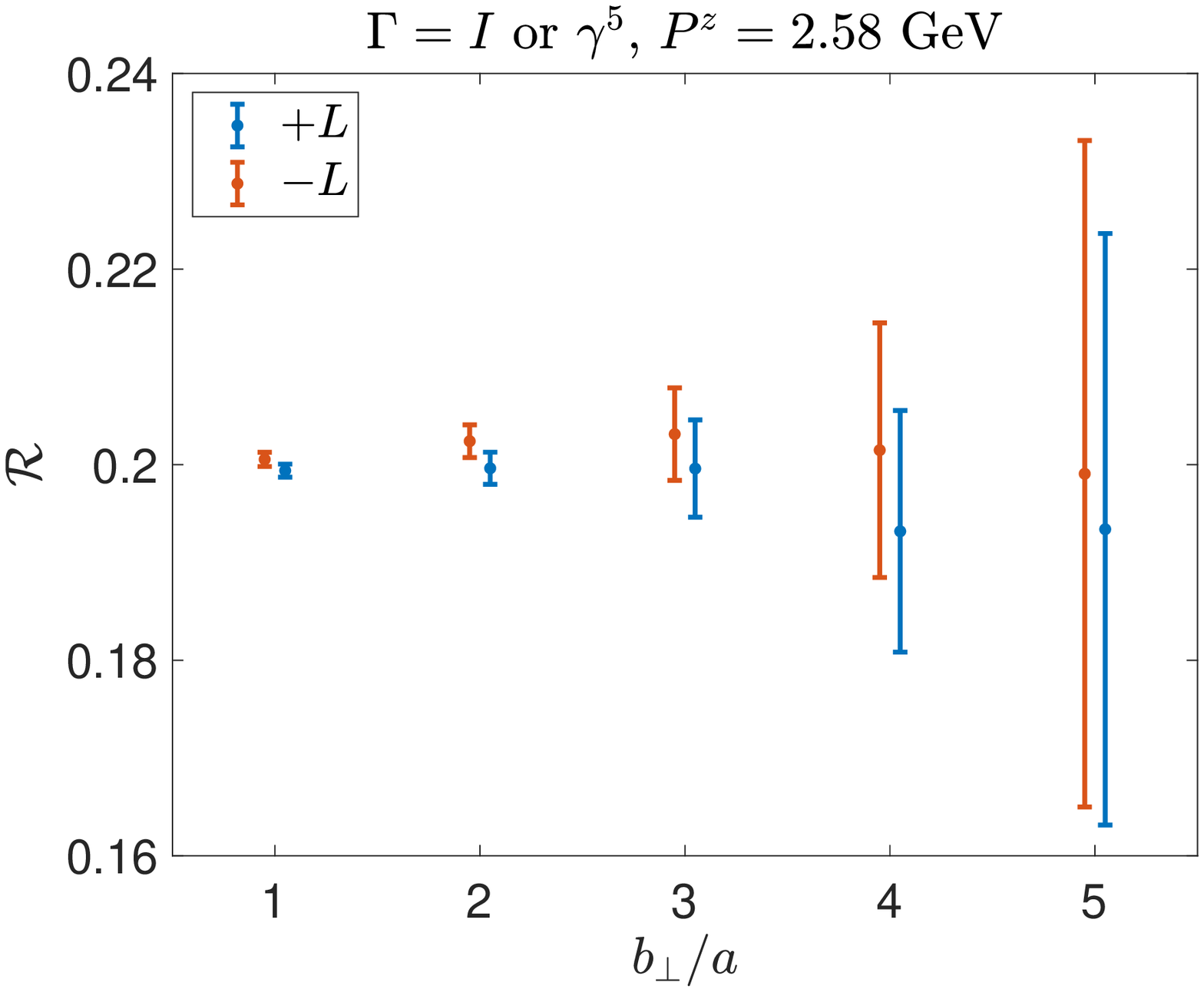}
\includegraphics[width=0.45\textwidth]{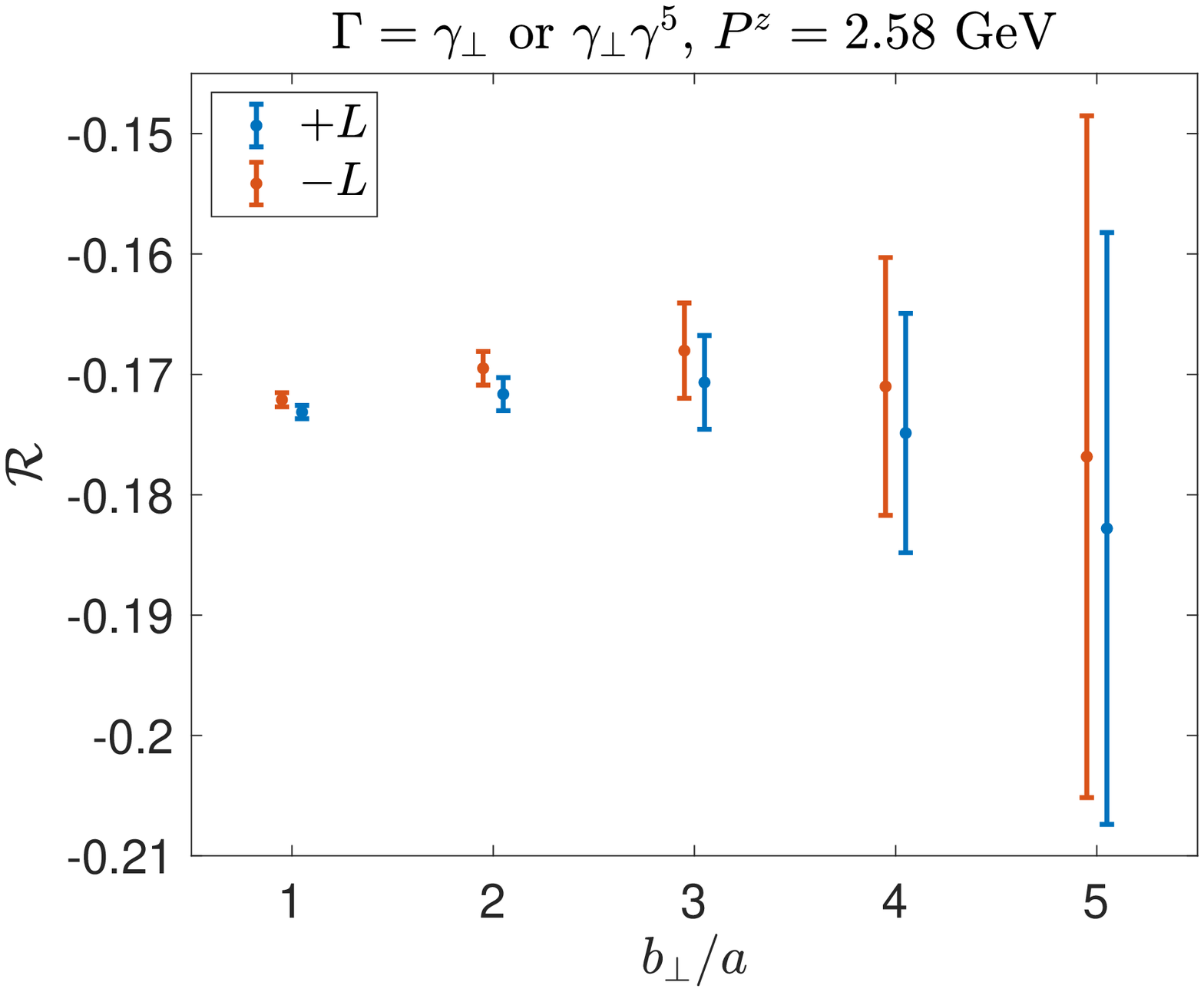}
\caption{One-loop QCD corrections to the denominator to extract  reduced soft function as defined in Eq.~\eqref{eq:ratio_denominator}.  The lattice data on quasi-TMDWFs from Lattice Parton Collaboration with $P^z=1.72~{\rm GeV}$, $2.15~{\rm GeV}$ and $2.58~{\rm GeV}$ under $b_\perp=a,2a,3a,4a,5a~(a=0.12~{\rm fm})$ is used~\cite{LPC:2022ibr}. The left panels show the results for $\Gamma=\Gamma'=I, \gamma_5$, while the right ones correspond to $\Gamma=\Gamma'=\gamma_\perp, \gamma_\perp\gamma_5$. }
\label{fig:latticeRatio}
\end{figure*}

\begin{figure*}[htb!]
\centering
\includegraphics[width=0.45\textwidth]{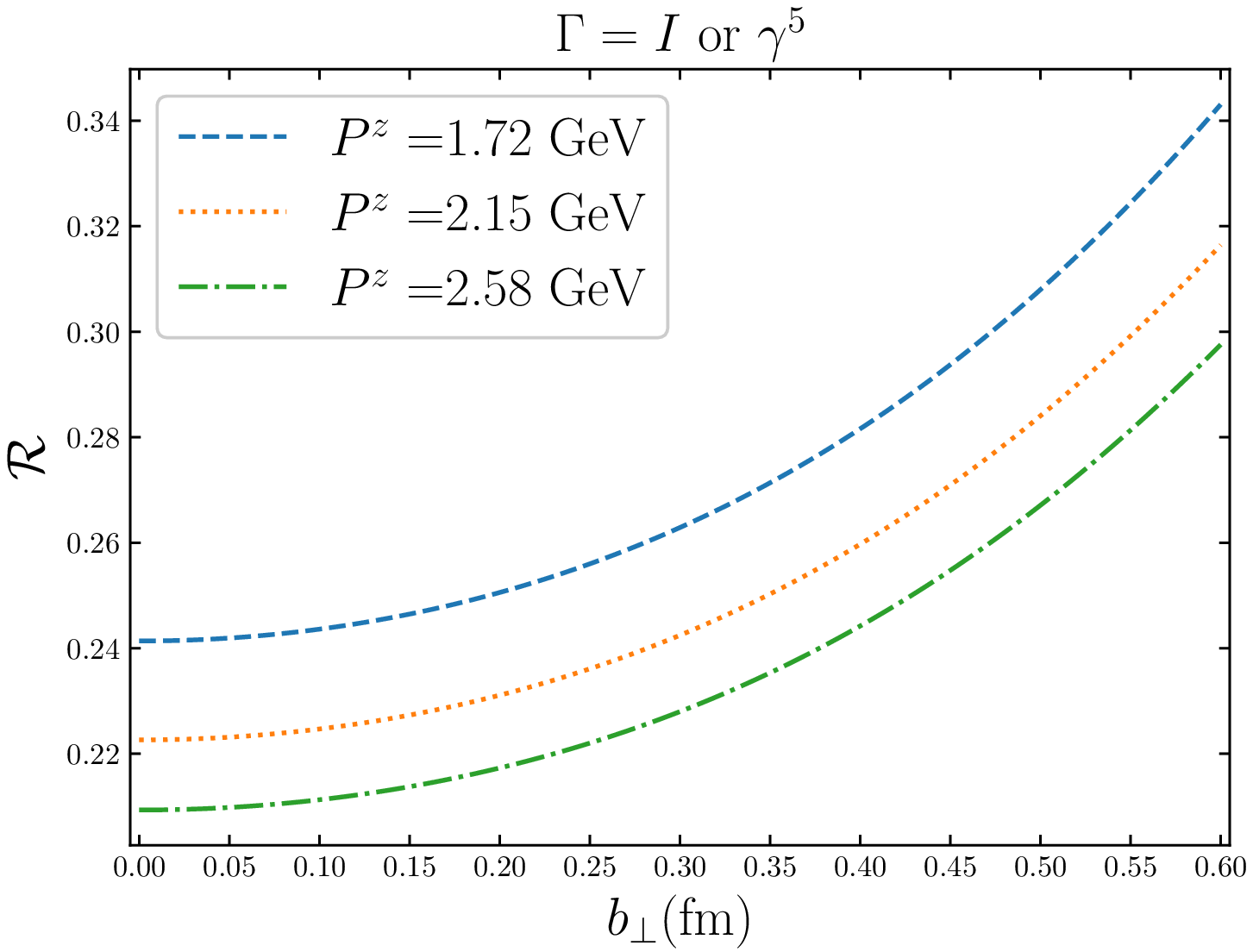}
\includegraphics[width=0.45\textwidth]{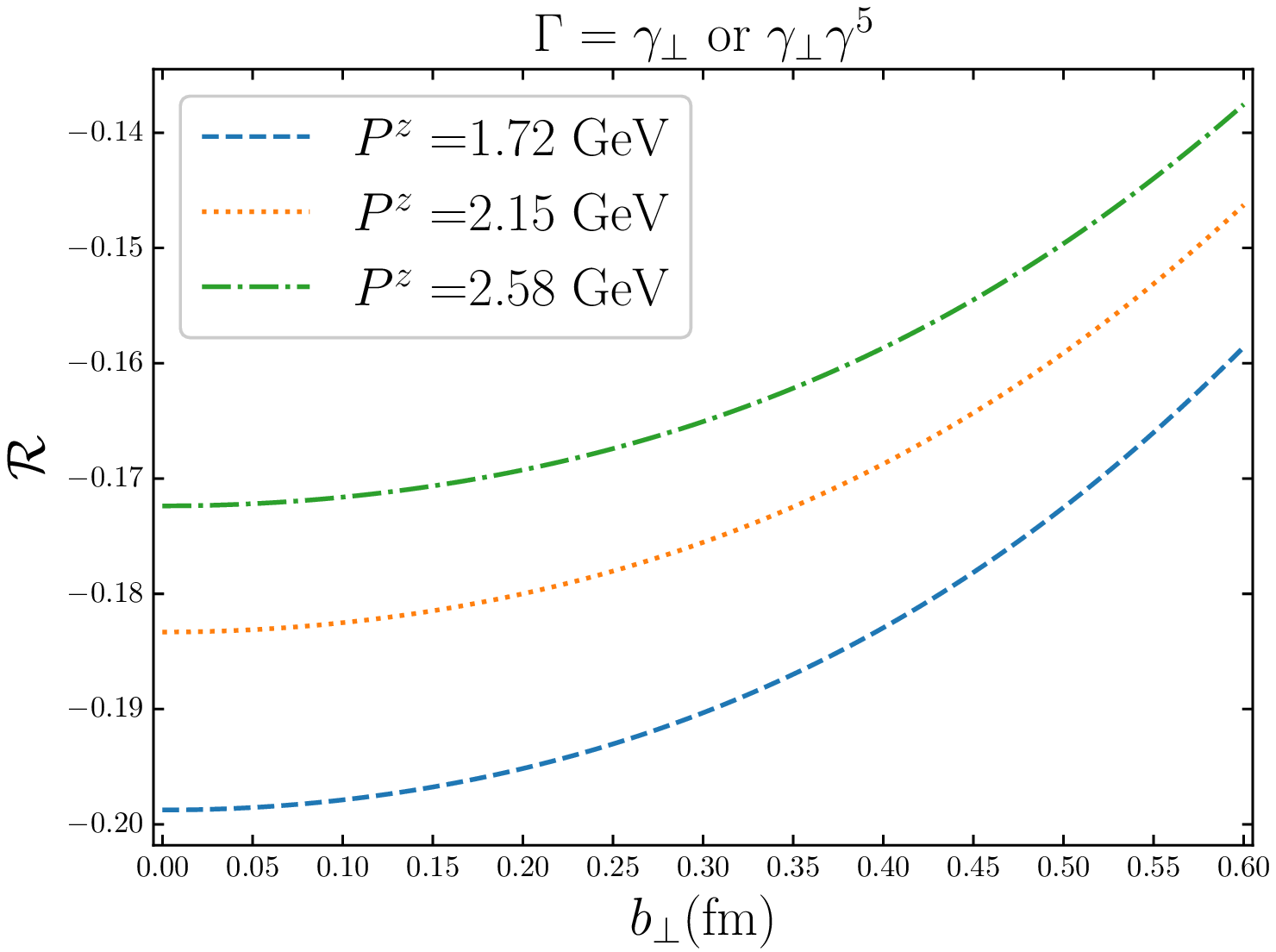}
\caption{Similar with Fig.~\ref{fig:latticeRatio} but using a phenomenological model for quasi-TMDWFs in Eq.~\eqref{eq:model_quasi_TMDWFs}.  }
\label{fig:Ratio_model}
\end{figure*}

In the previous section, we have validated the factorization of the form factor and determined  the hard kernel in perturbation theory. A direct use of the previous  results is that from Eq.~\eqref{eq:factorization_form_factor2}, one can express the reduced soft function as:
\begin{eqnarray}
S_r(b_\perp,\mu)=\frac{F(b_\perp,P_1,P_2,\mu)}{\mathcal H}.  \label{eq:extraction_reduced_soft_function}
\end{eqnarray}
where the denominator term  is 
\begin{eqnarray}\label{eq:lattice_calc}
\mathcal H &=&\int dx_1dx_2 H(x_1,x_2) \nonumber\\
&& \times \tilde{\Psi}^{\dagger }(x_2,b_\perp,P^z,\zeta_2^z) \tilde{\Psi}(x_1,b_\perp,P^z,\zeta_1^z).
\label{eq:denominator_convolution}
\end{eqnarray}
Once the form factor and quasi-TMDWFs are simulated on  the Lattice, the reduced soft functions can be determined from first-principles, and the first attempts can be found in Refs.~\cite{Shanahan:2020zxr,LatticeParton:2020uhz,Schlemmer:2021aij,Li:2021wvl,Shanahan:2021tst,LPC:2022ibr}. 

In the first analyses, the tree-level result  is used for the perturbative hard kernel $H$~\cite{LatticeParton:2020uhz,Li:2021wvl}, while a precision determination requests to include the radiative corrections.  
Based on  the lattice data on quasi-TMDWFs from Lattice Parton Collaboration with $P^z=1.72~{\rm GeV}$, $2.15~{\rm GeV}$ and $2.58~{\rm GeV}$ under $b_\perp=a,2a,3a,4a,5a~(a=0.12~{\rm fm})$ \cite{LPC:2022ibr}, we give an estimate of the impact from  ${\cal O}(\alpha_s)$ corrections on the reduced soft function. To be more explicit, we define the ratio
\begin{eqnarray}
\mathcal R=\frac{\mathcal H_{\rm 1}-{\mathcal H_{\rm 0}}}{\mathcal H_{\rm 0}}
\label{eq:ratio_denominator}
\end{eqnarray}
which directly manifests the effects on the denominator of the reduced soft function in Eq.~\eqref{eq:extraction_reduced_soft_function}.  Here in  ${\mathcal H_{\rm 0}}$,  the tree-level hard kernel is used, while in ${\mathcal H_{\rm 1}}$, the one-loop results are incorporated.  The corresponding results are shown in Fig.~\ref{fig:latticeRatio}.  The left panels show the results for $\Gamma=\Gamma'=I, \gamma_5$, while the right ones correspond to $\Gamma=\Gamma'=\gamma_\perp, \gamma_\perp\gamma_5$. 

A few remarks are given in order. 
\begin{itemize}
\item In the lattice simulation, the Wilson line can have two directions, denoted as $+L$ and $-L$. Results are consistent with each other within errors. 
\item Errors shown in the plots arise from the lattice data, which significantly increase with the increase of transverse separation. 

\item  From the figure, one can see that the magnitude of QCD corrections can reach  about $(20-30)\%$ for $\Gamma=\Gamma'=I,\gamma_5$, and  about $-(10-20)\%$ when  $\Gamma=\Gamma'=\gamma_\perp, \gamma_\perp\gamma_5$.  
However in the latter case, the QCD corrections to the denominator  are negative, which means the corresponding reduced soft function are enhanced.

\item The corrections to the denominator  will decrease with the increase of $P^z$.

\item The results are insensitive to the transverse separation $b_\perp$, though at large $b_\perp$, the errors are too large to make a decisive conclusion.  

\end{itemize}

It should be noted that the convolution in Eq.~\eqref{eq:denominator_convolution} involves both the longitudinal momentum fraction and the transverse separation. In general, the QCD corrections should contain the dependences on these two parameters, while results in Fig.~\ref{fig:latticeRatio} exhibit the dependences. A natural understanding of this feature is that the quasi-TMDWFs could be written as a factorized form, namely 
\begin{eqnarray}
\tilde{\Psi^{\prime}}(x,b_\perp) = \phi(x) \times\Sigma(b_\perp). 
\end{eqnarray}
Substituting this result into Eq.~\eqref{eq:denominator_convolution}, one can see that this dependence on $b_\perp$ cancels in the ratio in Eq.~\eqref{eq:ratio_denominator}.

As a comparison,  we adopt a phenomenological model for  quasi-TMDWF $\tilde{\Psi^{\prime}}(x,b_\perp)$~\cite{Lu:2007hr}:
\begin{eqnarray}
\tilde{\Psi^{\prime}}(x,b_\perp)&=&6x(1-x)\bigg[1+\frac{3a_2^\pi}{2}\bigg(5(2x-1)^2-1\bigg)\bigg]\nonumber\\
&&\times \exp\bigg[-\frac{x(1-x)b^2_\perp}{\alpha^2}\bigg],
\label{eq:model_quasi_TMDWFs}
\end{eqnarray}
where the longitudinal and transverse  distributions are entangled. We choose $\alpha=0.197~{\rm fm}$, and  the Gegenbauer moments $a_2^\pi=0.25$~\cite{Bakulev:2001pa}. 
 The behavior of ${\cal R}$ as a function of $b_\perp$ is shown in Fig.~\ref{fig:Ratio_model}.
Using the above model, we have also calculated the radiative corrections to the denominator and find that the results  are about $(20-30)\%$ for scalar or pseudo-scalar Lorentz structure, and reach  about $-(10-20)\%$ for vector or axial-vector structure.
A dramatic difference with the results in Fig.~\ref{fig:latticeRatio} is that these results show an explicit dependence on the transverse separation.

\section{Conclusion}\label{conslusion}

TMDPDFs and TMDWFs  are  important physical quantities characterizing  the distributions of constituents momentum in the hadron, and  reflect  the non-perturbative internal structure of hadrons.  In LaMET, the TMDWFs can be extracted from the  first-principle simulation  of a four-quark form factor and quasi distributions~\cite{Ji:2019sxk,Ji:2019ewn,Ji:2021znw}.

In the present work, a number of details are provided to understand the proposal in  Refs.~\cite{Ji:2019sxk,Ji:2019ewn,Ji:2021znw}. In particular we have explored the form factors of four kinds of four-quark  operators and calculated the one-loop perturbative corrections to these quantities. In the calculation of four-quark form factors, we have adopted a modern technique based on integration by part and differential equations that can be generalized to the analysis of other nonlocal TMD quantities in LaMET.  With the perturbative results, we  have  validated the TMD factorization of  form factors and quasi-TMDWFs at one-loop level, and then extracted the ${\cal O}(\alpha_s)$ hard function.  Converting the TMDWFs to quasi-TMDWFs,  the LaMET provided a ``two-step'' approach to the LFQ physics and achieves the goal of LFQ without performing the LFQ explicitly. Using the lattice data on quasi-TMDWFs and a phenomenological model, we have investigated the effects from the one-loop matching kernel and find that the magnitude of perturbative corrections  to the soft function depend on the operator to define the form factor,
but are less sensitive to the transverse separation. These results are helpful to precisely extract the soft functions and TMD wave functions from the first-principle in future.

\section*{Acknowledgment.}

We thank  Minhuan Chu, Yizhuang Liu,  Xiangdong Ji, Kai Yan,  Yibo Yang,  Jialu Zhang,  Jianhui Zhang  and Qi-An Zhang for valuable discussions.  We are grateful to  Yizhuang Liu and  Xiangdong Ji for carefully reading the manuscript.  We thank Lattice Parton Collaboration (LPC) for allowing us to use the lattice data on quasi-TMDWFs~\cite{LPC:2022ibr}. This work is supported in part by Natural Science Foundation of China under Grants No. 12147140, No. 11735010, No. 12125503 and No. 11905126, by Natural Science Foundation of Shanghai under grant No. 15DZ2272100, by the China Postdoctoral Science Foundation under Grant No. 2022M712088.

\begin{appendix}

\begin{widetext}
\section{Normalization of TMDWFs, and Trace Formalism and Fierz transformation}
\label{sec:Fierz_transformation}

Decay constant of a pion is parametrized by the matrix element as
\begin{align}\label{eq:localdecayconstant}
\left\langle 0\left|\overline{\psi}\left(0\right) \gamma^{\mu} \gamma^{5} \psi(0)\right| \pi\right\rangle=-if_{\pi}P^{\mu}.
\end{align}
This provides a formal normalization for the TMDWF and quasi-TMDWFs. 
For the TMDWFs defined in Eq.~(\ref{eq:lc-wave-function-define}),  
one can set  $\vec{b}_\perp=0$ and integrate  over the momentum fraction  $x$ 
\begin{eqnarray}
\int dx \psi^{\pm} \left(x, 0, \mu, \delta^{-}\right)&=& \frac{1}{-if_{\pi}}  \int dx\int \frac{d \lambda}{2 \pi} e^{-i (x-\frac{1}{2})P^{+} \lambda} \left\langle 0\left|\overline{\Psi}_{n}^{\pm}\left(\lambda \frac{n}{2}\right) \gamma^{+} \gamma^{5} \Psi_{n}^{\pm}\bigg(-\lambda \frac{n}{2}\bigg)\right| \pi\right\rangle \bigg|_{\delta^{-}}\nonumber\\
&=&\frac{1}{-if_{\pi} P^+} \left\langle 0\left|\overline{\psi} (0) \gamma^{+} \gamma^{5} \psi(0)\right| \pi\right\rangle
=1. 
\end{eqnarray}
From this equation, one can see that the TMDWF is formally normalized to 1.  Two remarks are given in order. 
\begin{itemize}
\item 
Firstly, if $\lambda=0$ and $\vec{b}_\perp=0$, the Wilson  line in ${\Psi}_{n}^{\pm}$ vanishes, and then the interpolating operator is local. 
\item  Secondly, it   should be noticed that the physical TMDWFs does not have to satisfy this normalization constraint. The TMDWF is valid in the hierarchy  $\Lambda_{\rm QCD}\ll 1/ b_\perp \ll 1/ \lambda\sim P^z$, however when integrating over the momentum fraction, this hierarchy is not   satisfied in all kinematics region. In addition, the renormalization procedure does not commute with the integration,  which also indicates that the above normalization only has a formal meaning. 
\end{itemize} 

At the parton level, the quark matrix element has been calculated in Eq.~\eqref{eq:formal_normalization}, and this can be implemented with a trace formalism.  
We consider a tree level matrix element:
\begin{eqnarray}\label{eq:AppendixSpinor}
\left\langle 0\left|
\bar{\psi}_{c}(0) \gamma^{\mu}\gamma^5 \psi_{b}(0) \right| q_{b}\left(x P\right) \bar{q}_{c}(\bar{x} P)\right\rangle
&=&\left\langle 0\left|
\bar{\psi}_{c}(0) \gamma^{\mu}\gamma^5 \psi_{b}(0) \frac{1}{\sqrt{2}}\left[  b_{\uparrow,b}^{\dagger}\left(xP\right)d_{\downarrow,c}^{\dagger}\left(\bar{x}P\right)- b_{\downarrow,b}^{\dagger}\left(xP\right)d_{\uparrow,c}^{ \dagger}\left(\bar{x}P\right) \right] \right| 0 \right\rangle  \nonumber\\
&=&\frac{1}{\sqrt{2}}{\rm{Tr}} \left\{ \left[ u_{\uparrow} \left(xP\right)\bar v_{\downarrow} \left(\bar x P\right)- u_{\downarrow}\left(xP\right) \bar v_{\uparrow}\left(\bar x P\right) \right]\gamma^\mu\gamma^5\right\}, 
\end{eqnarray}
where the arrow $\uparrow$ and $\downarrow$ denote the spin $+1/2$ and $-1/2$ for quark pair. Using the spinor 
\begin{eqnarray}
u^{\uparrow}(xP)=v^{\downarrow}(xP)=\sqrt{xP^z}\left(
\begin{array}{ccc}
1\\ 0\\1\\ 0
\end{array}
\right), u^{\downarrow}(xP)= v^{\uparrow}(xP)= \sqrt{xP^z}\left(
\begin{array}{ccc}
0\\1\\ 0 \\-1
\end{array}
\right)
\end{eqnarray} 
under the Dirac representation, one has  
\begin{eqnarray}
 \frac{1}{\sqrt{2}}   \left[ u_{\uparrow} \left(xP\right)\bar v_{\downarrow} \left(\bar x P\right)- u_{\downarrow}\left(xP\right) \bar v_{\uparrow}\left(\bar x P\right) \right] = c_1 \gamma_5 P\!\!\!\!\slash. 
\end{eqnarray}
with the coefficient $c_1=  \sqrt{{x \bar x}/{2}}$.   Since this factor $c_1$ appears both in the evaluation of tree level and one-loop matrix elements, we can take it to be 1. Thus one can employ the matrix element $
\left\langle 0\left|\overline{\psi}_{\bar q}\left(0\right) \gamma^{\mu} \gamma^{5} \psi_{q}(0)\right| q\bar{q}\right\rangle|_{\rm tree}=\frac{1}{2}\sum_s \bar v \gamma^\mu \gamma^5u \equiv 2P^{\mu}$.


Now we  derive the tree-level matching coefficient for the form factor, based on the Fierz transformation of four-quark operators: 
\begin{eqnarray}
\left( \bar{\psi}_a\psi_b \right) \left( \bar{\psi}_c\psi_d \right)&=&-\frac{1}{4}\bar{\psi}_c\psi_b \bar{\psi}_a\psi_d-\frac{1}{4}\bar{\psi}_c\gamma^\mu \psi_b  \bar{\psi}_a\gamma_\mu \psi_d-\frac{1}{8}\bar{\psi}_c\sigma^{\mu\nu} \psi_b  \bar{\psi}_a\sigma_{\mu\nu} \psi_d \nonumber\\
&&+\frac{1}{4}\bar{\psi}_c\gamma^\mu \gamma^5 \psi_b  \bar{\psi}_a\gamma_\mu \gamma^5 \psi_d-\frac{1}{4}\bar{\psi}_c \gamma^5 \psi_b  \bar{\psi}_a \gamma^5 \psi_d, \label{eq:scale spinor}\\
\left( \bar{\psi}_a \gamma^5 \psi_b \right) \left( \bar{\psi}_c \gamma^5 \psi_d \right)&=&-\frac{1}{4}\bar{\psi}_c\psi_b \bar{\psi}_a\psi_d+\frac{1}{4}\bar{\psi}_c\gamma^\mu \psi_b  \bar{\psi}_a\gamma_\mu \psi_d-\frac{1}{8}\bar{\psi}_c\sigma^{\mu\nu} \psi_b  \bar{\psi}_a\sigma_{\mu\nu} \psi_d \nonumber\\
&&-\frac{1}{4}\bar{\psi}_c\gamma^\mu \gamma^5 \psi_b  \bar{\psi}_a\gamma_\mu \gamma^5 \psi_d-\frac{1}{4}\bar{\psi}_c \gamma^5 \psi_b  \bar{\psi}_a \gamma^5 \psi_d,\label{eq:pseudoscale spinor}\\
\left( \bar{\psi}_a \gamma^\mu \psi_b \right) \left( \bar{\psi}_c\gamma_\mu  \psi_d \right)&=&-\bar{\psi}_c\psi_b \bar{\psi}_a\psi_d+\frac{1}{2}\bar{\psi}_c\gamma^\nu \psi_b  \bar{\psi}_a\gamma_\nu \psi_d+\frac{1}{2}\bar{\psi}_c\gamma^\nu \gamma^5 \psi_b  \bar{\psi}_a\gamma_\nu \gamma^5 \psi_d+\bar{\psi}_c \gamma^5 \psi_b  \bar{\psi}_a \gamma^5 \psi_d,\label{eq:vector spinor}\\
\left( \bar{\psi}_a \gamma^\mu\gamma^5 \psi_b \right) \left( \bar{\psi}_c\gamma_\mu \gamma^5 \psi_d \right)&=&\bar{\psi}_c\psi_b \bar{\psi}_a\psi_d+\frac{1}{2}\bar{\psi}_c\gamma^\nu \psi_b  \bar{\psi}_a\gamma_\nu \psi_d+\frac{1}{2}\bar{\psi}_c\gamma^\nu \gamma^5 \psi_b  \bar{\psi}_a\gamma_\nu \gamma^5 \psi_d-\bar{\psi}_c \gamma^5 \psi_b  \bar{\psi}_a \gamma^5 \psi_d,\label{eq:pseudovector spinor}
\\
\left( \bar{\psi}_a \sigma^{\mu\nu}\psi_b \right) \left( \bar{\psi}_c\sigma_{\mu\nu}\psi_d \right)&=&-3\bar{\psi}_c\psi_b \bar{\psi}_a\psi_d+\frac{1}{2}\bar{\psi}_c\sigma^{\rho\delta} \psi_b  \bar{\psi}_a\sigma_{\rho\delta} \psi_d-3\bar{\psi}_c \gamma^5 \psi_b  \bar{\psi}_a \gamma^5 \psi_d\label{eq:tensor spinor}.
\end{eqnarray}
Note that all  repeated  Lorentz indices in Eqs.~(\ref{eq:scale spinor}-\ref{eq:tensor spinor}) should be summed. 
To be complete, one should also include the color Fierz transformation: 
\begin{eqnarray}
\delta_{ij}\delta_{kl}=\frac{\delta_{il}\delta_{kj}}{N_c}+ \frac{T^{a}_{il}T^{a}_{kj}}{T_R},
\end{eqnarray}
where $i, j, k, l$ are the color indices of those quark fields $\psi_a,\psi_b, \psi_c, \psi_d$ respectively. 
The second term will vanish  at tree level   because the index $i$ and $j$ are anti-symmetry for $T^{a}_{il}$.

Taking $\Gamma=\Gamma'=I$ as an example, we evaluate the matrix element: 
\begin{eqnarray}
F(b_\perp,P_1,P_2,\mu)&=& \frac{\left\langle P_2\left|\left( \bar{\psi}_a\psi_b \right)(b) \left( \bar{\psi}_c\psi_d \right)(0) \right|P_1\right\rangle}{f^2_{\pi}P_1 \cdot P_2}\nonumber\\
&=&\frac{1}{4N_cf^2_{\pi}P_1 \cdot P_2}  \left\langle P_2\left| \bar{\psi}_c(0)\gamma^\mu \gamma^5 \psi_b (b) \bar{\psi}_a(b)\gamma_\mu \gamma^5 \psi_d(0)  \right|P_1\right\rangle.
\end{eqnarray}
At tree level, there is no interaction between the quarks, and thus the four quarks can be split into two groups, each of which is related to TMDWFs: 
\begin{eqnarray}
F(b_\perp,P_1,P_2,\mu)&=&\frac{1}{4N_cf^2_{\pi}P_1 \cdot P_2}  \left\langle P_2\left|  \bar{\psi}_c(0)\gamma^\mu \gamma^5 \psi_b (b)   \right|0\right\rangle \left\langle 0\left| \bar{\psi}_a(b)\gamma_\mu \gamma^5 \psi_d(0) \right|P_1\right\rangle\nonumber\\
&=&\frac{1}{4N_cf^2_{\pi}P_1 \cdot P_2}  (-if_\pi P_2^\mu \int d x_2 \psi(x_2, b, P_2) )^\dagger(-if_\pi {P_1}_{\mu} \int dx_1 \psi(x_1, b, P_1) ). 
\end{eqnarray}
Comparing with the factorization of form factor, one can easily derive the tree-level hard kernel:
\begin{eqnarray}
H_F^{(0)}= \frac{1}{4N_c}. 
\end{eqnarray}
This is also similar for the cases $\Gamma=\gamma_5$. For $\Gamma=\Gamma'=\gamma_\perp, \gamma_\perp\gamma_5$,  notice that the form factor that we defined in the main text does not include the summation, and there is a sign difference.

The tensor form factor $\left\langle P_2\left|\left( \bar{\psi}_a\sigma^{\mu\nu}\psi_b \right)(b) \left( \bar{\psi}_c\sigma_{\mu\nu}\psi_d \right)(0) \right|P_1\right\rangle$  seems to contribute with  a leading-twist
component $\left\langle P_2\left|\left( \bar{\psi}_a \sigma_{\perp}^{\mu\nu}\psi_b \right) \left( \bar{\psi}_c\sigma_{\mu\nu\perp}\psi_d \right)\right|P_1\right\rangle$. However, as  shown in Eq.~(\ref{eq:tensor spinor}) that the Fierz transformation for tensor Lorentz structure can not generate a axial-vector Lorentz structure. Therefore, the contribution of tensor current for form factor is zero.

In addition, based on Fierz transformations one can make uses of the combinations to eliminate the power-suppressed contributions: 
\begin{eqnarray}
\left( \bar{\psi}_a\psi_b \right) \left( \bar{\psi}_c\psi_d \right)-\left( \bar{\psi}_a \gamma^5 \psi_b \right) \left( \bar{\psi}_c \gamma^5 \psi_d \right) &=&-\frac{1}{2}\bar{\psi}_c\gamma^\mu \psi_b  \bar{\psi}_a\gamma_\mu \psi_d 
+\frac{1}{2}\bar{\psi}_c\gamma^\mu \gamma^5 \psi_b  \bar{\psi}_a\gamma_\mu \gamma^5 \psi_d,\\
\left( \bar{\psi}_a \gamma^\mu \psi_b \right) \left( \bar{\psi}_c\gamma_\mu  \psi_d \right)+ \left( \bar{\psi}_a \gamma^\mu\gamma^5 \psi_b \right) \left( \bar{\psi}_c\gamma_\mu \gamma^5 \psi_d \right)&=& \bar{\psi}_c\gamma^\nu \psi_b  \bar{\psi}_a\gamma_\nu \psi_d+ \bar{\psi}_c\gamma^\nu \gamma^5 \psi_b  \bar{\psi}_a\gamma_\nu \gamma^5 \psi_d. 
\end{eqnarray} 
These combinations have been used in Ref.~\cite{Li:2021wvl}.

\section{TMD wave function}\label{appendixTMDWF}
The real diagram at the one-loop QCD correction as shown in Fig.~\ref{fig:TMDWF-one-loop} (a) can be obtained as follows:
\begin{eqnarray}\label{eq:wave-func-a}
\psi_{\overline{q} q}^{\pm (1,a)}&=&\mu_{0}^{2\epsilon}\frac{i g^2C_F}{2}  \int\frac{d^d q}{(2\pi)^d}  \frac{\bar{v}\gamma^+\gamma^5 (x_0\slashed{P}-\slashed{q})\slashed{n} u}{(-q^+\pm i\frac{\delta}{2})[(x_0P-q)^2+i\epsilon](q^2+i\epsilon)}e^{-iq\cdot b } \delta\bigg[(x-x_0)P^++q^+\bigg]\nonumber\\
&=&\frac{\alpha_s C_F}{2 \pi} \frac{ x\theta(x_0-x)}{ x_0 (x-x_0 \pm i\frac{\delta^{-}}{2P^{+}})}\left(\frac{1}{\epsilon_{\rm IR}} + L_b\right).
\end{eqnarray}
The UV divergence is regularized by transverse coordinates deviations of the two quark fields. The contribution from the mirror diagram Fig. \ref{fig:TMDWF-one-loop} (b) can be obtained from Eq. (\ref{eq:wave-func-a}) with the replacement  $x \to (1-x)$  and  $x_0 \to (1- {x}_0)$: 
\begin{eqnarray}\label{eq:wave-func-b}
\psi_{\overline{q} q}^{\pm (1,b)}
&=&	\frac{\alpha_s C_F }{2\pi } \frac{(1-x) \theta(x-x_0) }{  (1-x_0)(x_0-x\pm i\frac{\delta^{-}}{2P^{+}}) }\left(\frac{1}{\epsilon_{\rm IR}} + L_b\right).
\end{eqnarray}

For the vertex diagram Fig. \ref{fig:TMDWF-one-loop} (c) we obtain:
\begin{eqnarray}\label{eq:wave-func-c}
\psi_{\overline{q} q}^{\pm (1,c)}(x, b_{\perp}, \mu, \delta^{-})&=&\mu_{0}^{2\epsilon} \frac{i g^2C_F}{2}  \int\frac{d^dq}{(2\pi)^d} \frac{\bar{v}\gamma_\mu(\bar{x}_0\slashed{P}+\slashed{q})\gamma^+\gamma^5(x_0\slashed{P}-\slashed{q})\gamma^\mu u}{[(\bar{x}_0P+q)^2+i\epsilon][(x_0P-q)^2+i\epsilon](q^2+i\epsilon)}e^{-iq\cdot b } \delta\bigg[(x-x_0)P^++q^+\bigg]\nonumber\\
&=&	-\frac{\alpha_s C_F}{2\pi } \Bigg(\frac{\bar{x}}{\bar{x}_0}\theta(x-x_0)+\frac{x}{x_0}\theta(x_0-x)\Bigg)\left(\frac{1}{\epsilon_{\rm{IR}}} +L_b-1\right). 
\end{eqnarray}
Using the plus function we obtain
\begin{eqnarray}
\psi_{\overline{q} q}^{\pm (1,c)}(x, b_{\perp}, \mu, \delta^{-})&=&\bigg[\psi_{\overline{q} q}^{\pm (1,c)}\bigg]_+ -\frac{\alpha_s C_F}{2\pi } \delta(x-x_0)\int^1_0 dx \Bigg(\frac{\bar{x}}{\bar{x}_0}\theta(x-x_0)+\frac{x}{x_0}\theta(x_0-x)\Bigg)\left(\frac{1}{\epsilon_{\rm{IR}}} +L_b-1\right)\nonumber\\
&=&\bigg[\psi_{\overline{q} q}^{\pm (1,c)}\bigg]_+ - \delta(x-x_0) \frac{\alpha_s C_F}{4\pi }  \left(\frac{1}{\epsilon_{\rm{IR}}} +L_b-1\right)
\end{eqnarray}
For Fig. \ref{fig:TMDWF-one-loop} (d) and Fig. \ref{fig:TMDWF-one-loop} (e),  we obtain
\begin{eqnarray}\label{eq:wave-func-d}
\psi_{\overline{q} q}^{\pm (1,d)}
&=&-\mu_{0}^{2\epsilon} \frac{i g^2C_F}{2P^+} \delta (x-x_0) \int\frac{d^dq}{(2\pi)^d} \frac{1}{2}\sum_s \frac{\bar{v}\gamma^+\gamma^5 (x_0\slashed{P}-\slashed{q})\slashed{n} u}{(-q^+\pm i\frac{\delta}{2})[(x_0P-q)^2+i\epsilon](q^2+i\epsilon)}  \nonumber\\
&=&\frac{\alpha_s C_FP^+ }{2\pi x_0} \delta (x-x_0) \int_0^{x_0} d y \frac{\theta(x_0-y)y}{(y-x_0) P^{+}\pm i\frac{\delta^{-}}{2}}\left(\frac{1}{\epsilon_{\rm UV}}-\frac{1}{\epsilon_{\rm IR}}\right),\label{eq:wave-func-e}\\
\psi_{\overline{q} q}^{\pm 1,(e)}&=&-\mu_{0}^{2\epsilon} \frac{i g^2C_F}{2P^+} \delta (x-x_0) \int\frac{d^dq}{(2\pi)^d} \frac{1}{2}\sum_s \frac{\bar{v}\slashed{n} (\bar{x}_0\slashed{P}+\slashed{q}) \gamma^+\gamma^5  u}{[(\bar{x}_0P+q)^2+i\epsilon](q^+\pm i\frac{\delta}{2})(q^2+i\epsilon)}  \nonumber\\
&=&\frac{\alpha_s C_F P^+}{2\pi \bar{x}_0}\delta(x-x_0)\int^{1}_{x_0} dy \frac{\theta(x-x_0)\bar{y}}{(x_0-y)P^{+}\pm i\frac{\delta^-}{2}}\left(\frac{1}{\epsilon_{\rm UV}}-\frac{1}{\epsilon_{\rm IR}}\right),
\end{eqnarray}
where $\bar{y}=1-y$.  The combination of Eq.~(\ref{eq:wave-func-a}) and Eq.~(\ref{eq:wave-func-d}) will cancel the IR divergence. After by taking the UV renormalization, the finite term can be obtained in the form of plus function:
\begin{eqnarray}\label{eq:wave-func-ad}
\psi_{\overline{q} q}^{\pm (1,a)}+\psi_{\overline{q} q}^{\pm (1,d)}&=&\bigg[\psi_{\overline{q} q}^{\pm (1,a)}\bigg]_+ +\delta(x-x_0) \frac{\alpha_s C_F P^+}{2 \pi}\int^{x_0}_0 dx\frac{x}{ x_0\bigg[(x-x_0)P^{+} \pm i\frac{\delta^{-}}{2}\bigg]}\left(\frac{1}{\epsilon_{\rm UV}}+L_b\right)\nonumber\\
&=&\bigg[\psi_{\overline{q} q}^{\pm (1,a)}\bigg]_+ +\delta(x-x_0)\frac{\alpha_s C_F}{2 \pi} \Bigg[1+\left(1\mp \frac{i\delta^-}{2 x_0 P^+}\right)\ln \frac{\mp i \delta^-}{2x_0P^+\mp i\delta^-}\Bigg]\left(\frac{1}{\epsilon_{\rm UV}}+L_b\right)\nonumber\\
&=&\bigg[\psi_{\overline{q} q}^{\pm (1,a)}\bigg]_+ +\delta(x-x_0)\frac{\alpha_s C_F}{2 \pi} \left(1+\frac{1}{2}\ln \frac{- {\delta^-}^2\mp i0}{4x^2{P^+}^2}\right)\left(\frac{1}{\epsilon_{\rm UV}}+L_b\right),
\end{eqnarray}
where we  take the limit $\delta^- \to 0^+$ in the last step.  Results for Fig.~(\ref{fig:TMDWF-one-loop}b, e)  are given similarly: 
\begin{eqnarray}\label{eq:wave-func-be}
\psi_{\overline{q} q}^{\pm (1,b)}+\psi_{\overline{q} q}^{\pm (1,e)}
&=&\bigg[\psi_{\overline{q} q}^{\pm (1,b)}\bigg]_+ +\delta(x-x_0)\frac{\alpha_s C_F}{2 \pi} \left(1+\frac{1}{2}\ln \frac{- {\delta^-}^2\mp i0}{4\bar x^2{P^+}^2}\right)\left(\frac{1}{\epsilon_{\rm UV}}+L_b\right).
\end{eqnarray}



\section{Four-quark Form factor}
\label{Appendix:analysis_of_cross_diagram}

For  Fig. \ref{fig:form_factor} ($c$), we obtain the amplitude: 
\begin{eqnarray}\label{eq:form_factor_iii-appendix}
F^{c}&=&\frac{-\mu_{0}^{2\epsilon}ig^2C_F}{4 P_1 \cdot P_2}  \int\frac{d^dq}{(2\pi)^d}  \frac{\bar{u}_a(x_2P_2)\Gamma (\slashed{q}+x_1\slashed{P}_1)  \gamma_\mu u_b(x_1P_1) \bar{v}_c(\bar{x}_1P_1) \Gamma (\slashed{q}+\bar{x}_2\slashed{P}_2) \gamma^\mu   v_d(\bar{x}_2P_2)}{[(q+x_1P_1)^2+i\epsilon][(q+\bar{x}_2P_2)^2+i\epsilon](q^2+i\epsilon)}e^{-iq\cdot b} .  
\end{eqnarray}
For $\Gamma=\gamma_5$ or $\Gamma=I$ we obtain
\begin{eqnarray}
F^{c} &=&(-F^0)\times(- \mu_{0}^{2\epsilon} ig^2C_F)\bigg\{ \int\frac{d^dq}{(2\pi)^d} \bigg[ \frac{ \frac{(D-2)}{2x_1\bar{x}_2 (P_1\cdot P_2)}\bigg((q+\bar{x}_2P_2)^2-q^2\bigg)\bigg((q+x_1P_1)^2-q^2\bigg) }{[(q+x_1P_1)^2+i\epsilon][(q+\bar{x}_2P_2)^2+i\epsilon](q^2+i\epsilon)}\nonumber\\
&&+\frac{ -(D-4)q^2+4x_1\bar{x}_2(P_1\cdot P_2)+2\bigg((q+\bar{x}_2P_2)^2-q^2\bigg)+ 2\bigg((q+x_1P_1)^2-q^2\bigg) }{[(q+x_1P_1)^2+i\epsilon][(q+\bar{x}_2P_2)^2+i\epsilon](q^2+i\epsilon)}\bigg] e^{-iq\cdot b} \nonumber\\
&=&(-F^0)\times(- \mu_{0}^{2\epsilon}ig^2C_F ) \int\frac{d^dq}{(2\pi)^d} \bigg[ \frac{(D-2)}{2x_1\bar{x}_2 (P_1\cdot P_2)} \bigg(\frac{1}{q^2+i\epsilon} -\frac{1}{(q+x_1P_1)^2+i\epsilon} -\frac{1}{(q+\bar{x}_2P_2)^2+i\epsilon} \nonumber\\
&& +\frac{ q^2 }{[(q+x_1P_1)^2+i\epsilon][(q+\bar{x}_2P_2)^2+i\epsilon]}\bigg) -\frac{ (D-4) }{[(q+x_1P_1)^2+i\epsilon][(q+\bar{x}_2P_2)^2+i\epsilon]}  \nonumber\\
&&+\frac{ 4x_1\bar{x}_2(P_1\cdot P_2) }{[(q+x_1P_1)^2+i\epsilon][(q+\bar{x}_2P_2)^2+i\epsilon](q^2+i\epsilon)} +\frac{2}{[(q+x_1P_1)^2+i\epsilon](q^2+i\epsilon)} -\frac{ 2 }{[(q+x_1P_1)^2+i\epsilon][(q+\bar{x}_2P_2)^2+i\epsilon]} \nonumber\\
&&+\frac{ 2 }{[(q+\bar{x}_2P_2)^2+i\epsilon](q^2+i\epsilon)} -\frac{ 2  }{[(q+x_1P_1)^2+i\epsilon][(q+\bar{x}_2P_2)^2+i\epsilon]} \bigg]e^{-iq\cdot b}\nonumber\\
&=&(-F^0)\times\bigg(\frac{\alpha_sC_F}{\pi}\bigg ) \bigg\{  \mu_{0}^{2\epsilon}\frac{16\pi^2P_1\cdot P_2}{2^d\pi^{d/2}} G_{1,1,1,0} -\frac{1}{\epsilon_{\rm IR} }- L_b -\frac{3
   }{2 b_\perp^2 P^{z2} x_1\bar{x}_2
   }   \bigg\}.\nonumber
\end{eqnarray}
In the above equation  the $G_{1,1,1,0}$ is the three-point loop integral  defined in Eq.~(\ref{eq:MaterIntegration_1}) and will  be calculated by DEs. For $\Gamma=\gamma_\perp$ or $\Gamma=\gamma_5\gamma_\perp$, we will encounter a similar result as
\begin{eqnarray}
F^{c} &=&(-F^0)\times \bigg(\frac{g^2C_F}{4\pi^2}\bigg ) \bigg\{  \mu_{0}^{2\epsilon}\frac{16\pi^2P_1\cdot P_2}{2^d\pi^{d/2}} G_{1,1,1,0} -\frac{1}{\epsilon_{\rm IR} } -L_b  -\frac{1
   }{ b_\perp^2 P^{z2} x_1\bar{x}_2
   }   \bigg\}.\nonumber
\end{eqnarray}
For Fig. \ref{fig:form_factor} ($d$), and $\Gamma=\gamma_5$ or $\Gamma=I$ we obtain
\begin{eqnarray}
F^{d} &=&(-F^0)\times \bigg(\frac{g^2C_F}{4\pi^2}\bigg ) \bigg\{  \mu_{0}^{2\epsilon}\frac{16\pi^2P_1\cdot P_2}{2^d\pi^{d/2}} G_{1,1,1,0} -\frac{1}{\epsilon_{\rm IR} } -L_b  -\frac{3}{2b_\perp^2 P^{z2} x_2
   \bar{x}_1}    \bigg\}.\nonumber
\end{eqnarray}
For $\Gamma=\gamma_\perp$ or $\Gamma=\gamma_\perp\gamma_5$ we obtain
\begin{eqnarray}
F^{d} &=&(-F^0)\times \bigg(\frac{g^2C_F}{4\pi^2}\bigg ) \bigg\{  \mu_{0}^{2\epsilon}\frac{16\pi^2P_1\cdot P_2}{2^d\pi^{d/2}} G_{1,1,1,0} -\frac{1}{\epsilon_{\rm IR} } -L_b  -\frac{1}{b_\perp^2 P^{z2} x_2
   \bar{x}_1}  \bigg\}.\nonumber
\end{eqnarray}

Therefore, by taking the large-momentum limit $P^z \to \infty$, $G_{1,1,1,0}$  can be determined as~\cite{Kai_Yan}: 
\begin{eqnarray}
G_{1,1,1,0}&=&\frac{e^{-\epsilon \gamma_E}}{(Q'^2)^{1+\epsilon}} \bigg(-\frac{1}{\epsilon^2}+\frac{1}{2} \ln^2 \frac{Q'^2b_{\perp}^2}{4}  +2\gamma_E \ln \frac{Q'^2b_{\perp}^2}{4} +2\gamma_E^2+\frac{\pi^2}{12} \bigg).
\end{eqnarray}
When the integration variable $q$ in Eq.~(\ref{eq:MaterIntegration_1}) goes to infinity, the power suppression and exponential oscillation cause $G_{1,0,0,0}$ without UV divergence. The divergence in above integral is purely infrared.

\section{quasi-TMDWFs}\label{Appendix:quasi-TMDWF}

From the definition
\begin{align}
\tilde{\psi}_{q \overline{q}}^{\pm}\left(x, b_{\perp}, \mu, \zeta^{z}\right)=\lim _{L \rightarrow \infty} \int \frac{d \lambda}{4 \pi} e^{-i x_{r} (-P^{z}) \lambda} \left\langle 0\left|\overline{\Psi}_{\mp n_{z}}\left(\frac{\lambda n_{z}}{2}+b\right) \gamma^{z} \gamma^{5} \Psi_{\mp n_{z}}\left(-\frac{\lambda n_{z}}{2}\right)\right|q \overline{q}\right\rangle,
\end{align}
one obtains the tree level result:  
\begin{eqnarray}\label{eq:qusai-wave-func-tree}
\tilde{\psi}_{\overline{q} q}^{\pm (0)}= \delta(x-x_0). 
\end{eqnarray} 

By taking pure dimensional regularization $d=4-2\epsilon$, the sail diagram  as shown in Fig.~\ref{fig:quasi-WF} (b) and Fig.~\ref{fig:quasi-WF} (c) in the large $L$ limit and large $P^z$ limit  gives: 
\begin{eqnarray}\label{eq:quasi-wave-func-sail}
\tilde{\psi}_{\overline{q} q}^{\pm (1,bc)}\left(x, b_{\perp}, \mu, \zeta^{z}\right)
&=&-\frac{g^2C_F}{2} \int \frac{d \lambda}{2 \pi} e^{i (x-1)P^{z} \lambda}\int_0^1 ds \mathcal{L}^{\prime}(s)_\mu \bigg[ e^{-i \bar{x}_0 \lambda P\cdot  n_{z}}\int \frac{d^d k}{(2\pi)^d}\bar v \gamma^{z} \gamma^{5}\frac{e^{-i (x_0P-k) \cdot \mathcal{L}(s)}}{(x_0P-k)^2+i\epsilon}  \frac{\slashed{k}}{k^2+i\epsilon}\gamma^\mu u \nonumber \\
&& +\int \frac{d^d k}{(2\pi)^d}\bar v \gamma^\mu \frac{e^{-i (\bar{x}_0P+k) \cdot \mathcal{L}(s)}}{(\bar{x}_0P+k)^2+i\epsilon}  \frac{\slashed{k}}{k^2+i\epsilon} e^{-i k \cdot(-\lambda n_{z}-b)}\gamma^{z} \gamma^{5}u\bigg]\nonumber\\
&=&\frac{\alpha_s C_F}{2\pi }\bigg\{\frac{1}{(x-x_0)}\bigg[\frac{x}{x_0} \bigg(\frac{1}{\epsilon_{\rm IR}}+L_b\bigg)\theta(x_0-x)\theta(x) \bigg]_++\frac{1}{2}\delta(x_0-x)\left(\frac{1}{\epsilon_{\rm UV}}+L_b\right) \nonumber\\
&&-\frac{\delta(x-x_0)}{2  }\Bigg[\frac{L_b^2}{2}+\left(\ln\frac{-\zeta^z \pm i0}{\mu^2}-1\right)L_b+\frac{1}{2}\Bigg(\ln^2\frac{-\zeta^z \pm i0}{\mu^2}-2\ln\frac{-\zeta^z \pm i0}{\mu^2}+4\Bigg)+\frac{\pi^2}{2}\Bigg]\nonumber\\
&&+\{x_0\to \bar x_0,x \to \bar x\}\bigg\}.
\end{eqnarray}
For the vertex diagram Fig. \ref{fig:quasi-WF} (a) we obtain:
\begin{eqnarray}\label{eq:quasi-wave-func-vertex}
\tilde{\psi}_{\overline{q} q}^{\pm (1,a)}\left(x, b_{\perp}, \mu, \zeta^{z}\right)&=&\mu_0^{2\epsilon}\frac{ig^2C_F}{2} \int\frac{d \lambda}{2 \pi} e^{i (x-1)P^{z} \lambda} \int\frac{d^dq}{(2\pi)^d} \bar{v} \gamma_\mu \frac{\bar{x}_0\slashed{P}+\slashed{q}}{(\bar{x}_0P+q)^2 } \gamma^z\gamma^5 \frac{x_0\slashed{P}-\slashed{q}}{(x_0P-q)^2 }\gamma^\mu \frac{1}{q^2+i\epsilon} u e^{-iq\cdot b } e^{i(\bar{x}_0P^z+q^z)\lambda}\nonumber\\
&=&	\frac{\alpha_s C_F}{2\pi} \bigg(  \frac{1}{\epsilon_{\rm IR}}+L_b-1 \bigg) \bigg[ \bigg(-\frac{x}{x_0}\theta(x_0-x)\theta(x) \bigg)_+ 
-\frac{1}{4}\delta(x-x_0)+\{x_0\to \bar x_0,x \to \bar x\}\bigg].
\end{eqnarray}
For self-energy diagram Fig. \ref{fig:quasi-WF} (d) in large $L$ limit we obtain
\begin{eqnarray}
\label{eq:quasi-wave-func-selfenergy}
\tilde{\psi}_{\overline{q} q}^{\pm (1,d)}\left(x, b_{\perp}, \mu, \zeta^{z}\right)
&=&\frac{1}{4}\bar{v}(\bar{x}_0P)e^{-i\bar{x}_0P\cdot( \lambda n_z+b_\perp)}\gamma^{z} \gamma^{5}\int \frac{d \lambda}{2 \pi} e^{i (x-1)P^{z} \lambda}\left[ig\int_\mathcal{L}ds_1\mathcal{L}^\prime_\mu(s_1)    \right]\left[ig\int_\mathcal{L}ds_2\mathcal{L}^\prime_\mu(s_2)    \right] \nonumber\\
&&\times u(x_0P)\int\frac{d^dq}{(2\pi)^d}\frac{-ig^{\mu\nu}}{q^2+i\epsilon}e^{-iq\cdot[\mathcal{L}(s_1)-\mathcal{L}(s_2)]}\nonumber\\
&=&\frac{\alpha_s C_F}{4\pi}\delta(x-x_0)  \bigg[\frac{6}{{\epsilon}_{\rm UV}} +4+6L_b +    \frac{4\pi L}{b_\perp}\bigg].
\end{eqnarray}

Then consider the Wilson loop $Z_E$ which is the denominator of the quasi-TMDWF we defined before limit. At tree-level, the operator of matrix element only involves a $N_c \times N_c$ unit matrix. Therefore, for the tree-level of soft function it is easy to obtain $Z_E\left(2 L, b_{\perp}, \mu\right)=1$.
At one-loop QCD correction as shown in Fig.~\ref{fig:Wilson-loop} , the Wilson loop could be written as
\begin{eqnarray}\label{eq:definitionwilsonloop}
Z_E&=&1+\frac{1}{N_c}
	\left\langle 0\left|\frac{1}{ 2!}\left[ig\int_\mathcal{C} d\vec{s}_1 \cdot A\left(\vec{s}_1\right)\right]\left[ig\int_\mathcal{C} d\vec{s}_2 \cdot A\left(\vec{s}_2\right)\right] \right|0\right\rangle\nonumber\\
&=&1+\frac{g^2C_F}{8\pi^2}\mu_0^{2\epsilon}\pi^\epsilon\Gamma(1-\epsilon)\int_\mathcal{C}ds_1 \int_\mathcal{C}ds_2\left[ \mathcal{C}^\prime_\mu(s_1)\mathcal{C}^{\prime\mu}(s_2)    \right] \left[-\left(  \mathcal{C}(s_1)-\mathcal{C}(s_2) \right)^2\right]^{\epsilon-1},
\end{eqnarray}
where the integral on the route $\mathcal{C}$ could be divide into four parts as $\int_\mathcal{C}=\int_{\mathcal{C}_1}+\int_{\mathcal{C}_2}+\int_{\mathcal{C}_3}+\int_{\mathcal{C}_4}$ and each part itself have route
\begin{eqnarray}
\mathcal{C}_1(s)&=&-Ln_z+2Lsn_z,\nonumber\\
\mathcal{C}_2(s)&=&Ln_z+bs,\nonumber\\
\mathcal{C}_3(s)&=&b+Ln_z-2Lsn_z,\nonumber\\
\mathcal{C}_4(s)&=&b-Ln_z-bs.
\end{eqnarray}
Here the variable $s$ ranges from $0$ to $1$. By computing this integral directly, we have
\begin{eqnarray}\label{eq:wilson-loop}
Z_E&=&1+\frac{\alpha_s C_F}{2\pi}\Bigg[2\bigg(  \frac{1}{\epsilon_{\rm UV}}+\ln{\frac{L^2\mu^2}{e^{-2\gamma_E}}+2}\bigg)+2\bigg(  \frac{1}{\epsilon_{\rm UV}}+\ln{\frac{b_\perp^2\mu^2}{4e^{-2\gamma_E}}+2}\bigg)+2\bigg(  \frac{4L}{b_\perp}\arctan\frac{2L}{b_\perp} +\ln{\frac{b_\perp^2}{b_\perp^2+4L^2}} \bigg)\nonumber\\
&&+2\bigg(  \frac{b_\perp}{L}\arctan\frac{b_\perp}{2L} +\ln{\frac{4L^2}{b_\perp^2+4L^2}} \bigg)\Bigg].
\end{eqnarray}
After taking the $L\to \infty$ limit, the final result of wilson loop can be achieved, 
\begin{eqnarray}
Z_E= 1+\frac{\alpha_s C_F}{4\pi}\bigg(  \frac{8}{\epsilon_{\rm UV}}+8L_b + \frac{8L\pi}{b_\perp}  +8 \bigg).
\end{eqnarray}
One should note that the contribution of transverse and longitudinal Wilson lines to exchange gluons is $0$. That means
\begin{eqnarray}
\frac{1}{N_c}
	\left\langle 0\left|\frac{1}{ 2!}\left[ig\int_{\mathcal{C}_i} d\vec{s}_1 \cdot A\left(\vec{s}_1\right)\right]\left[ig\int_{\mathcal{C}_{i+1}} d\vec{s}_2 \cdot A\left(\vec{s}_2\right)\right] \right|0\right\rangle=0,\\
\frac{1}{N_c}
	\left\langle 0\left|\frac{1}{ 2!}\left[ig\int_{\mathcal{C}_{i+1}} d\vec{s}_1 \cdot A\left(\vec{s}_1\right)\right]\left[ig\int_{\mathcal{C}_i} d\vec{s}_2 \cdot A\left(\vec{s}_2\right)\right] \right|0\right\rangle=0
\end{eqnarray}
for $\mathcal{C}_i\in\mathcal{C}_{1,2,3,4}$.

\section{Lorentz structures in TMDWFs}
\label{Appendix:threshold expansion}

In the definition of  quasi-TMDWFs, one has two options for the Lorentz structures in the interpolating operator: $\gamma^z\gamma_5$,  and $\gamma^0\gamma_5$. In the main text we have presented the result for $\gamma^z \gamma_5$, but here we will show that the short-distance results for $\gamma^0\gamma_5$, namely the hard kernel,  are the same at least at one-loop level.

It is easy to see that the tree-level  TMDWFs for both structures are the delta function. At one-loop level, it is also obvious that the virtual corrections in Fig.~(\ref{fig:quasi-WF}d) are the same for the two Lorentz structures.

For the diagram Fig.~(\ref{fig:quasi-WF}b, c) on $l$ direction (where $l=z$ or $0$), we obtain the matrix element: 
\begin{eqnarray}\label{eq:appendixsail}
\tilde{\psi}_{\overline{q} q}^{\pm (1,bc)}&=&-\frac{g^2C_F}{2} \int \frac{d \lambda}{2 \pi} e^{i (x-1)P^{z} \lambda}\int_0^1 ds \mathcal{C}^{\prime}(s)_\mu \bigg[ e^{-i \bar{x}_0 \lambda P\cdot  n_{z}}\int \frac{d^d k}{(2\pi)^d}\bar v \gamma^{l} \gamma^{5}\frac{e^{-i (x_0P-k) \cdot \mathcal{C}(s)}}{(x_0P-k)^2+i\epsilon}  \frac{\slashed{k}}{k^2+i\epsilon}\gamma^\mu u \nonumber \\
&& +\int \frac{d^4 k}{(2\pi)^4}\bar v \gamma^\mu \frac{e^{-i (\bar{x}_0P+k) \cdot \mathcal{C}(s)}}{(\bar{x}_0P+k)^2+i\epsilon}  \frac{\slashed{k}}{k^2+i\epsilon} e^{-i k \cdot(-\lambda n_{z}-b)}\gamma^{l} \gamma^{5}u\bigg],
\end{eqnarray}
where $\mathcal{C}(s)$ is the route of the Wilson line. The spinor structures in Eq.~(\ref{eq:appendixsail}) are 
\begin{eqnarray}
\bar v \gamma^\mu  \slashed{k} \gamma^{z} \gamma^{5} u &=& \bar v \gamma^\mu \slashed{k} \frac{\slashed{n}+\slashed{\bar{n}}}{\sqrt{2}} \gamma^{5}  u
=\bar v \gamma^\mu \slashed{k} \frac{\slashed{n}}{\sqrt{2}} \gamma^{5}  u ,\\
\bar v \gamma^\mu \slashed{k} \gamma^{0} \gamma^{5}  u &=& \bar v \gamma^\mu \slashed{k} \frac{\slashed{n}-\slashed{\bar{n}}}{\sqrt{2}} \gamma^{5}  u
=\bar v \gamma^\mu \slashed{k} \frac{\slashed{n}}{\sqrt{2}} \gamma^{5}  u . 
\end{eqnarray}
We find that the result of this diagram is independent of the Lorentz structure.

The vertex diagram Fig.~\ref{fig:quasi-WF} (a) on $l$ direction (where $l=z$ or $0$), we
have the amplitude: 
\begin{eqnarray}\label{eq:fig:quasi-WF_a}
\tilde{\psi}_{\overline{q} q}^{\pm (1,a)}&=&\mu_0^{2\epsilon}i\frac{g^2C_F}{2} (\bar{u}\gamma^l\gamma^5 v) \int\frac{d^dq}{(2\pi)^d} \frac{\frac{D-2}{P^z}[(\bar{x}_0P+q)^2q^l-(x_0P-q)^2q^l-P^lq^2]}{[(\bar{x}_0P+q)^2+i\epsilon][(x_0P-q)^2+i\epsilon](q^2+i\epsilon)}e^{-iq\cdot b } \delta\bigg[ (x-x_0)P^z+q^z \bigg]. 
\end{eqnarray}
A brutal-force evaluation of this amplitude indicates the equivalence for the two Lorentz structures, but in the following we adopt the expansion by regions technique. Explicitly, we will  demonstrate in this diagram only the collinear modes contribute. 

In the quasi-TMDWFs, there are three typical models according to the decomposition of the momentum $q=(q^+,q_\perp,q^-)$, \begin{itemize}
\item Hard mode  with  $q\sim(1,1,1)P^z$:

The amplitude in Eq.~\eqref{eq:fig:quasi-WF_a}  contains an exponential factor $e^{iq_\perp\cdot b_\perp}$, which is oscillating in the region $1/b_\perp\ll P^z$. After the integration over the $q$, the final result is power suppressed accordingly. 

\item Collinear mode with  $q\sim(Q,\Lambda_{\rm QCD},\Lambda^2_{\rm QCD}/Q)$: 

In this region, one can find that the amplitude is $\mathcal O(Q)$, and actually the amplitudes for both structures are reduced to the TMDWF: 
\begin{eqnarray}
\tilde{\psi}_{\overline{q} q}^{\pm (1,a)}&=&\mu_0^{2\epsilon}i\frac{g^2C_F}{2} (\bar{u}\gamma^l\gamma^5 v) \int\frac{d^dq}{(2\pi)^d} \frac{D-2}{P^z} \frac{-P^l q^2_\perp}{[(\bar{x}_0P+q)^2+i\epsilon][(x_0P-q)^2+i\epsilon](q^2+i\epsilon)}e^{-iq\cdot b } \sqrt{2} \delta \bigg[ (x-x_0)P^++q^+ \bigg]\nonumber. 
\end{eqnarray}

\item Soft mode  $q\sim(\Lambda_{\rm QCD},\Lambda_{\rm QCD},\Lambda_{\rm QCD})$: 

In this kinematics region, one can find the power of this amplitude is   $\mathcal O(\Lambda^3_{\rm QCD}/Q^2)$, and namely this amplitude is suppressed. 
\end{itemize}

This analysis indicates that the amplitude from  Fig.~\ref{fig:quasi-WF} (a) is independent of the Lorentz structure, and moreover we have checked that the expansion by regions technique can be used to demonstrate the multiplicative factorization of the quasi-TMDWFs. 
\end{widetext}

\end{appendix}


\begin{thebibliography}{}
\bibitem{Lepage:1979zb}
G.~P.~Lepage and S.~J.~Brodsky,
Phys. Lett. B \textbf{87}, 359-365 (1979)
doi:10.1016/0370-2693(79)90554-9.

\bibitem{Lepage:1980fj}
G.~P.~Lepage and S.~J.~Brodsky,
Phys. Rev. D \textbf{22}, 2157 (1980)
doi:10.1103/PhysRevD.22.2157.

\bibitem{Brodsky:1997de}
S.~J.~Brodsky, H.~C.~Pauli and S.~S.~Pinsky,
Phys. Rept. \textbf{301}, 299-486 (1998)
doi:10.1016/S0370-1573(97)00089-6
[arXiv:hep-ph/9705477 [hep-ph]].

\bibitem{Politzer:1982mf}
H.~D.~Politzer,
Phys. Lett. B \textbf{116}, 171-174 (1982)
doi:10.1016/0370-2693(82)91002-4.

\bibitem{Miransky:1985wzx}
V.~A.~Miransky,
Phys. Lett. B \textbf{165}, 401-404 (1985)
doi:10.1016/0370-2693(85)91254-7.

\bibitem{Alkofer:2008tt}
R.~Alkofer, C.~S.~Fischer, F.~J.~Llanes-Estrada and K.~Schwenzer,
Annals Phys. \textbf{324}, 106-172 (2009)
doi:10.1016/j.aop.2008.07.001
[arXiv:0804.3042 [hep-ph]].

\bibitem{Serna:2018dwk}
F.~E.~Serna, C.~Chen and B.~El-Bennich,
Phys. Rev. D \textbf{99}, no.9, 094027 (2019)
doi:10.1103/PhysRevD.99.094027
[arXiv:1812.01096 [hep-ph]].

\bibitem{Li:1994iu}
H.~n.~Li and H.~L.~Yu,
Phys. Rev. D \textbf{53}, 2480-2490 (1996)
doi:10.1103/PhysRevD.53.2480
[arXiv:hep-ph/9411308 [hep-ph]].

\bibitem{Keum:2000wi}
Y.~Y.~Keum, H.~N.~Li and A.~I.~Sanda,
Phys. Rev. D \textbf{63}, 054008 (2001)
doi:10.1103/PhysRevD.63.054008
[arXiv:hep-ph/0004173 [hep-ph]].

\bibitem{Keum:2000ph}
Y.~Y.~Keum, H.~n.~Li and A.~I.~Sanda,
Phys. Lett. B \textbf{504}, 6-14 (2001)
doi:10.1016/S0370-2693(01)00247-7
[arXiv:hep-ph/0004004 [hep-ph]].

\bibitem{Lu:2000em}
C.~D.~Lu, K.~Ukai and M.~Z.~Yang,
Phys. Rev. D \textbf{63}, 074009 (2001)
doi:10.1103/PhysRevD.63.074009
[arXiv:hep-ph/0004213 [hep-ph]].

\bibitem{Beneke:2001ev}
M.~Beneke, G.~Buchalla, M.~Neubert and C.~T.~Sachrajda,
Nucl. Phys. B \textbf{606}, 245-321 (2001)
doi:10.1016/S0550-3213(01)00251-6
[arXiv:hep-ph/0104110 [hep-ph]].

\bibitem{Shifman:1978bx}
M.~A.~Shifman, A.~I.~Vainshtein and V.~I.~Zakharov,
Nucl. Phys. B \textbf{147}, 385-447 (1979)
doi:10.1016/0550-3213(79)90022-1

\bibitem{Chernyak:1983ej}
V.~L.~Chernyak and A.~R.~Zhitnitsky,
Phys. Rept. \textbf{112}, 173 (1984)
doi:10.1016/0370-1573(84)90126-1

\bibitem{Gockeler:2005jz}
M.~Gockeler, R.~Horsley, D.~Pleiter, P.~E.~L.~Rakow, A.~Schafer, G.~Schierholz, W.~Schroers and J.~M.~Zanotti,
Nucl. Phys. B Proc. Suppl. \textbf{161}, 69-74 (2006)
doi:10.1016/j.nuclphysbps.2006.08.064
[arXiv:hep-lat/0510089 [hep-lat]].

\bibitem{Braun:2006dg}
V.~M.~Braun, M.~Gockeler, R.~Horsley, H.~Perlt, D.~Pleiter, P.~E.~L.~Rakow, G.~Schierholz, A.~Schiller, W.~Schroers and H.~Stuben, \textit{et al.}
Phys. Rev. D \textbf{74}, 074501 (2006)
doi:10.1103/PhysRevD.74.074501
[arXiv:hep-lat/0606012 [hep-lat]].

\bibitem{Boyle:2006pw}
P.~A.~Boyle \textit{et al.} [UKQCD],
Phys. Lett. B \textbf{641}, 67-74 (2006)
doi:10.1016/j.physletb.2006.07.033
[arXiv:hep-lat/0607018 [hep-lat]].

\bibitem{Arthur:2010xf}
R.~Arthur, P.~A.~Boyle, D.~Brommel, M.~A.~Donnellan, J.~M.~Flynn, A.~Juttner, T.~D.~Rae and C.~T.~C.~Sachrajda,
Phys. Rev. D \textbf{83}, 074505 (2011)
doi:10.1103/PhysRevD.83.074505
[arXiv:1011.5906 [hep-lat]].

\bibitem{Braun:2015axa}
V.~M.~Braun, S.~Collins, M.~G\"ockeler, P.~P\'erez-Rubio, A.~Sch\"afer, R.~W.~Schiel and A.~Sternbeck,
Phys. Rev. D \textbf{92}, no.1, 014504 (2015)
doi:10.1103/PhysRevD.92.014504
[arXiv:1503.03656 [hep-lat]].

\bibitem{Bali:2017ude}
G.~S.~Bali \textit{et al.} [RQCD],
Phys. Lett. B \textbf{774}, 91-97 (2017)
doi:10.1016/j.physletb.2017.08.077
[arXiv:1705.10236 [hep-lat]].

\bibitem{RQCD:2019osh}
G.~S.~Bali \textit{et al.} [RQCD],
JHEP \textbf{08}, 065 (2019)
doi:10.1007/JHEP08(2019)065
[arXiv:1903.08038 [hep-lat]].

\bibitem{Ji:2013dva}
X.~Ji,
Phys. Rev. Lett. \textbf{110}, 262002 (2013)
doi:10.1103/PhysRevLett.110.262002
[arXiv:1305.1539 [hep-ph]].

\bibitem{Ji:2014gla}
X.~Ji,
Sci. China Phys. Mech. Astron. \textbf{57}, 1407-1412 (2014)
doi:10.1007/s11433-014-5492-3
[arXiv:1404.6680 [hep-ph]].

\bibitem{Cichy:2018mum}
K.~Cichy and M.~Constantinou,
Adv. High Energy Phys. \textbf{2019}, 3036904 (2019)
doi:10.1155/2019/3036904
[arXiv:1811.07248 [hep-lat]].

\bibitem{Ji:2020ect}
X.~Ji, Y.~S.~Liu, Y.~Liu, J.~H.~Zhang and Y.~Zhao,
Rev. Mod. Phys. \textbf{93}, no.3, 035005 (2021)
doi:10.1103/RevModPhys.93.035005
[arXiv:2004.03543 [hep-ph]].

\bibitem{Li:1992nu}
H.~n.~Li and G.~F.~Sterman,
Nucl. Phys. B \textbf{381}, 129-140 (1992)
doi:10.1016/0550-3213(92)90643-P

\bibitem{Efremov:1979qk}
A.~V.~Efremov and A.~V.~Radyushkin,
Phys. Lett. B \textbf{94}, 245-250 (1980)
doi:10.1016/0370-2693(80)90869-2

\bibitem{Aznaurian:1979zz}
I.~G.~Aznaurian, S.~V.~Esaibegian and N.~L.~Ter-Isaakian,
Phys. Lett. B \textbf{90}, 151 (1980)
[erratum: Phys. Lett. B \textbf{92}, 371-371 (1980)]
doi:10.1016/0370-2693(80)90072-6

\bibitem{Li:1992ce}
H.~n.~Li,
Phys. Rev. D \textbf{48}, 4243-4254 (1993)
doi:10.1103/PhysRevD.48.4243

\bibitem{Duncan:1979hi}
A.~Duncan and A.~H.~Mueller,
Phys. Rev. D \textbf{21}, 1636 (1980)
doi:10.1103/PhysRevD.21.1636

\bibitem{Lepage:1979za}
G.~P.~Lepage and S.~J.~Brodsky,
Phys. Rev. Lett. \textbf{43}, 545-549 (1979)
[erratum: Phys. Rev. Lett. \textbf{43}, 1625-1626 (1979)]
doi:10.1103/PhysRevLett.43.545

\bibitem{Li:2012nk}
H.~n.~Li, Y.~L.~Shen and Y.~M.~Wang,
Phys. Rev. D \textbf{85}, 074004 (2012)
doi:10.1103/PhysRevD.85.074004
[arXiv:1201.5066 [hep-ph]].

\bibitem{Ebert:2019okf}
M.~A.~Ebert, I.~W.~Stewart and Y.~Zhao,
JHEP \textbf{09}, 037 (2019)
doi:10.1007/JHEP09(2019)037
[arXiv:1901.03685 [hep-ph]].

\bibitem{Ebert:2019tvc}
M.~A.~Ebert, I.~W.~Stewart and Y.~Zhao,
JHEP \textbf{03}, 099 (2020)
doi:10.1007/JHEP03(2020)099
[arXiv:1910.08569 [hep-ph]].

\bibitem{Ji:2019sxk}
X.~Ji, Y.~Liu and Y.~S.~Liu,
Nucl. Phys. B \textbf{955}, 115054 (2020)
doi:10.1016/j.nuclphysb.2020.115054
[arXiv:1910.11415 [hep-ph]].

\bibitem{Ji:2019ewn}
X.~Ji, Y.~Liu and Y.~S.~Liu,
Phys. Lett. B \textbf{811}, 135946 (2020)
doi:10.1016/j.physletb.2020.135946
[arXiv:1911.03840 [hep-ph]].

\bibitem{Ji:2021znw}
X.~Ji and Y.~Liu,
Phys. Rev. D \textbf{105}, no.7, 076014 (2022)
doi:10.1103/PhysRevD.105.076014
[arXiv:2106.05310 [hep-ph]].

\bibitem{Collins:1981va}
J.~C.~Collins and D.~E.~Soper,
Nucl. Phys. B \textbf{197}, 446-476 (1982)
doi:10.1016/0550-3213(82)90453-9

\bibitem{Shanahan:2020zxr}
P.~Shanahan, M.~Wagman and Y.~Zhao,
Phys. Rev. D \textbf{102}, no.1, 014511 (2020)
doi:10.1103/PhysRevD.102.014511
[arXiv:2003.06063 [hep-lat]].

\bibitem{LatticeParton:2020uhz}
Q.~A.~Zhang \textit{et al.} [Lattice Parton],
Phys. Rev. Lett. \textbf{125}, no.19, 192001 (2020)
doi:10.22323/1.396.0477
[arXiv:2005.14572 [hep-lat]].

\bibitem{Schlemmer:2021aij}
M.~Schlemmer, A.~Vladimirov, C.~Zimmermann, M.~Engelhardt and A.~Sch\"afer,
JHEP \textbf{08}, 004 (2021)
doi:10.1007/JHEP08(2021)004
[arXiv:2103.16991 [hep-lat]].

\bibitem{Li:2021wvl}
Y.~Li, S.~C.~Xia, C.~Alexandrou, K.~Cichy, M.~Constantinou, X.~Feng, K.~Hadjiyiannakou, K.~Jansen, C.~Liu and A.~Scapellato, \textit{et al.}
Phys. Rev. Lett. \textbf{128}, no.6, 062002 (2022)
doi:10.1103/PhysRevLett.128.062002
[arXiv:2106.13027 [hep-lat]].

\bibitem{Shanahan:2021tst}
P.~Shanahan, M.~Wagman and Y.~Zhao,
Phys. Rev. D \textbf{104}, no.11, 114502 (2021)
doi:10.1103/PhysRevD.104.114502
[arXiv:2107.11930 [hep-lat]].

\bibitem{LPC:2022ibr}
M.~H.~Chu \textit{et al.} [LPC],
[arXiv:2204.00200 [hep-lat]].

\bibitem{Echevarria:2015usa}
M.~G.~Echevarria, I.~Scimemi and A.~Vladimirov,
Phys. Rev. D \textbf{93}, no.1, 011502 (2016)
[erratum: Phys. Rev. D \textbf{94}, no.9, 099904 (2016)]
doi:10.1103/PhysRevD.93.011502
[arXiv:1509.06392 [hep-ph]].

\bibitem{Echevarria:2015byo}
M.~G.~Echevarria, I.~Scimemi and A.~Vladimirov,
Phys. Rev. D \textbf{93}, no.5, 054004 (2016)
doi:10.1103/PhysRevD.93.054004
[arXiv:1511.05590 [hep-ph]].

\bibitem{Wilson:1974sk}
K.~G.~Wilson,
Phys. Rev. D \textbf{10}, 2445-2459 (1974)
doi:10.1103/PhysRevD.10.2445

\bibitem{Lenz:1991sa}
F.~Lenz, M.~Thies, K.~Yazaki and S.~Levit,
Annals Phys. \textbf{208}, 1-89 (1991)
doi:10.1016/0003-4916(91)90342-6

\bibitem{Bauer:2000yr}
C.~W.~Bauer, S.~Fleming, D.~Pirjol and I.~W.~Stewart,
Phys. Rev. D \textbf{63}, 114020 (2001)
doi:10.1103/PhysRevD.63.114020
[arXiv:hep-ph/0011336 [hep-ph]].

\bibitem{Ji:2004wu}
X.~d.~Ji, J.~p.~Ma and F.~Yuan,
Phys. Rev. D \textbf{71}, 034005 (2005)
doi:10.1103/PhysRevD.71.034005
[arXiv:hep-ph/0404183 [hep-ph]].

\bibitem{Collins:2011ca}
J.~Collins,
Int. J. Mod. Phys. Conf. Ser. \textbf{4}, 85-96 (2011)
doi:10.1142/S2010194511001590
[arXiv:1107.4123 [hep-ph]].

\bibitem{Collins:2004nx}
J.~C.~Collins and A.~Metz,
Phys. Rev. Lett. \textbf{93}, 252001 (2004)
doi:10.1103/PhysRevLett.93.252001
[arXiv:hep-ph/0408249 [hep-ph]].

\bibitem{Echevarria:2012js}
M.~G.~Echevarr\'\i{}a, A.~Idilbi and I.~Scimemi,
Phys. Lett. B \textbf{726}, 795-801 (2013)
doi:10.1016/j.physletb.2013.09.003
[arXiv:1211.1947 [hep-ph]].

\bibitem{Echevarria:2011epo}
M.~G.~Echevarria, A.~Idilbi and I.~Scimemi,
JHEP \textbf{07}, 002 (2012)
doi:10.1007/JHEP07(2012)002
[arXiv:1111.4996 [hep-ph]].

\bibitem{Kotikov:1990kg}
A.~V.~Kotikov,
Phys. Lett. B \textbf{254}, 158-164 (1991)
doi:10.1016/0370-2693(91)90413-K

\bibitem{Kotikov:1991pm}
A.~V.~Kotikov,
Phys. Lett. B \textbf{267}, 123-127 (1991)
[erratum: Phys. Lett. B \textbf{295}, 409-409 (1992)]
doi:10.1016/0370-2693(91)90536-Y

\bibitem{Remiddi:1997ny}
E.~Remiddi,
Nuovo Cim. A \textbf{110}, 1435-1452 (1997)
doi:10.1007/BF03185566
[arXiv:hep-th/9711188 [hep-th]].

\bibitem{Gehrmann:1999as}
T.~Gehrmann and E.~Remiddi,
Nucl. Phys. B \textbf{580}, 485-518 (2000)
doi:10.1016/S0550-3213(00)00223-6
[arXiv:hep-ph/9912329 [hep-ph]].

\bibitem{Argeri:2007up}
M.~Argeri and P.~Mastrolia,
Int. J. Mod. Phys. A \textbf{22}, 4375-4436 (2007)
doi:10.1142/S0217751X07037147
[arXiv:0707.4037 [hep-ph]].

\bibitem{Henn:2013pwa}
J.~M.~Henn,
Phys. Rev. Lett. \textbf{110}, 251601 (2013)
doi:10.1103/PhysRevLett.110.251601
[arXiv:1304.1806 [hep-th]].

\bibitem{Henn:2013nsa}
J.~M.~Henn, A.~V.~Smirnov and V.~A.~Smirnov,
JHEP \textbf{03}, 088 (2014)
doi:10.1007/JHEP03(2014)088
[arXiv:1312.2588 [hep-th]].

\bibitem{Argeri:2014qva}
M.~Argeri, S.~Di Vita, P.~Mastrolia, E.~Mirabella, J.~Schlenk, U.~Schubert and L.~Tancredi,
JHEP \textbf{03}, 082 (2014)
doi:10.1007/JHEP03(2014)082
[arXiv:1401.2979 [hep-ph]].

\bibitem{Henn:2014qga}
J.~M.~Henn,
J. Phys. A \textbf{48}, 153001 (2015)
doi:10.1088/1751-8113/48/15/153001
[arXiv:1412.2296 [hep-ph]].



\bibitem{Kai_Yan}
Kai Yan, to appear. 

\bibitem{Collins:2017oxh}
J.~Collins and T.~C.~Rogers,
Phys. Rev. D \textbf{96}, no.5, 054011 (2017)
doi:10.1103/PhysRevD.96.054011
[arXiv:1705.07167 [hep-ph]].

\bibitem{Bateman:1953htf}
Harry B. and Arthur E. , Higher Transcendental Functions, Vol. I. New York: McGraw–Hill. 

\bibitem{Collins:2011zzd}
J.~Collins,
Camb. Monogr. Part. Phys. Nucl. Phys. Cosmol. \textbf{32}, 1-624 (2011)

\bibitem{Lu:2007hr}
C.~D.~Lu, W.~Wang and Y.~M.~Wang,
Phys. Rev. D \textbf{75}, 094020 (2007)
doi:10.1103/PhysRevD.75.094020
[arXiv:hep-ph/0702085 [hep-ph]].

\bibitem{Bakulev:2001pa}
A.~P.~Bakulev, S.~V.~Mikhailov and N.~G.~Stefanis,
Phys. Lett. B \textbf{508}, 279-289 (2001)
[erratum: Phys. Lett. B \textbf{590}, 309-310 (2004)]
doi:10.1016/S0370-2693(01)00517-2
[arXiv:hep-ph/0103119 [hep-ph]].
\end{thebibliography}
\end{document}